\documentclass[11pt,english]{article}
\usepackage[T1]{fontenc}
\usepackage[latin9]{inputenc}
\usepackage[letterpaper]{geometry}
\usepackage[active]{srcltx}
\usepackage{color}
\usepackage{babel}
\usepackage{booktabs}
\usepackage{mathrsfs}
\usepackage{mathtools}
\usepackage{amsmath}
\usepackage{amsthm}
\usepackage{amssymb}
\usepackage{graphicx}
\usepackage{setspace}
\usepackage{wasysym}
\usepackage[authoryear]{natbib}
\usepackage[unicode=true,pdfusetitle,
 bookmarks=true,bookmarksnumbered=false,bookmarksopen=false,
 breaklinks=false,pdfborder={0 0 0},pdfborderstyle={},backref=false,colorlinks=true]
 {hyperref}
\hypersetup{
 linkcolor=blue, citecolor=blue}

\makeatletter

\providecommand{\tabularnewline}{\\}

\theoremstyle{plain}
\newtheorem{assumption}{\protect\assumptionname}
\theoremstyle{plain}
\newtheorem{thm}{\protect\theoremname}
\theoremstyle{plain}
\newtheorem{lyxalgorithm}{\protect\algorithmname}
\theoremstyle{plain}
\newtheorem{lem}{\protect\lemmaname}

\allowdisplaybreaks

\usepackage{bbm}

\date{}

\newtheoremstyle{remboldstyle}
  {}{}{}{}{\bfseries}{.}{.5em}{{\thmname{#1 }}{\thmnumber{#2}}{\thmnote{ (#3)}}}
\theoremstyle{remboldstyle}
\newtheorem{rembold}{Remark}

\usepackage{appendix}
\appendixtitleon

\usepackage{xr}
\usepackage{scalerel,stackengine}
\stackMath
\newcommand\reallywidecheck[1]{%
\savestack{\tmpbox}{\stretchto{%
  \scaleto{%
    \scalerel*[\widthof{\ensuremath{#1}}]{\kern-.6pt\bigwedge\kern-.6pt}%
    {\rule[-\textheight/2]{1ex}{\textheight}}
  }{\textheight}%
}{0.5ex}}%
\stackon[1pt]{#1}{\scalebox{-1}{\tmpbox}}%
}
\date{}
\allowdisplaybreaks

\makeatletter
\renewenvironment{proof}[1][\proofname]{%
  \par\pushQED{\qed}\normalfont%
  \topsep6\p@\@plus6\p@\relax
  \trivlist\item[\hskip\labelsep\bfseries#1\@addpunct{.}]%
  \ignorespaces
}{%
  \popQED\endtrivlist\@endpefalse
}
\makeatother

\newtheoremstyle{ntnboldstyle}
  {}{}{}{}{\bfseries}{.}{.5em}{{\thmname{#1}}{\thmnote{(#2)}}}
\theoremstyle{ntnboldstyle}

\makeatletter
\def\@seccntformat#1{\@ifundefined{#1@cntformat}%
   {\csname the#1\endcsname\quad}  
   {\csname #1@cntformat\endcsname}
}
\let\oldappendix\appendix 
\renewcommand\appendix{%
    \oldappendix
    \newcommand{\section@cntformat}{\appendixname~\thesection\quad}
}
\makeatother

\makeatother

\providecommand{\algorithmname}{Algorithm}
\providecommand{\assumptionname}{Assumption}
\providecommand{\lemmaname}{Lemma}
\providecommand{\theoremname}{Theorem}

\begin{document}
\title{\textbf{Inference on Individual Treatment Effects in Nonseparable
Triangular Models}\let\thefootnote\relax\footnotetext{\today.}\thanks{We thank our coeditor, Xiaohong Chen, the associate editor and two
anonymous referees, whose comments have greatly improved the paper.
This research is supported by the National Natural Science Foundation
of China under grant 71903190, fund for building world-class universities
(disciplines) of Renmin University of China (Ma), and Japan Society
for the Promotion of Science KAKENHI Grant Number 21K01419 (Yu).}}
\author{Jun Ma\thanks{School of Economics, Renmin University of China, P.R. China. Email:
jun.ma@ruc.edu.cn}\and Vadim Marmer\thanks{Vancouver School of Economics, University of British Columbia, Canada.
Email: vadim.marmer@ubc.ca}\and Zhengfei Yu\thanks{Faculty of Humanities and Social Sciences, University of Tsukuba,
Japan. Email: yu.zhengfei.gn@u.tsukuba.ac.jp}}
\maketitle
\begin{abstract}
In nonseparable triangular models with a binary endogenous treatment
and a binary instrumental variable, \citet{vuong2017counterfactual}
and \citet*{feng2019estimation} respectively provide the identification
results for the individual treatment effects (ITEs) under the rank
invariance assumption and propose a uniformly consistent kernel density
estimator using estimated ITEs for the density of the ITE. This paper
establishes the asymptotic normality of the density estimator of \citet*{feng2019estimation}
and shows that the estimation error of the ITEs that vanishes at the
root $n$ rate has a non-negligible effect on the asymptotic distribution.
We propose asymptotically valid standard errors that account for estimated
ITEs, as well as a bias correction. Furthermore, we develop uniform
confidence bands for the density of the ITE using the jackknife multiplier
or nonparametric bootstrap critical values.
\end{abstract}

\section{Introduction}

Heterogeneous treatment effects have received increasing attention
in the causal inference and policy evaluation literature (\citealp{angrist2004treatment,Heckman1997,Heckman2006}).
There is a vast literature studying the causal effect of ceteris paribus
change of a treatment variable using triangular models (see, e.g.,
\citealp{Chesher2003,Chesher2005,d2015identification,Imbens2009,Jun2011,Newey1999,Torgovitsky2015,Vytlacil2007}
among others). In a triangular model, the outcome variable is generated
by an outcome equation, and a selection equation determines the endogenous
treatment variable. Recently, \citet[VX, hereafter]{vuong2017counterfactual}
and \citet[FVX, hereafter]{feng2019estimation} developed nonparametric
identification and estimation methods for individual treatment effects
(ITEs) in a triangular model with a nonseparable outcome equation,
a selection equation which is a latent index model (\citealp{Vytlacil2002}),
a binary endogenous treatment variable and a binary instrument under
the rank invariance assumption. VX assumes that disturbances in both
equations are scalar-valued and that the outcome is a strictly monotone
function of the disturbance. The outcome equation in VX satisfies
the rank invariance assumption (i.e., for given covariates, the ranks
of the two potential outcomes are the same). See, e.g., \citet{chernozhukov2020instrumental}
for discussion of this assumption. The triangular model considered
in VX and FVX is also closely related to the classical local average
treatment effect (LATE) model (see, e.g., \citealp{Abadie2002,Abadie2003,Froelich2013,imbens1994identification}
among others) and the instrumental variable quantile regression (IVQR)
model (see, e.g., \citealp{Chernozhukov2005} among others).\footnote{A more detailed literature review about the triangular model with
an endogenous treatment can be found in our online supplement available
at: \href{http://ruc-econ.github.io/ITE_Supp_Rev_V13.pdf}{ruc-econ.github.io/ITE\_Supp\_Rev\_V13.pdf}.}

Since the ITE is defined and estimated for each individual, it is
natural to focus on its probability density function (PDF) when assessing
the heterogeneity of a treatment. For that purpose, FVX uses the conventional
kernel density estimator applied to estimated ITEs. They show its
uniform consistency as well as derive its rate of convergence. However,
when it comes to inference for the density of the ITE, two theoretical
problems still need to be solved. First, asymptotically valid standard
errors for the density estimator should incorporate the uncertainty
stemming from the estimation of ITEs. Second, as researchers are often
interested in the shape of the distribution of the ITE, it is important
to have asymptotically valid uniform confidence bands (UCBs) for the
density of the ITE. Our paper contributes to the literature by providing
easy-to-implement solutions to both problems.

We first provide a sharper bound for the uniform rate of convergence
of the FVX density estimator and show that it attains the optimal
rate under seemingly minimal conditions. We then show the asymptotic
normality of the FVX estimator and derive an analytical formula for
the standard error of the density estimator that incorporates the
influence of ITEs' estimation in the first step of the FVX procedure.

The asymptotic normality result in this paper is non-trivial as the
asymptotic distribution of the FVX estimator is different from that
of the infeasible estimator based on true unobserved ITEs. FVX uses
nonparametrically estimated counterfactual mappings to generate pseudo
(i.e., estimated) ITEs in the first step of their procedure. In the
second step, they apply kernel density estimation to the pseudo ITEs
to construct an estimator for the PDF of the ITE. While the estimated
counterfactual mappings converge at the root $n$ rate, we show that
the first-step estimation errors' contribution to the density estimator's
asymptotic variance is non-negligible and can substantially dominate
that of the second stage. This phenomenon is due to discontinuities
in the linearization of the first-step estimator. At the same time,
the asymptotic bias is unaffected by the first-stage estimation errors
and equal to that of the infeasible estimator.

The paper's second contribution is to propose asymptotically valid
UCBs for the density of the ITE. The proposed UCB captures the uncertainty
about the entire estimated density function and, therefore, can be
used for inference about the shape of the ITE's distribution (e.g.,
the number and locations of the modes) and for comparisons between
the distributions of the ITE in different sub-populations. UCBs can
also be used for the specification of parametric models. Our bootstrap
UCBs have the desirable property of polynomial coverage error decay
rates. Following \citet{Calonico2014}, we also propose bias-corrected
UCBs using standard errors that incorporate additional variability
from the estimated bias (Section \ref{subsec:Bias-corrected-JMB}).
This approach is common in the recent literature, as it validates
the use of conventional data-driven bandwidth selectors for inference.
In addition, we consider an extension to inference on PDF conditional
on a sub-vector of covariates (Section \ref{subsec:Conditioning-on-sub-vectors}).

The FVX estimator and our UCBs require bandwidth selection. Our results
explicitly allow for random data-dependent bandwidths. Following the
literature (e.g., \citealp{hsiao2007consistent,Li:2010dy}) and the
standard practice in applied work, we assume that the data-dependent
bandwidth consistently estimates some deterministic bandwidth. Similarly
to the existing literature, we verify that the uniform rate of convergence
and asymptotic normality results are unaffected by the bandwidth estimation.
However, we go a step further and provide an explicit estimate of
the effect of bandwidth estimation on the coverage error of the UCBs.
The result is new for this literature and adds to our understanding
of the impact of bandwidth selection on inferential procedures.

In the empirical section of the paper, we use the FVX estimator with
our UCBs to study the effect of having more than two children on their
parents' labor income using the instrument proposed in \citet{angrist1998children}.
We show that the conditional distribution of the ITE is significantly
different between households with high-school-only and college-educated
mothers. In the latter case, the ITE's distribution is more dispersed.
However, it also places significantly more weight on positive effects.

From the perspective of nonparametric inference, our paper contributes
to the literature on inference in the presence of nonparametrically
generated variables. See, e.g., \citet{Mammen2012} and \citet{Ma2019}.
The asymptotic theory of the FVX estimator is different from the results
obtained in \citet{Mammen2012} and \citet{Ma2019}. E.g., among their
other results, \citet{Mammen2012} show that the influence of variables'
estimation can be made asymptotically negligible using a proper choice
of the bandwidth. However, in the case of the FVX estimator, the influence
of ITEs' estimation has a non-negligible effect regardless of the
bandwidth choice. \citet{Ma2019} show that in the context of first-price
auctions, the estimation of latent bidders' valuations contributes
to the asymptotic variance of the \citet{Guerre2000} estimator for
the density of valuations. Moreover, the variance of the \citet{Guerre2000}
estimator has a slower decay rate than that of the infeasible estimator
constructed using the true latent valuations. Here, we show that while
the variance of the FVX estimator has the same decay rate as that
of the infeasible estimator, the estimation of ITEs cannot be ignored.
Our paper also contributes to the literature on inference for nonparametrically
estimated functions. In this paper, we take the intermediate Gaussian
approximation approach to show the asymptotic validity of our UCBs
by using tools developed by \citet{Chernozhukov2014gaussian,Chernozhukov2014anti,Chernozhukov2016,Chen2020jackknife}.\footnote{Under certain conditions, it can often be shown that the suprema of
estimation errors can be approximated by the suprema of tight Gaussian
random elements using the theorems of \citet{Chen2020jackknife,Chernozhukov2014gaussian}.
Then theorems in \citet{Chen2020jackknife,Chernozhukov2014anti,Chernozhukov2016}
show that the distributions of these Gaussian suprema can be approximated
by bootstrapping.} This approach was recently applied, e.g., in \citet{chen2018optimal,Cheng2019,Kato2019,Ma2019}
among others, to show the asymptotic validity of bootstrap UCBs for
various nonparametric curves in different contexts.

The rest of the paper is organized as follows. Section \ref{sec:Model-and-identification}
reviews the model setup and identification of the distribution of
the ITE and discusses the nonparametric estimation of the density
of the ITE. Section \ref{sec:Asymptotic-properties} shows the density
estimator's uniform convergence rate and asymptotic normality. Section
\ref{sec:Robust-inference} provides standard errors for the density
estimator that can be used for asymptotically valid inference. It
also establishes the validity of the bootstrap UCBs. Section \ref{subsec:Jackknife-multiplier-bootstrap}
describes the algorithm for our proposed confidence band. Section
\ref{sec:Monte-Carlo-experiments} presents the results from Monte
Carlo experiments. Section \ref{sec:Empirical-illustrations} applies
our inference method to study the effect of family size on labor income.
The proofs of the theorems and statements of the technical lemmas
are presented in the appendices. The proofs of the lemmas, auxiliary
results, and additional simulation evidence are collected in the online
supplement (\href{http://ruc-econ.github.io/ITE_Supp_Rev_V13.pdf}{ruc-econ.github.io/ITE\_Supp\_Rev\_V13.pdf}).

\section{Model and the FVX estimator \label{sec:Model-and-identification}}

For completeness, first, we describe the model setup of VX and FVX,
and their estimator. Let $\mathbbm{1}\left(\cdot\right)$ denote the
indicator function. The outcome and selection equations are given
respectively by 
\begin{eqnarray}
Y & = & g\left(D,X,\epsilon\right)\nonumber \\
D & = & \mathbbm{1}\left(\eta\leq s\left(Z,X\right)\right),\label{eq: triangular model}
\end{eqnarray}
where $Y\in\mathbb{R}$ is a continuously distributed outcome variable,
$D\in\left\{ 0,1\right\} $ is an endogenous treatment variable, and
$X\in\mathscr{S}_{X}$ is a vector of observed explanatory variables
(covariates) with $\mathscr{S}_{V}$ denoting the support of the distribution
of a random vector $V$ (i.e., the smallest closed set $C$ satisfying
$\mathrm{Pr}\left[V\in C\right]=1$). $Z\in\left\{ 0,1\right\} $
is a binary instrumental variable that has no direct effect on $Y$
and, therefore, are excluded from the outcome equation. $\left(\epsilon,\eta\right)$
are unobserved scalar-valued disturbances conditionally independent
of $Z$ given $X$. $g$ and $s$ are unknown functions.

The functions $g\left(d,x,\cdot\right)$ and $s\left(\cdot,x\right)$
are assumed to be strictly increasing. The selection equation in (\ref{eq: triangular model})
has the form of a latent index selection model: treatment is assigned
if some latent index or utility $s\left(Z,X\right)$ crosses the threshold
$\eta$. The ITE is defined as 
\begin{equation}
\varDelta\coloneqq g\left(1,X,\epsilon\right)-g\left(0,X,\epsilon\right),\label{eq:Delta through Y}
\end{equation}
where ``$a\coloneqq b$'' is understood as ``$a$ is defined by
$b$''. Note that $\varDelta$ is random conditionally on $X$ due
to the unobserved $\epsilon$, i.e., the treatment effect varies among
individuals with the same observed characteristics. The unobserved
disturbances $\epsilon$ and $\eta$ are allowed to be correlated
conditionally on $X$.\footnote{The model allows the ITEs to be ``essentially heterogeneous'' (\citealp{Heckman2006})
since whether or not individuals who have the same observed characteristics
select into treatment can be correlated with the gain from treatment.} Denote $d'\coloneqq1-d$, and let $g^{-1}\left(d',x,\cdot\right)$
be the inverse function of $g\left(d',x,\cdot\right)$. Let $\mathscr{S}_{V\mid W=w}$
denote the support of the conditional distribution of $V$ given $W=w$.
For $y\in\mathscr{S}_{g\left(d',x,\epsilon\right)\mid X=x}$, define
the corresponding counterfactual mapping $\phi_{dx}\left(y\right)\coloneqq g\left(d,x,g^{-1}\left(d',x,y\right)\right)$,
i.e., $\phi_{dx}\left(y\right)$ is the outcome one would observe
instead of $y$ if the treatment status $d'$ were switched to $d$.
Using the counterfactual mappings $\left(\phi_{0x},\phi_{1x}\right)$,
we can write the ITE as 
\begin{equation}
\varDelta=D\left(Y-\phi_{0X}\left(Y\right)\right)+\left(1-D\right)\left(\phi_{1X}\left(Y\right)-Y\right).\label{eq:Delta}
\end{equation}
VX shows constructive nonparametric identification of the counterfactual
mappings. This result establishes the identification of the entire
distribution of $\varDelta$. FVX shows that $\phi_{dx}\left(y\right)$
is the unique minimizer of the strictly convex function $Q_{dx}\left(\cdot;y\right)$
defined below:
\begin{multline}
Q_{dx}\left(t;y\right)\coloneqq\\
\left(\mathrm{E}\left[\mathbbm{1}\left(D=d\right)\left|Y-t\right|\mid Z=d,X=x\right]-\mathrm{E}\left[\mathbbm{1}\left(D=d'\right)\mathrm{sgn}\left(Y-y\right)\mid Z=d,X=x\right]\times t\right)\\
-\left(\mathrm{E}\left[\mathbbm{1}\left(D=d\right)\left|Y-t\right|\mid Z=d',X=x\right]-\mathrm{E}\left[\mathbbm{1}\left(D=d'\right)\mathrm{sgn}\left(Y-y\right)\mid Z=d',X=x\right]\times t\right),\label{eq:Q_dx definition}
\end{multline}
where $\mathrm{sgn}\left(u\right)\coloneqq2\times\mathbbm{1}\left(u>0\right)-1$
denotes the left continuous sign function.

The econometrician observes $\{\left(Y_{i},D_{i},X_{i},Z_{i}\right):i=1,\ldots,n\}$,
a sample of observations on $\left(Y,D,X^{\top},Z\right)^{\top}$
generated by the model. Let $\widehat{Q}_{dx}^{\left(-i\right)}\left(t;y\right)$
denote the leave-$i$-out sample analogue of $Q_{dx}\left(t;y\right)$
constructed under the FVX assumption that $X$ is discretely distributed:
\begin{multline}
\widehat{Q}_{dx}^{\left(-i\right)}\left(t;y\right)\coloneqq\\
\frac{\sum_{j\neq i}\left\{ \mathbbm{1}\left(D_{j}=d,Z_{j}=d,X_{j}=x\right)\left|Y_{j}-t\right|-\mathbbm{1}\left(D_{j}=d',Z_{j}=d,X_{j}=x\right)\mathrm{sgn}\left(Y_{j}-y\right)t\right\} }{\sum_{j\neq i}\mathbbm{1}\left(Z_{j}=d,X_{j}=x\right)}\\
-\frac{\sum_{j\neq i}\left\{ \mathbbm{1}\left(D_{j}=d,Z_{j}=d',X_{j}=x\right)\left|Y_{j}-t\right|-\mathbbm{1}\left(D_{j}=d',Z_{j}=d',X_{j}=x\right)\mathrm{sgn}\left(Y_{j}-y\right)t\right\} }{\sum_{j\neq i}\mathbbm{1}\left(Z_{j}=d',X_{j}=x\right)}.\label{eq:Q_hat definition}
\end{multline}
The leave-$i$-out nonparametric estimator of $\phi_{dx}\left(y\right),d\in\left\{ 0,1\right\} $,
can be constructed as
\begin{equation}
\widehat{\phi}_{dx}^{\left(-i\right)}\left(y\right)\coloneqq\underset{t\in\left[\underline{y}_{dx},\overline{y}_{dx}\right]}{\mathrm{argmin}}\widehat{Q}_{dx}^{\left(-i\right)}\left(t;y\right),\label{eq:phi_hat definition}
\end{equation}
where we write $\mathscr{S}_{g\left(d,x,\epsilon\right)\mid X=x}=\left[\underline{y}_{dx},\overline{y}_{dx}\right]$.\footnote{As FVX, we assume that $\underline{y}_{dx}$ and $\overline{y}_{dx}$
are known. Lemma 1 of VX shows that the supports of the potential
outcomes, $\mathscr{S}_{g\left(d,x,\epsilon\right)\mid X=x}=\left[\underline{y}_{dx},\overline{y}_{dx}\right]$,
are identified by $\mathscr{S}_{g\left(d,x,\epsilon\right)\mid X=x}=\mathscr{S}_{Y\mid D=d,X=x}$.
In practical implementation, $\underline{y}_{dx}$ and $\overline{y}_{dx}$
can be estimated. See Section 3 of FVX for discussion.} One can now estimate the ITEs by replacing $\phi_{dx}(y)$ in (\ref{eq:Delta})
with its leave-$i$-out nonparametric estimator $\widehat{\phi}_{dx}^{\left(-i\right)}\left(y\right)$:
\begin{equation}
\widehat{\varDelta}_{i}=D_{i}\left(Y_{i}-\widehat{\phi}_{0X_{i}}^{\left(-i\right)}\left(Y_{i}\right)\right)+\left(1-D_{i}\right)\left(\widehat{\phi}_{1X_{i}}^{\left(-i\right)}\left(Y_{i}\right)-Y_{i}\right).\label{eq:pseudo ITE definition}
\end{equation}
The FVX estimator of $f_{\varDelta\mid X}\left(v\mid x\right)$, the
conditional density of $\varDelta$ given $X=x$, is the kernel density
estimator that uses $\widehat{\varDelta}_{i}$ in place of the true
unobserved ITEs:
\begin{equation}
\widehat{f}_{\varDelta\mid X}\left(v\mid x;b\right)\coloneqq\frac{\sum_{i=1}^{n}\frac{1}{b}K\left(\frac{\widehat{\varDelta}_{i}-v}{b}\right)\mathbbm{1}\left(X_{i}=x\right)}{\sum_{i=1}^{n}\mathbbm{1}\left(X_{i}=x\right)},\label{eq:f_hat definition}
\end{equation}
where $K\left(\cdot\right)$ and $b>0$ denote the kernel function
and bandwidth, respectively.

Let $\left[a\pm b\right]$ denote the interval $\left[a-b,a+b\right]$.
In this paper, we propose a bootstrap UCB defined by the following
collection of random intervals:
\begin{equation}
\mathit{CB}_{\mathsf{jmb}}\left(v\mid x;b,b_{\zeta}\right)\coloneqq\left[\widehat{f}_{\varDelta\mid X}\left(v\mid x;b\right)\pm z_{1-\alpha}^{\mathsf{jmb}}\sqrt{\frac{\widehat{V}\left(v\mid x;b,b_{\zeta}\right)}{nb}}\right],\label{eq:confidence band definition}
\end{equation}
where $b_{\zeta}$ is another bandwidth. Algorithm \ref{alg:multiplier}
in Section \ref{subsec:Jackknife-multiplier-bootstrap} provides detailed
step-by-step instructions for constructing the UCB. In (\ref{eq:confidence band definition})
above, the standard error $\sqrt{\widehat{V}\left(v\mid x;b,b_{\zeta}\right)/\left(nb\right)}$
uses the variance estimator $\widehat{V}\left(v\mid x;b,b_{\zeta}\right)$
defined in equation (\ref{eq:asy var}) in Section \ref{subsec:Standard-errors}.
The bootstrap critical value $z_{1-\alpha}^{\mathsf{jmb}}$ is defined
in equation (\ref{eq:z_pound definition}) in Section \ref{subsec:Jackknife-multiplier-bootstrap}.
The need for the second bandwidth $b_{\zeta}$ is discussed in Section
\ref{subsec:Standard-errors}. We provide data-dependent procedures
for selecting the two bandwidths $b$ and $b_{\zeta}$. See the discussions
following Assumption \ref{assu:h_hat} in Section \ref{subsec:Assumptions}
and Assumption \ref{assu:btw zeta} in Section \ref{subsec:Standard-errors}
respectively. Theorem \ref{thm:confidence band} in Section \ref{subsec:Jackknife-multiplier-bootstrap}
is our main result. It establishes that the proposed UCB covers $f_{\varDelta\mid X}\left(v\mid x\right)$
simultaneously over a range of $v$ values with a pre-specified confidence
level in large samples.

\section{Asymptotic properties of the FVX estimator\label{sec:Asymptotic-properties}}

In this section, we establish two new asymptotic results for the FVX
density estimator. Theorem \ref{thm:uniform rate} below shows that
under seemingly minimal conditions (see Assumptions \ref{assu: DGP1}
and \ref{assu: DGP4} ahead), the FVX estimator has the same uniform
rate of convergence as that of the infeasible kernel density estimator
that uses true ITEs, and attains the optimal uniform rate of convergence
(see \citealp{Stone1982}). Theorem \ref{thm:asymptotic normality}
shows that the FVX estimator is asymptotically normal. However, its
asymptotic variance is larger than that of the infeasible estimator.
We show that these results hold under either a deterministic bandwidth
or a data-dependent bandwidth that satisfies Assumption \ref{assu:h_hat}
below.

\subsection{Assumptions\label{subsec:Assumptions}}

The following assumption on the data generating process (DGP) is similar
to those in VX and FVX.
\begin{assumption}[DGP]
\label{assu: DGP1} (a) For all $\left(d,x\right)\in\mathscr{S}_{\left(D,X\right)}$,
$g\left(d,x,\cdot\right)$ is continuously differentiable and strictly
increasing. (b) $Z\in\left\{ 0,1\right\} $ is independent of $\left(\epsilon,\eta\right)$
conditionally on $X$. (c) For all $x\in\mathscr{S}_{X}$, $\mathrm{Pr}\left[D=1\mid Z=1,X=x\right]\ne\mathrm{Pr}\left[D=1\mid Z=0,X=x\right]$.
(d) The conditional distribution of $(\epsilon,\eta)$ given $X$
is absolutely continuous with respect to the Lebesgue measure, has
a compact support, and its PDF is continuous and bounded. (e) The
supports $\mathscr{S}_{\left(D,X\right)}$ and $\mathscr{S}_{\left(Z,X\right)}$
are equal to $\{0,1\}\times\mathscr{S}_{X}$. (f) For all $x\in\mathscr{S}_{X}$,
$s\left(0,x\right)<s\left(1,x\right)$. (g) For $D_{zx}\coloneqq\mathbbm{1}\left(\eta\leq s\left(z,x\right)\right)$,
the complier group is given by $D_{0x}<D_{1x}$. We assume that for
all $\left(d,x\right)\in\mathscr{S}_{\left(D,X\right)}$, $\mathscr{S}_{g\left(d,x,\epsilon\right)\mid X=x,D_{0x}<D_{1x}}=\mathscr{S}_{g\left(d,x,\epsilon\right)\mid X=x}$.
(h) For all $d\in\{0,1\}$, the conditional distribution of $g\left(d,x,\epsilon\right)$
given $X=x$ and $D_{0x}<D_{1x}$ has a bounded away from zero density
$f_{dx\mid C_{x}}$. (i) For all $x\in\mathscr{S}_{X}$ and $d\in\left\{ 0,1\right\} $,
the conditional distributions of $g\left(d,x,\epsilon\right)$ has
the support $\mathscr{S}_{g\left(d,x,\epsilon\right)\mid X=x}=\left[\underline{y}_{dx},\overline{y}_{dx}\right]$
with known boundaries $-\infty<\underline{y}_{dx}<\overline{y}_{dx}<+\infty$.
(j) The data $\left\{ W_{i}\coloneqq\left(Y_{i},D_{i},X_{i}^{\top},Z_{i}\right)^{\top}:i=1,\ldots,n\right\} $
are i.i.d. observations on $W=\left(Y,D,X^{\top},Z\right)^{\top}$.
(k) $X$ is discretely distributed and $\mathscr{S}_{X}$ is finite.
\end{assumption}
In the above assumption, the continuity and monotonicity conditions
in (a), the standard instrument exogeneity assumption in (b), the
instrument relevance condition in (c), the absolute continuity condition
in (d), (e,f), as well as the equality of the supports condition in
(g) are imposed for identification. The assumption that $\epsilon$
is scalar-valued and the condition in (a) impose rank invariance on
the potential outcomes. See Section 2.1 of VX. Parts (d,e) of the
assumption are mild regularity conditions. Parts (c,f) and the latent
index assumption on the selection equation imply that $\mathrm{Pr}\left[D=1\mid Z=1,X=x\right]>\mathrm{Pr}\left[D=1\mid Z=0,X=x\right]$.
Under (c,f) and the latent index assumption, we have $D_{0x}\leq D_{1x}$,
for all $x\in\mathscr{S}_{X}$. Clearly, the model satisfies the LATE
independence and monotonicity assumptions (see, e.g., \citealp[Section 4]{Vytlacil2002}).
See \citet{Kitagawa2015} for testable implications.

In part (g), it is assumed that conditionally on $X=x$, the support
of the conditional distribution of $g\left(d,x,\epsilon\right)$ in
the complier group $D_{0x}<D_{1x}$ is the same as that of the conditional
distribution of $g\left(d,x,\epsilon\right)$ given $X=x$. VX argues
that (g) is satisfied if the conditional distribution of $\left(\epsilon,\eta\right)$
given $X=x$ has a rectangular support, for all $x\in\mathscr{S}_{X}$.
Note that the identification result $\mathscr{S}_{g\left(d,x,\epsilon\right)\mid X=x}=\mathscr{S}_{Y\mid D=d,X=x}$
in VX, together with (a), implies that $\mathscr{S}_{\epsilon\mid X=x}=\mathscr{S}_{\epsilon\mid D=d,X=x}$,
for all $d\in\left\{ 0,1\right\} $. The rest of the conditions are
imposed for estimation. As in FVX, part (k) restricts the estimation
framework to discretely distributed covariates $X$. Under these assumptions,
the conditional distribution of $Y$ or $\epsilon$ given $\left(D,X,Z\right)$
is absolutely continuous with respect to the Lebesgue measure and
admits a continuous and bounded Lebesgue density. Under Assumption
\ref{assu: DGP1}, $f_{dx\mid C_{x}}$ is also continuous and bounded.\footnote{Let $F_{dx\mid C_{x}}$ denote the conditional CDF of $g\left(d,x,\epsilon\right)$
given $X=x$ and $D_{0x}<D_{1x}$. Then it is clear that $F_{dx\mid C_{x}}\left(y\right)=\mathrm{Pr}\left[\epsilon\leq g^{-1}\left(d,x,y\right)\mid s\left(0,x\right)<\eta\leq s\left(1,x\right),X=x\right]$.}

Theorem 1 of FVX shows the asymptotic properties of the estimated
counterfactual mappings $\widehat{\phi}_{dx}$ under Assumption \ref{assu: DGP1}.
Let $\varDelta_{x}\left(\cdot\right)\coloneqq g\left(1,x,\cdot\right)-g\left(0,x,\cdot\right)$,
which is continuously differentiable under Assumption \ref{assu: DGP1}.
FVX assumes in their Assumption 5(i) that the conditional density
of $\varDelta=\varDelta_{x}\left(\epsilon\right)$ given $X=x$ exists
and is $P$-times continuously differentiable. Without imposing further
restrictions, Assumptions \ref{assu: DGP1} alone does not guarantee
that the distribution of $\varDelta=\varDelta_{x}\left(\epsilon\right)$
is absolutely continuous with respect to the Lebesgue measure.\footnote{E.g., the distribution has a mass point if $\varDelta_{x}\left(\cdot\right)$
is constant on some sub-interval of $\mathscr{S}_{\epsilon\mid X=x}$.} The assumption below provides mild sufficient conditions for the
existence and differentiability of the Lebesgue density of the ITE
$\varDelta$ given $X=x$ (see Lemma \ref{lem:differentiability of the PDF}
in Appendix \ref{sec:Appendix A}). Let $F_{V\mid W}\left(\cdot\mid w\right)$
and $f_{V\mid W}\left(\cdot\mid w\right)$ denote the conditional
cumulative distribution function (CDF) and PDF of $V$ given $W=w$,
respectively.
\begin{assumption}
[Existence and differentiability of the conditional PDF of the ITE]\label{assu: DGP4}(a)
For all $x\in\mathscr{S}_{X}$, the conditional CDF $F_{\epsilon\eta\mid X}\left(\cdot\mid x\right)$
of $\left(\epsilon,\eta\right)$ given $X=x$ and $g\left(d,x,\cdot\right)$
are both $\left(P+1\right)$-times continuously differentiable. (b)
There is a partition of $\mathscr{S}_{\epsilon\mid X=x}=\left[\underline{\epsilon}_{x},\overline{\epsilon}_{x}\right]$,
$\underline{\epsilon}_{x}=\epsilon_{x,0}<\epsilon_{x,1}<\cdots<\epsilon_{x,m}=\overline{\epsilon}_{x}$
with $\mathscr{S}_{\epsilon\mid X=x}=\bigcup_{j=1}^{m}\left[\epsilon_{x,j-1},\epsilon_{x,j}\right]$,
such that $\varDelta_{x}\left(\cdot\right)$ is piecewise monotone:
for all $j=1,...,m$, the restriction of $\varDelta_{x}\left(\cdot\right)$
on $\left[\epsilon_{x,j-1},\epsilon_{x,j}\right]$, $\left.\varDelta_{x,j}\left(\cdot\right)\coloneqq\varDelta_{x}\right|_{\left[\epsilon_{x,j-1},\epsilon_{x,j}\right]}$
is strictly monotone. (c) Let $\varDelta_{x}\left(\left(\epsilon_{x,j-1},\epsilon_{x,j}\right)\right)\coloneqq\left\{ \varDelta_{x}\left(e\right):e\in\left(\epsilon_{x,j-1},\epsilon_{x,j}\right)\right\} $
denote the image of $\varDelta_{x,j}(\cdot)$. We assume that $\varDelta_{x}\left(\left(\epsilon_{x,0},\epsilon_{x,1}\right)\right)=\cdots=\varDelta_{x}\left(\left(\epsilon_{x,m-1},\epsilon_{x,m}\right)\right)$.
\end{assumption}
The smoothness assumption imposed by (a) with $P\geq1$ is stronger
than that imposed by Assumption \ref{assu: DGP1}(a,d). Under (a),
$f_{dx\mid C_{x}}$ is $P$-times continuously differentiable. The
piecewise monotonicity condition in (b) is easily satisfied if $\varDelta_{x}$
has finitely many local extrema on $\left[\underline{\epsilon}_{x},\overline{\epsilon}_{x}\right]$.\footnote{Since it was assumed in Assumption \ref{assu: DGP1} that $\varDelta_{x}$
is continuously differentiable, this condition is satisfied if the
set of zeros of the continuous derivative function $\varDelta_{x}'$,
$\left\{ e\in\left[\underline{\epsilon}_{x},\overline{\epsilon}_{x}\right]:\varDelta_{x}'\left(e\right)=0\right\} $,
contains only isolated points.} Parts (a,b) of the assumption guarantee the existence of the Lebesgue
density $f_{\varDelta\mid X}\left(\cdot\mid x\right)$. Note that
the knowledge of the partition in (b) is not required for estimation
or inference. Part (c) rules out discontinuities in the interior of
$\mathscr{S}_{\varDelta\mid X=x}$. See the proof of Lemma \ref{lem:differentiability of the PDF}
in Appendix \ref{sec:Appendix A} for more details. We are unaware
of any weaker conditions that could be imposed on $\varDelta_{x}$
to guarantee the existence and differentiability of the conditional
PDF of $\varDelta=\varDelta_{x}\left(\epsilon\right)$ given $X=x$.

Application of kernel-based nonparametric techniques is complicated
by the bandwidth selection issue. A common practice in applied work
is using a data-dependent bandwidth approximating some underlying
deterministic bandwidth. We allow the bandwidth $\widehat{h}=\widehat{h}_{n}$
used in the implementation to be data-dependent and, following \citet{Li:2010dy},
assume that $\widehat{h}$ is a consistent estimator of some deterministic
bandwidth sequence $h=h_{n}\downarrow0$ in the sense that $\widehat{h}_{n}/h_{n}\rightarrow_{p}1$.
To simplify the notation, we suppress the dependence of the bandwidths
on $n$. Formally, we make the following assumption.
\begin{assumption}
\label{assu:h_hat}$\mathrm{Pr}\left[\left|\widehat{h}/h-1\right|>\varepsilon_{n}\right]\leq\delta_{n}$
for some deterministic bandwidth $h$ and positive sequences $\varepsilon_{n},\delta_{n}\downarrow0$.\footnote{Assumption \ref{assu:h_hat} is equivalent to requiring $\widehat{h}/h\rightarrow_{p}1$.
It is clear that Assumption \ref{assu:h_hat} implies $\widehat{h}/h\rightarrow_{p}1$.
On the other hand, if $\widehat{h}/h\rightarrow_{p}1$, $\varepsilon_{n}=\delta_{n}\downarrow0$
can be taken to be the Ky Fan metric between $\widehat{h}/h$ and
$1$, which converges to 0 as $n\uparrow\infty$. See \citet[Theorem 9.2.2]{dudley2002real}.}
\end{assumption}
The deterministic bandwidth assumption ($\varepsilon_{n}=\delta_{n}=0$)
is nested as a special case. Clearly, $\widehat{h}/h-1=O_{p}\left(\varepsilon_{n}\right)$,
under Assumption \ref{assu:h_hat}. E.g., as in FVX, one can consider
a feasible version of the Silverman rule-of-thumb (ROT) bandwidth
by setting $\widehat{h}=\widehat{h}^{\mathsf{rot}}\coloneqq C_{K}\cdot\widehat{\sigma}_{\varDelta\mid X=x}\cdot n_{x}^{-1/5}$,
where $\widehat{\sigma}_{\varDelta\mid X=x}$ is the sample analogue
of the standard deviation $\sigma_{\varDelta\mid X=x}\coloneqq\sqrt{\mathrm{Var}\left[\varDelta\mid X=x\right]}$
computed using the pseudo (estimated) ITEs, $C_{K}$ is a known constant
that depends only on the kernel function $K$, and $n_{x}\coloneqq\sum_{i=1}^{n}\mathbbm{1}\left(X_{i}=x\right)$.\footnote{The ROT bandwidth is a parametric estimator of the asymptotic mean
integrated squared error (AMISE) optimal bandwidth for $\widetilde{f}_{\varDelta\mid X}\left(v\mid x;h\right)$
defined by (\ref{eq:f_til definition}) under $P=2$. See \citet{li2007nonparametric}.
Theorem \ref{thm:asymptotic normality} shows that the asymptotic
mean squared error (AMSE) of $\widehat{f}_{\varDelta\mid X}\left(v\mid x;h\right)$
is given by $\mathrm{AMSE}\left(v\mid x\right)\coloneqq f_{\varDelta\mid X}^{\left(P\right)}\left(v\mid x\right)^{2}\mu_{K,P}^{2}h^{2P}+\mathscr{V}\left(v\mid x\right)/\left(nh\right)$.
Hence, the bandwidth that minimizes the AMISE $\int_{I_{x}}\mathrm{AMSE}\left(v\mid x\right)\mathrm{d}v$
is also a multiple of $n^{-1/\left(2P+1\right)}$.} Let $a_{n}\wasypropto b_{n}$ denote $a_{n}=C\cdot b_{n}$ for some
constant $C>0$. Let $p_{x}\coloneqq\mathrm{Pr}\left[X=x\right]$.
By Lemma \ref{lem:Lemma 1 phi linear representation}, one can show
that such a choice satisfies Assumption \ref{assu:h_hat} with $h=C_{K}\cdot\left(\sigma_{\varDelta\mid X=x}p_{x}^{-1/5}\right)\cdot n^{-1/5}$,
$\varepsilon_{n}\wasypropto\sqrt{\mathrm{log}\left(n\right)/n}$ and
$\delta_{n}\wasypropto n^{-1}$. We also assume that the kernel function
$K$ is of order $P\geq2$.
\begin{assumption}
\label{assu:kernel}(a) $K$ is symmetric, compactly supported on
$\left[-1,1\right]$ and twice continuously differentiable on $\mathbb{R}$
with Lipschitz derivatives. (b) $\int K\left(u\right)\mathrm{d}u=1$.
(c) $\int u^{k}K\left(u\right)\mathrm{d}u=0$, for all $k=1,2,...,P-1$
($P\geq2$), and $\int u^{P}K\left(u\right)\mathrm{d}u\ne0$.
\end{assumption}

\subsection{Rate of convergence and asymptotic distribution}

In this section, we first derive a linearization for the FVX estimator.
The result is given by equation (\ref{eq:linearization f_hat - f three terms})
below and serves as the basis for establishing the asymptotic properties
of the FVX estimator. The latter are presented below as Theorems \ref{thm:uniform rate}
and \ref{thm:asymptotic normality} (the rate of convergence and asymptotic
distribution, respectively).

Let $\widetilde{f}_{\varDelta\mid X}\left(v\mid x;b\right)$ denote
the infeasible estimator of the density $f_{\varDelta\mid X}\left(v\mid x\right)$
that uses the true latent ITEs:
\begin{equation}
\widetilde{f}_{\varDelta\mid X}\left(v\mid x;b\right)\coloneqq\frac{\sum_{i=1}^{n}\frac{1}{b}K\left(\frac{\varDelta_{i}-v}{b}\right)\mathbbm{1}\left(X_{i}=x\right)}{\sum_{i=1}^{n}\mathbbm{1}\left(X_{i}=x\right)}.\label{eq:f_til definition}
\end{equation}
Let $I_{x}$ denote an inner closed sub-interval of $\mathscr{S}_{\Delta\mid X=x}$,
$\underline{h}\coloneqq\left(1-\varepsilon_{n}\right)h$ and $\overline{h}\coloneqq\left(1+\varepsilon_{n}\right)h$.
Under Assumption \ref{assu:h_hat}, $\widehat{h}\in\left[\underline{h},\overline{h}\right]$
with probability $1-\delta_{n}$. For $\left(v,b\right)\in I_{x}\times\left[\underline{h},\overline{h}\right]$,
we decompose the estimation error $\widehat{f}_{\varDelta\mid X}\left(v\mid x;b\right)-f_{\varDelta\mid X}\left(v\mid x\right)$
into that of the infeasible estimator $\widetilde{f}_{\varDelta\mid X}\left(v\mid x;b\right)-f_{\varDelta\mid X}\left(v\mid x\right)$
and the difference between the feasible and infeasible estimators
$\widehat{f}_{\varDelta\mid X}\left(v\mid x;b\right)-\widetilde{f}_{\varDelta\mid X}\left(v\mid x;b\right)$.
We show that the former satisfies
\begin{multline}
\widetilde{f}_{\varDelta\mid X}\left(v\mid x;b\right)-f_{\varDelta\mid X}\left(v\mid x\right)=p_{x}^{-1}\left(\widetilde{f}_{\varDelta X}\left(v,x;b\right)-m_{\varDelta X}\left(v,x;b\right)\right)\\
+p_{x}^{-1}\left(m_{\varDelta X}\left(v,x;b\right)-f_{\varDelta X}\left(v,x\right)\right)+O_{p}\left(\sqrt{\frac{\mathrm{log}\left(n\right)}{n}}\right),\label{eq:f_til - f linearization}
\end{multline}
where $\widetilde{f}_{\varDelta X}\left(v,x;b\right)\coloneqq\widetilde{f}_{\varDelta\mid X}\left(v\mid x;b\right)\widehat{p}_{x}$
with $\widehat{p}_{x}\coloneqq n^{-1}\sum_{i=1}^{n}\mathbbm{1}\left(X_{i}=x\right)$
is the infeasible estimator of the joint density function $f_{\varDelta X}\left(v,x\right)\coloneqq f_{\varDelta\mid X}\left(v\mid x\right)p_{x}$,
$m_{\varDelta X}\left(v,x;b\right)\coloneqq\mathrm{E}\left[\widetilde{f}_{\varDelta X}\left(v,x;b\right)\right]$,
and the remainder term is uniform in $\left(v,b\right)\in I_{x}\times\left[\underline{h},\overline{h}\right]$.
Note that $m_{\varDelta X}\left(v,x;b\right)-f_{\varDelta X}\left(v,x\right)$
is the bias term that depends on the bandwidth $b$. Let $\mu_{K,P}\coloneqq\left(\int u^{P}K\left(u\right)\mathrm{d}u\right)/P!$,
and let $f_{\varDelta\mid X}^{\left(P\right)}\left(v\mid x\right)\coloneqq\partial^{P}f_{\varDelta\mid X}\left(v\mid x\right)/\partial v^{P}$
denote the derivatives of the conditional PDF. It follows from standard
arguments for kernel density estimators (see, e.g., \citealp{Newey:1994jb})
that 
\begin{equation}
m_{\varDelta X}\left(v,x;b\right)-f_{\varDelta X}\left(v,x\right)=f_{\varDelta\mid X}^{\left(P\right)}\left(v\mid x\right)p_{x}\mu_{K,P}b^{P}+o\left(h^{P}\right),\label{eq:infeasible decompose}
\end{equation}
uniformly in $\left(v,b\right)\in I_{x}\times\left[\underline{h},\overline{h}\right]$.
For a deterministic bandwidth sequence $h$ such that $nh\uparrow\infty$,
it follows from standard arguments that $\sqrt{nh}\left(\widetilde{f}_{\varDelta X}\left(v,x;h\right)-m_{\varDelta X}\left(v,x;h\right)\right)$
is asymptotically normal.

Let $K'$ denote the derivative of the kernel function $K$. Denote
$\widehat{f}_{\varDelta X}\left(v,x;b\right)\coloneqq\widehat{f}_{\varDelta X}\left(v\mid x;b\right)\widehat{p}_{x}$.
We approximate $\widehat{f}_{\varDelta X}\left(v,x;b\right)-\widetilde{f}_{\varDelta X}\left(v,x;b\right)$
by $\left(nb^{2}\right)^{-1}\sum_{i=1}^{n}K'\left(\left(\varDelta_{i}-v\right)/b\right)\left(\widehat{\varDelta}_{i}-\varDelta_{i}\right)\mathbbm{1}\left(X_{i}=x\right)$.
The first-stage estimation errors $\widehat{\phi}_{dX_{i}}^{\left(-i\right)}\left(Y_{i}\right)-\phi_{dX_{i}}\left(Y_{i}\right)$
in $\widehat{\varDelta}_{i}-\varDelta_{i}$ can be approximated using
its linear representation (see Theorem 1 of FVX and Lemma \ref{lem:Lemma 1 phi linear representation}).
After recalling that $f_{dx\mid C_{x}}$ is the conditional PDF of
$g\left(d,x,\epsilon\right)$ given $X=x$ in the complier group,
we define:
\begin{eqnarray}
\zeta_{dx}\left(y\right) & \coloneqq & f_{dx\mid C_{x}}\left(y\right)\left(\mathrm{Pr}\left[D=d\mid Z=1,X=x\right]-\mathrm{Pr}\left[D=d\mid Z=0,X=x\right]\right),\nonumber \\
R_{d'x}\left(y\right) & \coloneqq & \mathrm{Pr}\left[Y\leq\phi_{dx}\left(y\right),D=d\mid X=x\right]+\mathrm{Pr}\left[Y\leq y,D=d'\mid X=x\right],\nonumber \\
q_{dx}\left(W_{i},W_{j}\right) & \coloneqq & \frac{\mathbbm{1}\left(D_{i}=d',X_{i}=x\right)}{\zeta_{dx}\left(\phi_{dx}\left(Y_{i}\right)\right)}\nonumber \\
 &  & \times\left\{ \mathbbm{1}\left(Y_{j}\leq\phi_{dx}\left(Y_{i}\right),D_{j}=d\right)+\mathbbm{1}\left(Y_{j}\leq Y_{i},D_{j}=d'\right)-R_{d'x}\left(Y_{i}\right)\right\} ,\label{eq:q_dx definition}\\
q_{x}\left(W_{i},W_{j}\right) & \coloneqq & q_{1x}\left(W_{i},W_{j}\right)-q_{0x}\left(W_{i},W_{j}\right),\nonumber \\
p_{zx} & \coloneqq & \mathrm{Pr}\left[Z=z,X=x\right],\nonumber \\
\pi_{x}\left(Z_{i},X_{i}\right) & \coloneqq & \frac{\mathbbm{1}\left(Z_{i}=0,X_{i}=x\right)}{p_{0x}}-\frac{\mathbbm{1}\left(Z_{i}=1,X_{i}=x\right)}{p_{1x}}.\label{eq:pi_x definition}
\end{eqnarray}
Using the above definitions, we can write that the difference between
the feasible and infeasible estimators as a \textit{U}-statistic with
a kernel that depends on the bandwidth: 
\begin{equation}
\widehat{f}_{\varDelta\mid X}\left(v\mid x;b\right)-\widetilde{f}_{\varDelta\mid X}\left(v\mid x;b\right)=p_{x}^{-1}\frac{1}{n\left(n-1\right)}\sum_{i=1}^{n}\sum_{j\neq i}\mathcal{G}_{x}\left(W_{i},W_{j},v;b\right)+O_{p}\left(\frac{\mathrm{log}\left(n\right)}{nh^{2}}+\frac{\mathrm{log}\left(n\right)^{3/4}}{n^{3/4}h}\right),\label{eq:f_hat - f_til U stat expansion}
\end{equation}
uniformly in $\left(v,b\right)\in I_{x}\times\left[\underline{h},\overline{h}\right]$,
where
\begin{equation}
\mathcal{G}_{x}\left(W_{i},W_{j},v;b\right)\coloneqq\frac{1}{b^{2}}K'\left(\frac{\varDelta_{i}-v}{b}\right)q_{x}\left(W_{i},W_{j}\right)\pi_{x}\left(Z_{j},X_{j}\right).\label{eq:G kernel definition}
\end{equation}
By Assumption \ref{assu: DGP1}(b) and (\ref{eq:q_dx simplify}) in
Appendix \ref{sec:Appendix A}, $\mathcal{G}_{x}^{\left[2\right]}\left(w,v;b\right)\coloneqq\mathrm{E}\left[\mathcal{G}_{x}\left(w,W,v;b\right)\right]=0$,
for all $w$ and $\mathrm{E}\left[\mathcal{G}_{x}\left(W_{i},W_{j},v;b\right)\right]=0$,
for all $\ensuremath{i\neq j}$. The leading term (or the H$\acute{\mathrm{a}}$jek
projection) in the Hoeffding decomposition of the \textit{U}-statistic
is given by $\mathcal{G}_{x}^{\left[1\right]}\left(w,v;b\right)\coloneqq\mathrm{E}\left[\mathcal{G}_{x}\left(W,w,v;b\right)\right]$.
Therefore, the Hoeffding decomposition is given by
\begin{multline}
\frac{1}{n\left(n-1\right)}\sum_{i=1}^{n}\sum_{j\neq i}\mathcal{G}_{x}\left(W_{i},W_{j},v;b\right)=\frac{1}{n}\sum_{i=1}^{n}\mathcal{G}_{x}^{\left[1\right]}\left(W_{i},v;b\right)\\
+\frac{1}{n\left(n-1\right)}\sum_{i=1}^{n}\sum_{j\neq i}\left\{ \mathcal{G}_{x}\left(W_{i},W_{j},v;b\right)-\mathcal{G}_{x}^{\left[1\right]}\left(W_{j},v;b\right)\right\} .\label{eq:G Hoeffding}
\end{multline}

By definition and since $\mathcal{G}_{x}^{\left[2\right]}\left(w,v;b\right)=0$,
the \textit{U}-statistic $\left(n\left(n-1\right)\right)^{-1}\sum_{i=1}^{n}\sum_{j\neq i}\mathcal{G}_{x}\left(W_{i},W_{j},v;b\right)$
is non-degenerate (\citealp{Chen2020jackknife}) if $\mathrm{Var}\left[\mathcal{G}_{x}^{\left[1\right]}\left(W,v;b\right)\right]>0$.
In the proof of Lemma \ref{lem:lemma 3}, we show that the condition
holds for all $\left(v,b\right)\in I_{x}\times\left[\underline{h},\overline{h}\right]$
when $n$ is sufficiently large. We also show that the second term
on the right-hand side of (\ref{eq:G Hoeffding}) is dominated by
the first term. See Remark \ref{rmk: proof of normality} below. Combining
the result with (\ref{eq:f_til - f linearization}) and (\ref{eq:f_hat - f_til U stat expansion}),
we can write the estimation error of the feasible estimator as
\begin{multline}
\widehat{f}_{\varDelta\mid X}\left(v\mid x;b\right)-f_{\varDelta\mid X}\left(v\mid x\right)=p_{x}^{-1}\left(\widetilde{f}_{\varDelta X}\left(v,x;b\right)-m_{\varDelta X}\left(v,x;b\right)\right)+p_{x}^{-1}\frac{1}{n}\sum_{i=1}^{n}\mathcal{G}_{x}^{\left[1\right]}\left(W_{i},v;b\right)\\
+p_{x}^{-1}\left(m_{\varDelta X}\left(v,x;b\right)-f_{\varDelta X}\left(v,x\right)\right)+O_{p}\left(\sqrt{\frac{\mathrm{log}\left(n\right)}{n}}+\frac{\mathrm{log}\left(n\right)}{nh^{2}}+\frac{\mathrm{log}\left(n\right)^{3/4}}{n^{3/4}h}\right),\label{eq:linearization f_hat - f three terms}
\end{multline}
where the remainder term is uniform in $\left(v,b\right)\in I_{x}\times\left[\underline{h},\overline{h}\right]$.

\sloppy The asymptotic variance of the FVX estimator $\widehat{f}_{\varDelta\mid X}\left(v\mid x;h\right)$
under the deterministic bandwidth sequence is determined by the sum
of $\left(\widetilde{f}_{\varDelta X}\left(v,x;h\right)-m_{\varDelta X}\left(v,x;h\right)\right)/p_{x}$
and $n^{-1}\sum_{i=1}^{n}\mathcal{G}_{x}^{\left[1\right]}\left(W_{i},v;h\right)/p_{x}$.
The first term comes from the infeasible estimator $\widetilde{f}_{\varDelta X}\left(v,x;h\right)$
as in equation (\ref{eq:f_til - f linearization}). The second term
reflects the influence of the estimation of ITEs in the first stage.
We show that these terms are both of order $O_{p}\left(\sqrt{\mathrm{log}\left(n\right)/\left(nh\right)}\right)$
uniformly in $v\in I_{x}$ and independent of each other. Consequently,
the first-stage estimation errors unambiguously add to the asymptotic
variance of $\widehat{f}_{\varDelta\mid X}\left(v\mid x;h\right)$
and their contribution must be taken into account for valid inference.\footnote{\label{fn:phi_hat - phi}Let $\widehat{\phi}_{dx}\left(y\right)$
be the leave-in version of $\widehat{\phi}_{dx}^{\left(-i\right)}\left(y\right)$
(i.e., $\widehat{\phi}_{dx}\left(y\right)$ minimizes the sample analogue
of $Q_{dx}\left(\cdot;y\right)$). The crucial observation is that
the linearization of $\widehat{\phi}_{dx}\left(y\right)-\phi_{dx}\left(y\right)$
derived in FVX (also see Lemma \ref{lem:Lemma 1 phi linear representation})
is discontinuous in both $W_{i}$ and $y$. As a result, the difference
between the feasible and infeasible estimators $\widehat{f}_{\varDelta\mid X}\left(v\mid x;h\right)-\widetilde{f}_{\varDelta\mid X}\left(v\mid x;h\right)$
converges at a rate slower than $n^{-1/2}$. One can show that the
difference would be of order $O_{p}(n^{-1/2})$ if the linearization
were smooth. See the proof of Lemma \ref{lem:lemma 2} for more details
on how the linearization is involved in the \textit{U}-statistic representation
given by $\left(n\left(n-1\right)\right)^{-1}\sum_{i=1}^{n}\sum_{j\neq i}\mathcal{G}_{x}\left(W_{i},W_{j},v;b\right)$.} Equation (\ref{eq:infeasible decompose}) implies that $m_{\varDelta X}\left(v,x;h\right)-f_{\varDelta X}\left(v,x\right)=O\left(h^{P}\right)$
uniformly in $v\in I_{x}$. These results provide the uniform rate
of convergence of $\widehat{f}_{\varDelta\mid X}\left(\cdot\mid x;h\right)$.
The bias expansion in (\ref{eq:infeasible decompose}) and the linearization
in (\ref{eq:linearization f_hat - f three terms}) are also valid
for a continuum $\left[\underline{h},\overline{h}\right]$ of bandwidths.
Since $\mathrm{Pr}\left[\widehat{h}\in\left[\underline{h},\overline{h}\right]\right]>1-\delta_{n}$
under Assumption \ref{assu:h_hat}, (\ref{eq:infeasible decompose})
and (\ref{eq:linearization f_hat - f three terms}) with $b$ replaced
by $\widehat{h}$ still hold. We show that the first two terms on
the right-hand side of the decomposition in (\ref{eq:linearization f_hat - f three terms})
are of the same $O_{p}\left(\sqrt{\mathrm{log}\left(n\right)/\left(nh\right)}\right)$
order uniformly in $\left(v,b\right)\in I_{x}\times\left[\underline{h},\overline{h}\right]$.
These results show that the uniform convergence rate remains the same
if $h$ is replaced by its estimator $\widehat{h}$.

We summarize the above results in Theorem \ref{thm:uniform rate}
below, which is a refinement of Theorem 2 in FVX. In Appendix \ref{sec:Appendix A},
we prove a stronger version (Theorem \ref{thm: uniform rate appendix})
of Theorem \ref{thm:uniform rate}. The latter establishes the non-asymptotic
deviation bounds for the uniform estimation errors of $\widehat{f}_{\varDelta\mid X}\left(\cdot\mid x;h\right)$
and $\widehat{f}_{\varDelta\mid X}\left(\cdot\mid x;\widehat{h}\right)$
and is used in the proof of Theorem \ref{thm:confidence band}.

For a function $f:A\rightarrow\mathbb{R}$, let $\left\Vert f\right\Vert _{A}\coloneqq\mathrm{sup}_{x\in A}\left|f\left(x\right)\right|$
denote the sup-norm of $f$ on $A$. For a subset $A$ in the Euclidean
space, let $\mathrm{Int}\left(A\right)$ denote the interior of $A$.
\begin{thm}
\label{thm:uniform rate}Suppose that Assumptions \ref{assu: DGP1}-\ref{assu:kernel}
hold, and the deterministic bandwidth $h$ is such that $\mathrm{log}\left(n\right)/\left(nh^{3}\right)\downarrow0$.
Then, for any $x\in\mathscr{S}_{X}$ and compact $I_{x}\subseteq\mathrm{Int}\left(\mathscr{S}_{\Delta\mid X=x}\right)$,
\[
\left\Vert \widehat{f}_{\varDelta\mid X}\left(\cdot\mid x;h\right)-f_{\varDelta\mid X}\left(\cdot\mid x\right)\right\Vert _{I_{x}}=O_{p}\left(\sqrt{\frac{\mathrm{log}\left(n\right)}{nh}}+h^{P}\right)
\]
and
\[
\left\Vert \widehat{f}_{\varDelta\mid X}\left(\cdot\mid x;\widehat{h}\right)-f_{\varDelta\mid X}\left(\cdot\mid x\right)\right\Vert _{I_{x}}=O_{p}\left(\sqrt{\frac{\mathrm{log}\left(n\right)}{nh}}+h^{P}\right).
\]
\end{thm}
\begin{rembold}In comparison, Theorem 2 of FVX has a slower $O_{p}\left(\sqrt{\mathrm{log}\left(n\right)/\left(nh^{2}\right)}+h^{P}\right)$
convergence rate. Theorem \ref{thm:uniform rate} implies that the
FVX and infeasible estimators of $f_{\varDelta\mid X}\left(\cdot\mid x\right)$
have the same uniform convergence rate. Moreover, the convergence
rate is unaffected by the estimation of the bandwidth. The optimal
bandwidth rate that leads to the fastest possible convergence rate
is of order $\left(\mathrm{log}\left(n\right)/n\right)^{1/\left(2P+1\right)}$.
Hence, both the FVX and infeasible estimators attain the optimal uniform
convergence rate $\left(\mathrm{log}\left(n\right)/n\right)^{P/\left(2P+1\right)}$.
Note that under our smoothness conditions, any uniformly consistent
estimator cannot converge uniformly at a rate faster than $\left(\mathrm{log}\left(n\right)/n\right)^{P/\left(2P+1\right)}$
(see \citealp{Stone1982}).\end{rembold}

The next theorem establishes the asymptotic normality of the FVX estimator
and quantifies the contribution of the first-stage estimation errors
to the asymptotic variance. By using (\ref{eq:infeasible decompose})
and the linearization (\ref{eq:linearization f_hat - f three terms})
for a single bandwidth $h$, we show that for any fixed $v\in I_{x}$,
asymptotic normality holds for $\sqrt{nh}\left(\widehat{f}_{\varDelta\mid X}\left(v\mid x;h\right)-f_{\varDelta\mid X}\left(v\mid x\right)-f_{\varDelta\mid X}^{\left(P\right)}\left(v\mid x\right)\mu_{K,P}h^{P}\right)$.
By using the uniform-in-bandwidth approximation ((\ref{eq:f_til - f linearization})
- (\ref{eq:f_hat - f_til U stat expansion})) of $\widehat{f}_{\varDelta\mid X}\left(v\mid x;b\right)-f_{\varDelta\mid X}\left(v\mid x\right)$
and an asymptotic equivalence result (Lemma \ref{lem:lemma random btw}),
we show that the same normality result holds if $h$ is replaced by
its estimator $\widehat{h}$. The result is analogous to those in
\citet{Li:2010dy}.

Let $f_{\epsilon DX}\left(e,d,x\right)\coloneqq f_{\epsilon\mid DX}\left(e\mid d,x\right)\mathrm{Pr}\left[D=d,X=x\right]$
denote the joint density of $(\epsilon,D,X^{\top})^{\top}$.
\begin{thm}
\label{thm:asymptotic normality}Suppose that Assumptions \ref{assu: DGP1}-\ref{assu:kernel}
hold, and $h\wasypropto n^{-\lambda}$ with $1/\left(2P+1\right)\leq\lambda<1/3$.
Then, for any $x\in\mathscr{S}_{X}$ and $v$ in a compact sub-interval
$I_{x}$ of $\mathscr{S}_{\Delta\mid X=x}$,
\[
\sqrt{nh}\left(\widehat{f}_{\varDelta\mid X}\left(v\mid x;h\right)-f_{\varDelta\mid X}\left(v\mid x\right)-f_{\varDelta\mid X}^{\left(P\right)}\left(v\mid x\right)\mu_{K,P}h^{P}\right)\to_{d}\mathrm{N}\left(0,\mathscr{V}\left(v\mid x\right)\right)
\]
and if $\varepsilon_{n}=o\left(\mathrm{log}\left(n\right)^{-1/2}\right)$
in Assumption \ref{assu:h_hat},
\[
\sqrt{n\widehat{h}}\left(\widehat{f}_{\varDelta\mid X}\left(v\mid x;\widehat{h}\right)-f_{\varDelta\mid X}\left(v\mid x\right)-f_{\varDelta\mid X}^{\left(P\right)}\left(v\mid x\right)\mu_{K,P}\widehat{h}^{P}\right)\to_{d}\mathrm{N}\left(0,\mathscr{V}\left(v\mid x\right)\right),
\]
where $\mathscr{V}\left(v\mid x\right)\coloneqq p_{x}^{-2}\left(\mathscr{V}_{1}\left(v,x\right)+\mathscr{V}_{2}\left(v,x\right)\right)$,
\begin{eqnarray*}
\mathscr{V}_{1}\left(v,x\right) & \coloneqq & f_{\varDelta X}\left(v,x\right)\int K\left(u\right)^{2}\mathrm{d}u,\\
\mathscr{V}_{2}\left(v,x\right) & \coloneqq & \left\{ \sum_{j=1}^{m}\left(\frac{f_{\epsilon DX}\left(\varDelta_{x,j}^{-1}\left(v\right),0,x\right)}{\zeta_{1x}\left(g\left(1,x,\varDelta_{x,j}^{-1}\left(v\right)\right)\right)}-\frac{f_{\epsilon DX}\left(\varDelta_{x,j}^{-1}\left(v\right),1,x\right)}{\zeta_{0x}\left(g\left(0,x,\varDelta_{x,j}^{-1}\left(v\right)\right)\right)}\right)^{2}\frac{f_{\epsilon\mid X}\left(\varDelta_{x,j}^{-1}\left(v\right)\mid x\right)}{\left|\varDelta_{x,j}'\left(\varDelta_{x,j}^{-1}(v)\right)\right|^{3}}\right\} \\
 &  & \times\left(p_{1x}^{-1}+p_{0x}^{-1}\right)\int K\left(u\right)^{2}\mathrm{d}u,
\end{eqnarray*}
and $\varDelta_{x,j}'$ is the derivative of $\varDelta_{x,j}$ defined
in Assumption \ref{assu: DGP4}.
\end{thm}
\begin{rembold}Under the same assumptions, the infeasible kernel
estimator that uses the true ITEs satisfies
\[
\sqrt{nh}\left(\widetilde{f}_{\varDelta\mid X}\left(v\mid x;h\right)-f_{\varDelta\mid X}\left(v\mid x\right)-f_{\varDelta\mid X}^{\left(P\right)}\left(v\mid x\right)\mu_{K,P}h^{P}\right)\rightarrow_{d}\mathrm{N}\left(0,p_{x}^{-2}\mathscr{V}_{1}\left(v,x\right)\right).
\]
Note that the estimation of ITEs does not affect the leading bias
term.\end{rembold}

\begin{rembold}\label{rmk: proof of normality}In the proof of Lemma
\ref{lem:lemma 3}, we show that $\mathrm{Var}\left[\mathcal{G}_{x}^{\left[1\right]}\left(W,v;b\right)\right]=b^{-1}\left(\mathscr{V}_{2}\left(v,x\right)+o\left(1\right)\right)$
uniformly in $\left(v,b\right)\in I_{x}\times\left[\underline{h},\overline{h}\right]$.
The proof of this result and derivation of the form of $\mathscr{V}_{2}\left(v,x\right)$
crucially rely on Assumption \ref{assu: DGP4}(b,c). It is clear from
the definition of $\zeta_{dx}$, the fact that $\mathscr{S}_{\epsilon\mid X=x}=\mathscr{S}_{\epsilon\mid D=d,X=x}$,
and Assumption \ref{assu: DGP1}(g) that for all $j$,
\begin{equation}
\left(\frac{f_{\epsilon DX}\left(\varDelta_{x,j}^{-1}\left(v\right),0,x\right)}{\zeta_{1x}\left(g\left(1,x,\varDelta_{x,j}^{-1}\left(v\right)\right)\right)}-\frac{f_{\epsilon DX}\left(\varDelta_{x,j}^{-1}\left(v\right),1,x\right)}{\zeta_{0x}\left(g\left(0,x,\varDelta_{x,j}^{-1}\left(v\right)\right)\right)}\right)^{2}\frac{f_{\epsilon\mid X}\left(\varDelta_{x,j}^{-1}\left(v\right)\mid x\right)}{\left|\varDelta_{x,j}'\left(\varDelta_{x,j}^{-1}(v)\right)\right|^{3}}>0\label{eq:V_2 > 0}
\end{equation}
and therefore, $\mathscr{V}_{2}\left(v,x\right)>0$. Moreover, in
the proof of Lemma \ref{lem:lemma random btw}, we show that $\sqrt{nb}\left(n^{-1}\sum_{i=1}^{n}\mathcal{G}_{x}^{\left[1\right]}\left(W_{i},v;b\right)\right)-\sqrt{nh}\left(n^{-1}\sum_{i=1}^{n}\mathcal{G}_{x}^{\left[1\right]}\left(W_{i},v;h\right)\right)=o_{p}\left(1\right)$,
uniformly in $b\in\left[\underline{h},\overline{h}\right]$. Further,
in the proof of Theorem \ref{thm:asymptotic normality}, we show that
$\sqrt{nh}\left(n^{-1}\sum_{i=1}^{n}\mathcal{G}_{x}^{\left[1\right]}\left(W_{i},v;h\right)\right)\rightarrow_{d}\mathrm{N}\left(0,\mathscr{V}_{2}\left(v,x\right)\right)$
and the second term on the right-hand side of (\ref{eq:G Hoeffding})
is $O_{p}\left(\left(nh^{2}\right)^{-1}\right)$. Therefore, in (\ref{eq:G Hoeffding}),
$n^{-1}\sum_{i=1}^{n}\mathcal{G}_{x}^{\left[1\right]}\left(W_{i},v;b\right)$
dominates the second term under our assumption on the rate of $h$.\end{rembold}

The $p_{x}^{-2}\mathscr{V}_{1}\left(v,x\right)$ term in the asymptotic
variance of the FVX estimator is the asymptotic variance of the infeasible
estimator. The $p_{x}^{-2}\mathscr{V}_{2}\left(v,x\right)$ term is
due to the estimation of the ITEs. Thus, the estimation of the ITEs
increases the variance (but not the bias). To illustrate the effect
of estimation of the ITEs numerically, consider the DGP used for the
Monte Carlo simulations in Section \ref{sec:Monte-Carlo-experiments}
with no controls $X$. The treatment status $D$ is determined by
the index model in (\ref{eq:model simulation}) with coefficients
$(\gamma_{0},\gamma_{1})=(-0.5,0.5)$. The kernel function $K$ is
taken to be the triweight kernel. In this case for $v=2$, $\mathscr{V}_{1}\left(v\right)=0.16$
and $\mathscr{V}_{2}\left(v\right)=2.30$. Hence, the contribution
of the ITE estimation errors to the asymptotic variance of the FVX
estimator can be substantial and even exceed the asymptotic variance
of the infeasible estimator.

\section{Inference\label{sec:Robust-inference}}

In this section, we discuss the construction of asymptotically valid
standard errors as well as construction of asymptotically valid UCBs
for $\left\{ f_{\varDelta\mid X}\left(v\mid x\right):v\in I_{x}\right\} $.
We maintain Assumptions \ref{assu: DGP1}, \ref{assu: DGP4} and \ref{assu:kernel}
with $P=2$. We also maintain the assumption that, as in the practical
implementation of many nonparametric econometric methods, the bandwidth
is data-driven and satisfies Assumption \ref{assu:h_hat}.

\subsection{Standard errors\label{subsec:Standard-errors}}

Inference for $f_{\varDelta\mid X}\left(v\mid x\right)$ requires
a consistent estimator of the asymptotic variance term $\mathscr{V}\left(v\mid x\right)$
defined in Theorem \ref{thm:asymptotic normality}. By the same arguments
as those used to establish Theorems \ref{thm:uniform rate} and \ref{thm:asymptotic normality},
one can show that
\begin{multline}
\sqrt{n\widehat{h}}\left(\widehat{f}_{\varDelta\mid X}\left(v\mid x;\widehat{h}\right)-f_{\varDelta\mid X}\left(v\mid x\right)\right)=\frac{1}{\sqrt{n}}\sum_{i=1}^{n}p_{x}^{-1}\left(\mathcal{U}_{x}^{\left[1\right]}\left(W_{i},v;h\right)-\mu_{\mathcal{U}_{x}}\left(v;h\right)\right)\\
+O_{p}\left(\varepsilon_{n}\sqrt{\mathrm{log}\left(n\right)}+\frac{\mathrm{log}\left(n\right)}{\sqrt{nh^{3}}}+\left(\frac{\mathrm{log}\left(n\right)^{3}}{nh^{2}}\right)^{1/4}+\sqrt{\mathrm{log}\left(n\right)h}+\sqrt{nh^{5}}\right),\label{eq:bias corrected linearization}
\end{multline}
where
\begin{eqnarray*}
\mathcal{U}_{x}^{\left[1\right]}\left(w,v;b\right) & \coloneqq & \mathrm{E}\left[\mathcal{U}_{x}\left(W,w,v;b\right)\right],\\
\mathcal{U}_{x}\left(W_{i},W_{j},v;b\right) & \coloneqq & \frac{1}{\sqrt{b}}K\left(\frac{\varDelta_{j}-v}{b}\right)\mathbbm{1}\left(X_{i}=x\right)+\sqrt{b}\cdot\mathcal{G}_{x}\left(W_{i},W_{j},v;b\right),\\
\mu_{\mathcal{U}_{x}}\left(v;b\right) & \coloneqq & \mathrm{E}\left[\mathcal{U}_{x}\left(W_{1},W_{2},v;b\right)\right].
\end{eqnarray*}
Note that $\varDelta_{j}$ can be expressed as a function of $W_{j}$
(see (\ref{eq:Delta})). Also note that the second H$\acute{\mathrm{a}}$jek
projection term $\mathcal{U}_{x}^{\left[2\right]}\left(w,v;b\right)\coloneqq\mathrm{E}\left[\mathcal{U}_{x}\left(w,W,v;b\right)\right]$
is constant and equal to $\mu_{\mathcal{U}_{x}}\left(v;b\right)=\sqrt{b}\cdot m_{\varDelta X}\left(v,x;b\right)$.
Since $\epsilon$ is conditionally independent of $Z$ given $X$,
one can show that the finite-sample variance of the right-hand side
term in (\ref{eq:bias corrected linearization}) is given by $V\left(v\mid x;h\right)\coloneqq p_{x}^{-2}\mathrm{Var}\left[\mathcal{U}_{x}^{\left[1\right]}\left(W,v;h\right)\right]=p_{x}^{-2}V\left(v,x;h\right)$
(see (\ref{eq:cross term zero})), where $V\left(v,x;b\right)\coloneqq V_{1}\left(v,x;b\right)+V_{2}\left(v,x;b\right)$
and
\begin{eqnarray*}
V_{1}\left(v,x;b\right) & \coloneqq & \mathrm{E}\left[\frac{1}{b}K\left(\frac{\varDelta-v}{b}\right)^{2}\mathbbm{1}\left(X=x\right)\right]-b\cdot m_{\varDelta X}\left(v,x;b\right)^{2},\\
V_{2}\left(v,x;b\right) & \coloneqq & \mathrm{E}\left[\frac{1}{b^{3}}K'\left(\frac{\varDelta_{3}-v}{b}\right)q_{x}\left(W_{3},W_{1}\right)K'\left(\frac{\varDelta_{2}-v}{b}\right)q_{x}\left(W_{2},W_{1}\right)\mathbbm{1}\left(X_{1}=x\right)\right]\\
 &  & \times p_{x}^{-1}\left(p_{1x}^{-1}+p_{0x}^{-1}\right).
\end{eqnarray*}

The plug-in estimator of the $V_{1}\left(v,x;b\right)$ term is given
by
\begin{equation}
\widehat{V}_{1}\left(v,x;b\right)\coloneqq\frac{1}{n}\sum_{i=1}^{n}\frac{1}{b}K\left(\frac{\widehat{\varDelta}_{i}-v}{b}\right)^{2}\mathbbm{1}\left(X_{i}=x\right)-b\cdot\widehat{f}_{\varDelta X}\left(v,x;b\right)^{2}.\label{eq:V_hat_dagger definition}
\end{equation}
Denote $\widehat{p}_{zx}\coloneqq n^{-1}\sum_{i=1}^{n}\mathbbm{1}\left(Z_{i}=z,X_{i}=x\right)$
and $\widehat{p}_{z\mid x}\coloneqq\widehat{p}_{zx}/\widehat{p}_{x}$.
Let
\begin{eqnarray}
\widehat{\zeta}_{1x}\left(y;b_{\zeta}\right) & \coloneqq & \frac{\sum_{i=1}^{n}\frac{1}{b_{\zeta}}K_{\zeta}\left(\frac{Y_{i}-y}{b_{\zeta}}\right)D_{i}\left(\frac{Z_{i}-\widehat{p}_{1\mid x}}{\widehat{p}_{1\mid x}\widehat{p}_{0\mid x}}\right)\mathbbm{1}\left(X_{i}=x\right)}{\sum_{i=1}^{n}\mathbbm{1}\left(X_{i}=x\right)}\nonumber \\
\widehat{\zeta}_{0x}\left(y;b_{\zeta}\right) & \coloneqq & \frac{\sum_{i=1}^{n}\frac{1}{b_{\zeta}}K_{\zeta}\left(\frac{Y_{i}-y}{b_{\zeta}}\right)\left(1-D_{i}\right)\left(\frac{\widehat{p}_{0\mid x}-\left(1-Z_{i}\right)}{\widehat{p}_{1\mid x}\widehat{p}_{0\mid x}}\right)\mathbbm{1}\left(X_{i}=x\right)}{\sum_{i=1}^{n}\mathbbm{1}\left(X_{i}=x\right)}\label{eq:reweighted kernel estimators}
\end{eqnarray}
be the reweighted kernel estimator proposed by \citet{Abadie2002},
where $b_{\zeta}>0$ is the bandwidth and $K_{\zeta}\left(\cdot\right)$
is a second-order kernel. Let
\[
\widehat{R}_{d'x}\left(y\right)\coloneqq\frac{\sum_{i=1}^{n}\left\{ \mathbbm{1}\left(Y_{i}\leq\widehat{\phi}_{dx}\left(y\right),D_{i}=d,X_{i}=x\right)+\mathbbm{1}\left(Y_{i}\leq y,D_{i}=d',X_{i}=x\right)\right\} }{\sum_{i=1}^{n}\mathbbm{1}\left(X_{i}=x\right)}
\]
be the plug-in nonparametric estimator of $R_{d'x}$. The $V_{2}\left(v,x;b\right)$
term can be estimated by a \textit{U}-statistic with an estimated
kernel: 
\begin{multline}
\widehat{V}_{2}\left(v,x;b,b_{\zeta}\right)\coloneqq\frac{1}{n\left(n-1\right)\left(n-2\right)}\sum_{i=1}^{n}\sum_{j\neq i}\sum_{k\neq i,\,k\neq j}\frac{1}{b^{3}}K'\left(\frac{\widehat{\varDelta}_{j}-v}{b}\right)\widehat{q}_{x}\left(W_{j},W_{i};b_{\zeta}\right)\\
\times K'\left(\frac{\widehat{\varDelta}_{k}-v}{b}\right)\widehat{q}_{x}\left(W_{k},W_{i};b_{\zeta}\right)\mathbbm{1}\left(X_{i}=x\right)\widehat{p}_{x}^{-1}\left(\widehat{p}_{1x}^{-1}+\widehat{p}_{0x}^{-1}\right),\label{eq:V_hat_ddager definition}
\end{multline}
where $\widehat{q}_{x}\left(W_{i},W_{j};b_{\zeta}\right)\coloneqq\widehat{q}_{1x}\left(W_{i},W_{j};b_{\zeta}\right)-\widehat{q}_{0x}\left(W_{i},W_{j};b_{\zeta}\right)$,
$\widehat{q}_{dx}\left(W_{i},W_{j};b_{\zeta}\right)$ is the plug-in
nonparametric estimator of $q_{dx}\left(W_{i},W_{j}\right)$ defined
in (\ref{eq:q_dx definition}) constructed by replacing $\zeta_{dx}$,
$\phi_{dx}$, and $R_{d'x}$ with their nonparametric estimators $\widehat{\zeta}_{dx}\left(\cdot;b_{\zeta}\right)$,
$\widehat{\phi}_{dx},$ and $\widehat{R}_{d'x}$ respectively.\footnote{It is known that the kernel estimator $\widehat{\zeta}_{dx}\left(y;b_{\zeta}\right)$
is asymptotically biased if $y$ is near the boundaries of the support
$\left[\underline{y}_{dx},\overline{y}_{dx}\right]$. As \citet{Guerre2000},
we can trim off the estimated counterfactual outcomes $\widehat{\phi}_{dx}\left(Y_{i}\right)$
that lie in the boundary region $\left[\underline{y}_{dx},\underline{y}_{dx}+b_{\zeta}\right)\cup\left(\overline{y}_{dx}-b_{\zeta},\overline{y}_{dx}\right]$
by multiplying $\widehat{\zeta}_{dx}\left(\widehat{\phi}_{dx}\left(Y_{i}\right);b_{\zeta}\right)^{-1}$
in $\widehat{q}_{dx}\left(W_{j},W_{i};b_{\zeta}\right)$ by a trimming
factor $\mathbbm{1}\left(\underline{y}_{dx}+b_{\zeta}\leq\widehat{\phi}_{dx}\left(Y_{i}\right)\leq\overline{y}_{dx}-b_{\zeta}\right)$.
It can be shown that the effect of the trimming factor is asymptotically
negligible. All of our asymptotic results remain true, and the finite-sample
performances may improve when trimming is used.}

Let 
\begin{equation}
\widehat{V}\left(v\mid x;b,b_{\zeta}\right)\coloneqq\widehat{p}_{x}^{-2}\left(\widehat{V}_{1}\left(v,x;b\right)+\widehat{V}_{2}\left(v,x;b,b_{\zeta}\right)\right).\label{eq:asy var}
\end{equation}
For estimating $V\left(v\mid x;h\right)$, we set $b=\widehat{h}$
in $\widehat{V}\left(v\mid x;b,b_{\zeta}\right)$, where $\widehat{h}$
satisfies Assumption \ref{assu:h_hat}. Similarly, we set the second
bandwidth $b_{\zeta}=\widehat{h}_{\zeta}$, where $\widehat{h}_{\zeta}$
is a random bandwidth that satisfies the following assumption similar
to Assumption \ref{assu:h_hat}.
\begin{assumption}
\label{assu:btw zeta}$\mathrm{Pr}\left[\left|\widehat{h}_{\zeta}/h_{\zeta}-1\right|>\varepsilon_{n}^{\zeta}\right]\leq\delta_{n}^{\zeta}$
for some deterministic bandwidth $h_{\zeta}$ and positive sequences
$\varepsilon_{n}^{\zeta},\delta_{n}^{\zeta}\downarrow0$.
\end{assumption}
Suppose $\widehat{h}_{\zeta}$ is the Silverman ROT bandwidth of the
form $\widehat{h}_{\zeta}^{\mathsf{rot}}=C_{K_{\zeta}}\cdot\widehat{\sigma}_{Y\mid X=x}\cdot n_{x}^{-1/5}$,
where $\widehat{\sigma}_{Y\mid X=x}$ is the sample analogue of $\sigma_{Y\mid X=x}\coloneqq\sqrt{\mathrm{Var}\left[Y\mid X=x\right]}$
and $C_{K_{\zeta}}$ is a constant that depends on $K_{\zeta}$. In
this case, Assumption \ref{assu:btw zeta} is satisfied with $\varepsilon_{n}^{\zeta}\wasypropto\sqrt{\mathrm{log}\left(n\right)/n}$
and $\delta_{n}^{\zeta}\wasypropto n^{-1}$.\footnote{One may use estimators $\left(\widehat{\zeta}_{0x}\left(y;b_{\zeta,0}\right),\widehat{\zeta}_{1x}\left(y;b_{\zeta,1}\right)\right)$with
different bandwidths $\left(b_{\zeta,0},b_{\zeta,1}\right)$. By easily
modifying the proofs, we get results similar to Theorems \ref{thm:variance estimator}
and \ref{thm:confidence band} under two data-dependent bandwidths
$\left(\widehat{h}_{\zeta,0},\widehat{h}_{\zeta,1}\right)$ that satisfy
the same assumption for $\widehat{h}_{\zeta}$. The ROT bandwidths
can be set as $\widehat{h}_{\zeta,d}^{\mathsf{rot}}=C_{K_{\zeta}}\cdot\widehat{\sigma}_{Y\mid X=x,D=d}\cdot n_{dx}^{-1/5}$,
where $n_{dx}\coloneqq\sum_{i=1}^{n}\mathbbm{1}\left(D_{i}=d,X_{i}=x\right)$,
for $d=0,1$, and $\widehat{\sigma}_{Y\mid X=x,D=d}$ denotes the
sample analogue of $\sigma_{Y\mid X=x,D=d}\coloneqq\sqrt{\mathrm{Var}\left[Y\mid X=x,D=d\right]}$.} Theorem \ref{thm:variance estimator} below provides a uniform convergence
rate for $\widehat{V}\left(v\mid x;\widehat{h},\widehat{h}_{\zeta}\right)$.
In Appendix \ref{sec:Proofs B}, Theorem \ref{thm:variance theorem appendix}
presents a non-asymptotic deviation bound for the uniform estimation
error of $\widehat{V}\left(v\mid x;\widehat{h},\widehat{h}_{\zeta}\right)$,
which implies the result of Theorem \ref{thm:variance estimator}.
The stronger result of Theorem \ref{thm:variance theorem appendix}
is used in the proof of Theorem \ref{thm:confidence band} below.
\begin{thm}
\label{thm:variance estimator}Suppose that Assumptions \ref{assu: DGP1}-\ref{assu:btw zeta}
hold with $P=2$, the third-order derivative functions in Assumption
\ref{assu: DGP4}(a) are Lipschitz continuous, $h\wasypropto n^{-\lambda}$
with $0<\lambda<1/4$, and $h_{\zeta}\wasypropto n^{-\lambda_{\zeta}}$
with $0<\lambda_{\zeta}<1$. Then, for any $x\in\mathscr{S}_{X}$
and compact $I_{x}\subseteq\mathrm{Int}\left(\mathscr{S}_{\Delta\mid X=x}\right)$,
\[
\left\Vert \widehat{V}\left(\cdot\mid x;\widehat{h},\widehat{h}_{\zeta}\right)-V\left(\cdot\mid x;h\right)\right\Vert _{I_{x}}=O_{p}\left(\sqrt{\frac{\mathrm{log}\left(n\right)}{nh_{\zeta}}}+h_{\zeta}^{2}+\frac{\mathrm{log}\left(n\right)}{nh^{4}}+\sqrt{\frac{\mathrm{log}\left(n\right)}{nh^{2}}}+\varepsilon_{n}h\right).
\]
\end{thm}
\begin{rembold}While the estimator $\widehat{V}_{2}\left(v,x;b,b_{\zeta}\right)$
may be negative in finite samples, its modification $\reallywidecheck{V}_{2}\left(v,x;b,b_{\zeta}\right)$
defined below is always non-negative: 
\[
\reallywidecheck{V}_{2}\left(v,x;b,b_{\zeta}\right)\coloneqq\frac{1}{n}\sum_{i=1}^{n}\frac{1}{b}\left\{ \frac{1}{n}\sum_{j=1}^{n}\frac{1}{b}K'\left(\frac{\widehat{\varDelta}_{j}-v}{b}\right)\widehat{q}_{x}\left(W_{j},W_{i};b_{\zeta}\right)\right\} ^{2}\mathbbm{1}\left(X_{i}=x\right)\widehat{p}_{x}^{-1}\left(\widehat{p}_{1x}^{-1}+\widehat{p}_{0x}^{-1}\right).
\]
One can show that the difference $\left\Vert \reallywidecheck{V}_{2}\left(\cdot,x;\widehat{h},\widehat{h}_{\zeta}\right)-\widehat{V}_{2}\left(\cdot,x;\widehat{h},\widehat{h}_{\zeta}\right)\right\Vert _{I_{x}}$
is of a smaller order than $\left\Vert \widehat{V}_{2}\left(\cdot,x;\widehat{h},\widehat{h}_{\zeta}\right)-V_{2}\left(\cdot,x;h\right)\right\Vert _{I_{x}}$.\end{rembold}

A pointwise $1-\alpha$ asymptotic confidence interval for $f_{\Delta\mid X}\left(v\mid x\right)$
can be constructed as
\begin{equation}
\left[\widehat{f}_{\varDelta\mid X}\left(v\mid x;\widehat{h}\right)\pm z_{1-\alpha/2}\sqrt{\frac{\widehat{V}\left(v\mid x;\widehat{h},\widehat{h}_{\zeta}\right)}{n\widehat{h}}}\right],\label{eq:pointwise CI}
\end{equation}
where $z_{1-\alpha/2}$ is the $1-\alpha/2$ quantile of the standard
normal distribution, and $h$ satisfies $nh^{5}\downarrow0$. However,
if one is interested in constructing valid confidence bands for the
density function, the $z_{1-\alpha/2}$ critical value must be replaced
with a bigger one determined by the distribution of the supremum of
the estimation errors along the domain, as interpolations of the pointwise
confidence intervals (\ref{eq:pointwise CI}) over the domain are
invalid in the uniform sense. In the section below, we discuss the
construction of valid UCBs.

\subsection{Jackknife multiplier bootstrap UCB\label{subsec:Jackknife-multiplier-bootstrap}}

Let $S\left(v\mid x;b\right)$ and $Z\left(v\mid x;b,b_{\zeta}\right)$
denote the non-studentized and studentized estimation errors, respectively:
\begin{equation}
S\left(v\mid x;b\right)\coloneqq\sqrt{nb}\left(\widehat{f}_{\varDelta\mid X}\left(v\mid x;b\right)-f_{\varDelta\mid X}\left(v\mid x\right)\right)\textrm{ and }Z\left(v\mid x;b,b_{\zeta}\right)\coloneqq\frac{S\left(v\mid x;b\right)}{\sqrt{\widehat{V}\left(v\mid x;b,b_{\zeta}\right)}}.\label{eq:S Z definition}
\end{equation}
Moreover, recall the expansion of the estimation error in (\ref{eq:bias corrected linearization}).
An asymptotically valid $1-\alpha$ UCB simultaneously covers $\left\{ f_{\varDelta\mid X}\left(v\mid x\right):v\in I_{x}\right\} $
with a pre-specified asymptotic coverage $1-\alpha$. To construct
a valid UCB, one has to replace the standard normal quantile $z_{1-\alpha/2}$
in the pointwise confidence interval (\ref{eq:pointwise CI}) with
a critical value approximating the $1-\alpha$ quantile of the distribution
of $\left\Vert Z\left(\cdot\mid x;\widehat{h},\widehat{h}_{\zeta}\right)\right\Vert _{I_{x}}$.
In this section, we discuss the validity of the computationally fast
jackknife multiplier bootstrap (JMB). Appendix \ref{sec:Nonparametric-bootstrap}
provides the algorithm and theoretical results for the nonparametric
bootstrap.

We consider the problem of estimating the distribution of $\left\Vert S\left(\cdot\mid x;\widehat{h}\right)\right\Vert _{I_{x}}$
or $\left\Vert Z\left(\cdot\mid x;\widehat{h},\widehat{h}_{\zeta}\right)\right\Vert _{I_{x}}$
using the linearization (\ref{eq:bias corrected linearization}),
under the ``undersmoothing'' assumption $nh^{5}\downarrow0$ to
ensure that the bias is asymptotically negligible in comparison to
the standard deviation. The conventional bandwidth selectors that
estimate the AMISE-optimal bandwidth such as the Silverman ROT method
or cross-validation (using the pseudo ITEs in our case) violate the
undersmoothing assumption. The undersmoothing assumption requires
that the selected bandwidth should vanish at a faster rate. In practical
implementation of undersmoothing for many nonparametric econometric
techniques, a commonly used strategy is to shrink a conventional approximately
AMISE-optimal data-driven bandwidth by an ad hoc amount.

The JMB approach of \citet[ CK hereafter,]{Chen2020jackknife} approximates
the distribution of the supremum (with respect to $v$) of $n^{-1/2}\sum_{i=1}^{n}\left(\mathcal{U}_{x}^{\left[1\right]}\left(W_{i},v;h\right)-\mu_{\mathcal{U}_{x}}\left(v;h\right)\right)$
with that of the Gaussian multiplier process (e.g., \citealp{Chernozhukov2014anti})
that uses the jackknife estimator of $\mathcal{U}_{x}^{\left[1\right]}\left(W_{i},v;h\right)$.
However unlike in CK, in our case the kernel $\mathcal{U}_{x}$ involves
the unknown nonparametric objects ($\phi_{dx},\zeta_{dx},R_{d'x}$),
unknown probabilities ($p_{0x}$ and $p_{1x}$), and latent ITEs.
Therefore, we use the estimated version of $\mathcal{U}_{x}$ that
replaces the unknown objects with their nonparametric estimators:
\begin{equation}
\widehat{\mathcal{U}}_{x}\left(W_{j},W_{i},v;b,b_{\zeta}\right)\coloneqq\frac{1}{\sqrt{b}}K\left(\frac{\widehat{\varDelta}_{i}-v}{b}\right)\mathbbm{1}\left(X_{i}=x\right)+\frac{1}{b^{3/2}}K'\left(\frac{\widehat{\varDelta}_{j}-v}{b}\right)\widehat{q}_{x}\left(W_{j},W_{i};b_{\zeta}\right)\widehat{\pi}_{x}\left(Z_{i},X_{i}\right),\label{eq:estimated kernel}
\end{equation}
where $\widehat{\pi}_{x}\left(Z_{i},X_{i}\right)$ is constructed
by replacing $\left(p_{0x},p_{1x}\right)$ with $\left(\widehat{p}_{0x},\widehat{p}_{1x}\right)$
in the definition of $\pi_{x}\left(Z_{i},X_{i}\right)$ in (\ref{eq:pi_x definition}).
Let $\left(\nu_{1},...,\nu_{n}\right)$ denote i.i.d. standard normal
random variables that are drawn independently from the data. Let $\left\{ \widehat{S}_{\mathsf{jmb}}\left(\cdot\mid x;b,b_{\zeta}\right):v\in I_{x}\right\} $
be the feasible JMB process, where
\begin{eqnarray}
\widehat{\mathcal{U}}_{x}^{\left[1\right]}\left(W_{i},v;b,b_{\zeta}\right) & \coloneqq & \frac{1}{n-1}\sum_{j\neq i}\widehat{\mathcal{U}}_{x}\left(W_{j},W_{i},v;b,b_{\zeta}\right),\nonumber \\
\widehat{S}_{\mathsf{jmb}}\left(v\mid x;b,b_{\zeta}\right) & \coloneqq & \frac{1}{\sqrt{n}}\sum_{i=1}^{n}\nu_{i}\widehat{p}_{x}^{-1}\left\{ \widehat{\mathcal{U}}_{x}^{\left[1\right]}\left(W_{i},v;b,b_{\zeta}\right)-\sqrt{b}\cdot\widehat{f}_{\varDelta X}\left(v,x;b\right)\right\} .\label{eq:Z bootstrap analogue}
\end{eqnarray}
We show in Appendix \ref{sec:Proofs B} (the proof of Theorem \ref{thm:confidence band appendix})
that the distribution of $\left\Vert Z\left(\cdot\mid x;\widehat{h},\widehat{h}_{\zeta}\right)\right\Vert _{I_{x}}$
can be approximated by the conditional distribution of $\left\Vert \widehat{Z}_{\mathsf{jmb}}\left(\cdot\mid x;\widehat{h},\widehat{h}_{\zeta}\right)\right\Vert _{I_{x}}$
given the original sample $W_{1}^{n}\coloneqq\left\{ W_{1},...,W_{n}\right\} $,
where 
\[
\widehat{Z}_{\mathsf{jmb}}\left(v\mid x;b,b_{\zeta}\right)\coloneqq\frac{\widehat{S}_{\mathsf{jmb}}\left(v\mid x;b,b_{\zeta}\right)}{\sqrt{\widehat{V}\left(v\mid x;b,b_{\zeta}\right)}}.
\]
Let $\mathrm{Pr}_{\mid W_{1}^{n}}\left[\cdot\right]$ and $\mathrm{E}_{\mid W_{1}^{n}}\left[\cdot\right]$
denote the conditional probability and expectation respectively given
$W_{1}^{n}$, and

\begin{equation}
z_{1-\alpha}^{\mathsf{jmb}}\coloneqq\mathrm{inf}\left\{ t\in\mathbb{R}:\mathrm{Pr}_{\mid W_{1}^{n}}\left[\left\Vert \widehat{Z}_{\mathsf{jmb}}\left(\cdot\mid x;\widehat{h},\widehat{h}_{\zeta}\right)\right\Vert _{I_{x}}\leq t\right]\geq1-\alpha\right\} \label{eq:z_pound definition}
\end{equation}
be the $1-\alpha$ quantile of the conditional distribution of $\left\Vert \widehat{Z}_{\mathsf{jmb}}\left(\cdot\mid x;\widehat{h},\widehat{h}_{\zeta}\right)\right\Vert _{I_{x}}$
given $W_{1}^{n}$. Recall that $\mathit{CB}_{\mathsf{jmb}}\left(v\mid x;b,b_{\zeta}\right)$
is defined by (\ref{eq:confidence band definition}). The JMB confidence
band is given by the family of random intervals $\left\{ \mathit{CB}_{\mathsf{jmb}}\left(v\mid x;\widehat{h},\widehat{h}_{\zeta}\right):v\in I_{x}\right\} $.
Note that one can approximate $z_{1-\alpha}^{\mathsf{jmb}}$ to any
degree of accuracy by Monte Carlo simulations, and that the width
of $\mathit{CB}_{\mathsf{jmb}}\left(v\mid x;\widehat{h},\widehat{h}_{\zeta}\right)$
varies with $\widehat{V}\left(v\mid x;\widehat{h},\widehat{h}_{\zeta}\right)$.\footnote{\label{fn:constant width}Alternatively, a constant-width UCB (e.g.,
\citealp{Cheng2019}) $\left[\widehat{f}_{\varDelta\mid X}\left(v\mid x;\widehat{h}\right)\pm s_{1-\alpha}^{\mathsf{jmb}}/\sqrt{n\widehat{h}}\right]$
is based on the critical value $s_{1-\alpha}^{\mathsf{jmb}}\coloneqq\mathrm{inf}\left\{ t\in\mathbb{R}:\mathrm{Pr}_{\mid W_{1}^{n}}\left[\left\Vert \widehat{S}_{\mathsf{jmb}}\left(\cdot\mid x;\widehat{h},\widehat{h}_{\zeta}\right)\right\Vert _{I_{x}}\leq t\right]\geq1-\alpha\right\} $
that approximates the $1-\alpha$ quantile of the distribution of
$\left\Vert S\left(\cdot\mid x;\widehat{h}\right)\right\Vert _{I_{x}}$.
However, such a UCB cannot exploit the fact that the tails of a density
function approach zero, and therefore the variable-width UCB is preferred
in this context.} The following algorithm summarizes the construction of the JMB confidence
band for the density of the ITE.
\begin{lyxalgorithm}[JMB confidence band]
\label{alg:multiplier}\textbf{Step 1}: Compute the pseudo ITEs using
(\ref{eq:phi_hat definition})-(\ref{eq:pseudo ITE definition}).
\textbf{Step 2}: Select the covariates' value $x$, the number of
grid points $G$, and a grid $I_{x}^{G}\coloneqq\left\{ v_{1}<v_{2}<\ldots<v_{G}\right\} $
over which the density is estimated. \textbf{Step 3}: Select a kernel
$K$, use the ROT bandwidth $\widehat{h}^{\mathsf{rot}}$ with undersmoothing
(e.g., $\widehat{h}=\widehat{h}^{\mathsf{rot}}n^{-1/40}$), and for
all $v\in I_{x}^{G}$ compute the kernel density estimator $\widehat{f}_{\varDelta\mid X}\left(v\mid x;\widehat{h}\right)$
using (\ref{eq:f_hat definition}) and the pseudo ITEs from Step 1.
\textbf{Step 4}: Select a kernel $K_{\zeta}$, use the ROT bandwidth
$\widehat{h}_{\zeta}^{\mathsf{rot}}$, and for all $v\in I_{x}^{G}$
compute the variance estimator $\widehat{V}\left(v\mid x;\widehat{h},\widehat{h}_{\zeta}\right)$
using (\ref{eq:V_hat_dagger definition}), (\ref{eq:V_hat_ddager definition}),
and (\ref{eq:asy var}). \textbf{Step 5}: Select the number of bootstrap
repetitions $B$, for $r=1,...,B$ generate i.i.d. standard normal
random variables $\left(\nu_{1}^{\left(r\right)},...,\nu_{n}^{\left(r\right)}\right)$,
and for all $v\in I_{x}^{G}$ compute $\widehat{Z}_{\mathsf{jmb}}^{\left(r\right)}\left(v\mid x;\widehat{h},\widehat{h}_{\zeta}\right)=\widehat{S}_{\mathsf{jmb}}^{\left(r\right)}\left(v\mid x;\widehat{h},\widehat{h}_{\zeta}\right)/\sqrt{\widehat{V}\left(v\mid x;\widehat{h},\widehat{h}_{\zeta}\right)}$
using (\ref{eq:Z bootstrap analogue}) with $\left(\nu_{1},...,\nu_{n}\right)$
replaced by $\left(\nu_{1}^{\left(r\right)},...,\nu_{n}^{\left(r\right)}\right)$.
\textbf{Step 6}: Select the coverage level $1-\alpha$ and compute
the critical value $z_{1-\alpha}^{\mathsf{jmb}}$:
\begin{equation}
z_{1-\alpha}^{\mathsf{jmb}}=\mathrm{inf}\left\{ t\in\mathbb{R}:\frac{1}{B}\sum_{r=1}^{B}\mathbbm{1}\left(\underset{v\in I_{x}^{G}}{\mathrm{max}}\left|\widehat{Z}_{\mathsf{jmb}}^{\left(r\right)}\left(v\mid x;\widehat{h},\widehat{h}_{\zeta}\right)\right|\leq t\right)\geq1-\alpha\right\} .\label{eq:JMB critical value}
\end{equation}
\textbf{Step 7}: Compute the JMB confidence band $\mathit{CB}_{\mathsf{jmb}}$
using (\ref{eq:confidence band definition}) over $v\in I_{x}^{G}$.
\end{lyxalgorithm}
Theorem \ref{thm:confidence band} below shows that the proposed JMB
confidence band is asymptotically valid and its coverage error decays
at a polynomial rate. The result rules out coverage probability errors
with logarithmic decay rates (see, e.g., \citealp{Chernozhukov2014anti}
for discussion).
\begin{thm}
\label{thm:confidence band}Suppose that Assumptions \ref{assu: DGP1}-\ref{assu:btw zeta}
hold with $P=2$, the third-order derivative functions in Assumption
\ref{assu: DGP4}(a) are Lipschitz continuous, $h\wasypropto n^{-\lambda}$
with $1/5<\lambda<1/4$, $h_{\zeta}\wasypropto n^{-\lambda_{\zeta}}$
with $1/8<\lambda_{\zeta}<1/2$, $\varepsilon_{n}=O\left(n^{-\varrho_{\varepsilon}}\right)$,
$\delta_{n}=O\left(n^{-\varrho_{\delta}}\right)$, and $\delta_{n}^{\zeta}=O\left(n^{-\varrho_{\delta}^{\zeta}}\right)$
for some $\varrho_{\varepsilon},\varrho_{\delta},\varrho_{\delta}^{\zeta}>0$.
Then, for all $x\in\mathscr{S}_{X}$ and any compact $I_{x}\subseteq\mathrm{Int}\left(\mathscr{S}_{\Delta\mid X=x}\right)$,
\[
\mathrm{Pr}\left[f_{\varDelta\mid X}\left(v\mid x\right)\in\mathit{CB}_{\mathsf{jmb}}\left(v\mid x;\widehat{h},\widehat{h}_{\zeta}\right),\,\textrm{for all }v\in I_{x}\right]=\left(1-\alpha\right)+O\left(n^{-\varrho}\right)
\]
for some $\varrho>0$.
\end{thm}
\begin{rembold}In the proof of Theorem \ref{thm:confidence band},
we explicitly derive an estimate of the coverage probability error,
which is presented in Theorem \ref{thm:confidence band appendix}
in Appendix \ref{sec:Proofs B}. We show that replacing the deterministic
bandwidths $\left(h,h_{\zeta}\right)$ with estimators satisfying
Assumptions \ref{assu:h_hat} and \ref{assu:btw zeta} incurs an additional
$O\left(\mathrm{log}\left(n\right)\varepsilon_{n}+\delta_{n}+\delta_{n}^{\zeta}\right)$
coverage probability error term. This result shows how the noise in
the estimated bandwidth translates into an error in the coverage probability.
The JMB confidence band achieves a potentially higher coverage accuracy
if $\left(\widehat{h},\widehat{h}_{\zeta}\right)$ converge to their
population counterparts at a fast rate, which is the case for the
simple and feasible Silverman ROT approach we recommended.\end{rembold}

\begin{rembold}\label{Rmk: truncation} Since density functions are
non-negative, the lower bound of the UCB can be truncated to zero
to avoid negative values. Thus, the lower bound of $\mathit{CB}_{\mathsf{jmb}}\left(v\mid x;\widehat{h},\widehat{h}_{\zeta}\right)$
can be replaced with $\mathrm{max}\left\{ 0,\widehat{f}_{\varDelta\mid X}\left(v\mid x;\widehat{h}\right)-z_{1-\alpha}^{\mathsf{jmb}}\sqrt{\widehat{V}\left(v\mid x;\widehat{h},\widehat{h}_{\zeta}\right)/\left(n\widehat{h}\right)}\right\} $
without affecting the coverage properties of the UCB. Hence, the result
of Theorem \ref{thm:confidence band} continues to hold with the modified
lower bound.\footnote{We thank the associate editor for suggesting the modification to us.}\end{rembold}

Our main focus is on the multiplier bootstrap approach, as it is computationally
fast even with large sample sizes. The more commonly used nonparametric
bootstrap would require re-calculation of the bootstrap versions of
the estimated ITEs at every bootstrap repetition, which can be computationally
burdensome. However, the constant-width version of the nonparametric
bootstrap confidence band has the advantage of fewer tuning parameters
as it does not require estimation of $\zeta_{dx}$ and therefore does
not need the second bandwidth $b_{\zeta}$. The validity of the nonparametric
bootstrap approach is discussed in Appendix \ref{sec:Nonparametric-bootstrap}.

\subsection{Bias-corrected JMB UCB\label{subsec:Bias-corrected-JMB}}

In this section, we discuss the bias correction approach to inference
that can accommodate conventional bandwidth selectors such as ROT
bandwidths that decay at the $n^{-1/5}$ rate as in \citet{Calonico2014}.
We assume that the third-order derivatives in Assumption \ref{assu: DGP4}
with $P=2$ are Lipschitz continuous. Theorem \ref{thm:asymptotic normality}
implies that in large samples and when $P=2$, $\widehat{f}_{\varDelta\mid X}\left(v\mid x;h\right)-f_{\varDelta\mid X}^{(2)}\left(v\mid x\right)\mu_{K,2}h^{2}$
is approximately distributed as $\mathrm{N}\left(f_{\varDelta\mid X}\left(v\mid x\right),\mathscr{V}\left(v\mid x\right)/\left(nh\right)\right)$.
We use an estimator of the density derivative $f_{\varDelta\mid X}^{(2)}\left(v\mid x\right)$
to remove the bias. Let 
\[
\widehat{f}_{\varDelta\mid X}^{(2)}\left(v\mid x;b_{\mathsf{b}}\right)\coloneqq\frac{\sum_{i=1}^{n}\frac{1}{b_{\mathsf{b}}^{3}}K_{\mathsf{b}}^{(2)}\left(\frac{\widehat{\varDelta}_{i}-v}{b_{\mathsf{b}}}\right)\mathbbm{1}\left(X_{i}=x\right)}{\sum_{i=1}^{n}\mathbbm{1}\left(X_{i}=x\right)}
\]
denote the kernel estimator of $f_{\varDelta\mid X}^{(2)}\left(v\mid x\right)$
using a bandwidth $b_{\mathsf{b}}$ and a second-order kernel $K_{\mathsf{b}}$
($K_{\mathsf{b}}^{(2)}$ denotes the second derivative of $K_{\mathsf{b}}$).
We assume that $K_{\mathsf{b}}^{(2)}$ satisfies Assumption \ref{assu:kernel}(a).
The bias-corrected estimator of $f_{\Delta\mid X}\left(v\mid x\right)$
is given by 
\begin{equation}
\widehat{f}_{\varDelta\mid X}^{\mathsf{bc}}\left(v\mid x;b,b_{\mathsf{b}}\right)\coloneqq\widehat{f}_{\varDelta\mid X}\left(v\mid x;b\right)-\widehat{f}_{\varDelta\mid X}^{(2)}\left(v\mid x;b_{\mathsf{b}}\right)\mu_{K,2}b^{2}=\frac{\sum_{i=1}^{n}\frac{1}{b}M\left(\frac{\widehat{\varDelta}_{i}-v}{b};b,b_{\mathsf{b}}\right)\mathbbm{1}\left(X_{i}=x\right)}{\sum_{i=1}^{n}\mathbbm{1}\left(X_{i}=x\right)},\label{eq:debiased estimator definition}
\end{equation}
where $M\left(u;b,b_{\mathsf{b}}\right)\coloneqq K\left(u\right)-\left(b/b_{\mathsf{b}}\right)^{3}\mu_{K,2}K_{\mathsf{b}}^{(2)}\left(\left(b/b_{\mathsf{b}}\right)u\right)$.
Typical choices for the bandwidth $b_{\mathsf{b}}$ used for bias
correction include the estimation-optimal bandwidth for the second-order
density derivative (e.g., \citealp{Xu2017}) or the same bandwidth
as that used for estimating the density \citep[e.g.,][]{Cheng2019}.
We assume that $b_{\mathsf{b}}$ is some random bandwidth $\widehat{h}_{\mathsf{b}}$
that consistently estimates some target bandwidth $h_{\mathsf{b}}$
in the sense that $\widehat{h}_{\mathsf{b}}/h_{\mathsf{b}}\rightarrow_{p}1$.
We assume that $h/h_{\mathsf{b}}\rightarrow\varsigma\in\left[0,\infty\right)$
as in \citet{Calonico2014}. In practice, we can take $\widehat{h}_{\mathsf{b}}$
to be the feasible Silverman ROT bandwidth for the second-order density
derivative that uses the pseudo ITEs and aims to approximate the AMISE-optimal
bandwidth (in this case, $h_{\mathsf{b}}\wasypropto n^{-1/9}$). Alternatively,
one can set $\widehat{h}_{\mathsf{b}}=\widehat{h}$.

One can show that the bias of $\widehat{f}_{\varDelta\mid X}^{\mathsf{bc}}\left(v\mid x;h,h_{\mathsf{b}}\right)$
is of the order $O\left(h^{2}h_{\mathsf{b}}\right)$, and its standard
deviation is of the order $\left(nh\right)^{-1/2}$.\footnote{Under Assumptions \ref{assu: DGP1}-\ref{assu:kernel} with $P=2$,
the bias part of the bias-corrected estimator in (\ref{eq:debiased estimator definition})
is $o\left(h^{2}\right)$. Lipschitz continuity ensures that the bias
part is $O\left(h^{2}h_{\mathsf{b}}\right)$, if $h=O\left(h_{\mathsf{b}}\right)$.} Hence, when the bandwidth $h$ is chosen AMISE-optimally so that
$h\wasypropto n^{-1/5}$, the bias of the bias-corrected estimator
$\widehat{f}_{\varDelta\mid X}^{\mathsf{bc}}\left(v\mid x;h,h_{\mathsf{b}}\right)$
is of smaller order than its standard deviation. Robust standard errors
(\citealp{Calonico2014}) have to take into account the additional
stochastic variability coming from the bias correction, i.e., estimation
of $f_{\varDelta\mid X}^{(2)}\left(v\mid x\right)$.

\sloppy Let $\widehat{V}^{\mathsf{bc}}\left(v\mid x;b,b_{\zeta},b_{\mathsf{b}}\right)$
and $\widehat{S}_{\mathsf{jmb}}^{\mathsf{bc}}\left(v\mid x;b,b_{\zeta},b_{\mathsf{b}}\right)$
be defined similarly to $\widehat{V}\left(v\mid x;b,b_{\zeta}\right)$
and $\widehat{S}_{\mathsf{jmb}}\left(v\mid x;b,b_{\zeta}\right)$
respectively with $K\left(\cdot\right)$ replaced by the bias-correcting
kernel $M\left(\cdot;b,b_{\mathsf{b}}\right)$. Define $\widehat{Z}_{\mathsf{jmb}}^{\mathsf{bc}}\left(v\mid x;b,b_{\zeta},b_{\mathsf{b}}\right)\coloneqq\widehat{S}_{\mathsf{jmb}}^{\mathsf{bc}}\left(v\mid x;b,b_{\zeta},b_{\mathsf{b}}\right)/\sqrt{\widehat{V}^{\mathsf{bc}}\left(v\mid x;b,b_{\zeta},b_{\mathsf{b}}\right)}$,
and let $z_{1-\alpha}^{\mathsf{jmb},\mathsf{bc}}$ be as in (\ref{eq:z_pound definition})
with $\left\Vert \widehat{Z}_{\mathsf{jmb}}\left(\cdot\mid x;\widehat{h},\widehat{h}_{\zeta}\right)\right\Vert _{I_{x}}$
replaced by $\left\Vert \widehat{Z}_{\mathsf{jmb}}^{\mathsf{bc}}\left(\cdot\mid x;\widehat{h},\widehat{h}_{\zeta},\widehat{h}_{\mathsf{b}}\right)\right\Vert _{I_{x}}$.
The bias-corrected JMB UCB is given by 
\begin{equation}
\mathit{CB}_{\mathsf{jmb}}^{\mathsf{bc}}\left(v\mid x;\widehat{h},\widehat{h}_{\zeta},\widehat{h}_{\mathsf{b}}\right)\coloneqq\left[\widehat{f}_{\varDelta\mid X}^{\mathsf{bc}}\left(v\mid x;\widehat{h},\widehat{h}_{\mathsf{b}}\right)\pm z_{1-\alpha}^{\mathsf{jmb},\mathsf{bc}}\sqrt{\frac{\widehat{V}^{\mathsf{bc}}\left(v\mid x;\widehat{h},\widehat{h}_{\zeta},\widehat{h}_{\mathsf{b}}\right)}{n\widehat{h}}}\right].\label{eq:bias corrected CB}
\end{equation}
The bias-corrected UCB can be computed by replacing $\widehat{f}_{\varDelta\mid X}\left(v\mid x;\widehat{h}\right)$,
$\widehat{V}\left(v\mid x;\widehat{h},\widehat{h}_{\zeta}\right)$,
and $\widehat{Z}_{\mathsf{jmb}}\left(v\mid x;\widehat{h},\widehat{h}_{\zeta}\right)$
in Algorithm \ref{alg:multiplier} with their respective bias-corrected
versions $\widehat{f}_{\varDelta\mid X}^{\mathsf{bc}}\left(v\mid x;\widehat{h},\widehat{h}_{\mathsf{b}}\right)$,
$\widehat{V}^{\mathsf{bc}}\left(v\mid x;\widehat{h},\widehat{h}_{\zeta},\widehat{h}_{\mathsf{b}}\right)$,
and $\widehat{Z}_{\mathsf{jmb}}^{\mathsf{bc}}\left(v\mid x;\widehat{h},\widehat{h}_{\zeta},\widehat{h}_{\mathsf{b}}\right)$.
We show that the conclusion of Theorem \ref{thm:confidence band}
(asymptotic validity with a polynomial rate) holds for $\mathit{CB}_{\mathsf{jmb}}^{\mathsf{bc}}$
under the assumption that $h_{\mathsf{b}}\wasypropto n^{-\lambda_{\mathsf{b}}}$,
$\lambda_{\mathsf{b}}\leq\lambda$, $\left(1-2\lambda_{\mathsf{b}}\right)/5<\lambda<1/4$,
and $\mathrm{Pr}\left[\left|\widehat{h}_{\mathsf{b}}/h_{\mathsf{b}}-1\right|>\varepsilon_{n}^{\mathsf{b}}\right]\leq\delta_{n}^{\mathsf{b}}$,
where $\varepsilon_{n}^{\mathsf{b}}=O\left(n^{-\varrho_{\varepsilon}^{\mathsf{b}}}\right)$
and $\delta_{n}^{\mathsf{b}}=O\left(n^{-\varrho_{\delta}^{\mathsf{b}}}\right)$
for some $\varrho_{\varepsilon}^{\mathsf{b}},\varrho_{\delta}^{\mathsf{b}}>0$.
Note that now the ROT bandwidth without undersmoothing satisfies the
rate requirement and thus can be used in Step 3. In the online supplement,
we sketch how to show this result.

\subsection{Conditioning on sub-vectors of the covariates\label{subsec:Conditioning-on-sub-vectors}}

In applications, researchers are often interested in the unconditional
PDF of the ITE, or the conditional PDF of the ITE after conditioning
only on some of the covariates. See, e.g., the application in Section
\ref{sec:Empirical-illustrations}. This section discusses how our
results can be applied in such cases.

\sloppy Partition the vector of covariates as $X=\left(X_{1}^{\top},X_{2}^{\top}\right)^{\top}$
(similarly, $X_{i}=\left(X_{1,i}^{\top},X_{2,i}^{\top}\right)^{\top}$).
Let $f_{\varDelta\mid X_{1}}\left(\cdot\mid x_{1}\right)$ denote
the conditional density of the ITE for some fixed $x_{1}\in\mathscr{S}_{X_{1}}$.
The estimator of the conditional density is given by 
\begin{equation}
\widehat{f}_{\varDelta\mid X_{1}}\left(v\mid x_{1};b\right)\coloneqq\frac{\sum_{i=1}^{n}\frac{1}{b}K\left(\frac{\widehat{\varDelta}_{i}-v}{b}\right)\mathbbm{1}\left(X_{1,i}=x_{1}\right)}{\sum_{i=1}^{n}\mathbbm{1}\left(X_{1,i}=x_{1}\right)}.\label{eq:f_hat x_dag definition}
\end{equation}
Let $\widehat{q}_{X_{i}}\left(W_{i},W_{j};b_{\zeta}\right)$ be defined
similarly to $\widehat{q}_{x}\left(W_{i},W_{j};b_{\zeta}\right)$
by the same formula with $x$ replaced by $X_{i}$. Note that in the
definition of $\widehat{q}_{X_{i}}\left(W_{i},W_{j};b_{\zeta}\right)$,
$b_{\zeta}$ may depend on $X_{i}$. Next, let $\widehat{q}_{x_{1}}\left(W_{i},W_{j};b_{\zeta}\right)\coloneqq\mathbbm{1}\left(X_{1,i}=x_{1}\right)\widehat{q}_{X_{i}}\left(W_{i},W_{j};b_{\zeta}\right)$,
$\widehat{p}_{x_{1}}\coloneqq n^{-1}\sum_{i=1}^{n}\mathbbm{1}\left(X_{1,i}=x_{1}\right)$,
$\widehat{f}_{\varDelta X_{1}}\left(v,x_{1};b\right)\coloneqq\widehat{f}_{\varDelta\mid X_{1}}\left(v\mid x_{1};b\right)\widehat{p}_{x_{1}}$,
and
\begin{eqnarray*}
\widehat{\mathcal{U}}_{x_{1}}\left(W_{j},W_{i},v;b,b_{\zeta}\right) & \coloneqq & \frac{1}{\sqrt{b}}K\left(\frac{\widehat{\varDelta}_{i}-v}{b}\right)\mathbbm{1}\left(X_{1,i}=x_{1}\right)\\
 &  & +\frac{1}{b^{3/2}}K'\left(\frac{\widehat{\varDelta}_{j}-v}{b}\right)\widehat{q}_{x_{1}}\left(W_{j},W_{i};b_{\zeta}\right)\widehat{\pi}_{X_{j}}\left(Z_{i},X_{i}\right),\\
\widehat{\mathcal{U}}_{x_{1}}^{\left[1\right]}\left(W_{i},v;b,b_{\zeta}\right) & \coloneqq & \frac{1}{n-1}\sum_{j\neq i}\widehat{\mathcal{U}}_{x_{1}}\left(W_{j},W_{i},v;b,b_{\zeta}\right),\\
\widehat{S}_{\mathsf{jmb}}\left(v\mid x_{1};b,b_{\zeta}\right) & \coloneqq & \frac{1}{\sqrt{n}}\sum_{i=1}^{n}\nu_{i}\widehat{p}_{x_{1}}^{-1}\left\{ \widehat{\mathcal{U}}_{x_{1}}^{\left[1\right]}\left(W_{i},v;b,b_{\zeta}\right)-\sqrt{b}\cdot\widehat{f}_{\varDelta X_{1}}\left(v,x_{1};b\right)\right\} .
\end{eqnarray*}
Define also
\begin{eqnarray}
\widehat{V}_{1}\left(v,x_{1};b\right) & \coloneqq & \frac{1}{n}\sum_{i=1}^{n}\frac{1}{b}K\left(\frac{\widehat{\varDelta}_{i}-v}{b}\right)^{2}\mathbbm{1}\left(X_{1,i}=x_{1}\right)-b\cdot\widehat{f}_{\varDelta X_{1}}\left(v,x_{1};b\right)^{2},\nonumber \\
\widehat{V}_{2}\left(v,x_{1};b,b_{\zeta}\right) & \coloneqq & \frac{1}{n\left(n-1\right)\left(n-2\right)}\sum_{i=1}^{n}\sum_{j\neq i}\sum_{k\neq i,\,k\neq j}\frac{1}{b^{3}}K'\left(\frac{\widehat{\varDelta}_{j}-v}{b}\right)\widehat{q}_{x_{1}}\left(W_{j},W_{i};b_{\zeta}\right)\nonumber \\
 &  & \times K'\left(\frac{\widehat{\varDelta}_{k}-v}{b}\right)\widehat{q}_{x_{1}}\left(W_{k},W_{i};b_{\zeta}\right)\left(\frac{1}{\widehat{p}_{0X_{i}}}+\frac{1}{\widehat{p}_{1X_{i}}}\right)\frac{\mathbbm{1}\left(X_{i}=X_{j}=X_{k}\right)}{\widehat{p}_{X_{i}}},\label{eq:V_hat x_dag definition}
\end{eqnarray}
$\widehat{V}\left(v\mid x_{1};b,b_{\zeta}\right)\coloneqq\widehat{p}_{x_{1}}^{-2}\left(\widehat{V}_{1}\left(v,x_{1};b\right)+\widehat{V}_{2}\left(v,x_{1};b,b_{\zeta}\right)\right)$.
For bandwidth selection, we can set $b$ to be the pseudo-ITE-based
version of the Silverman ROT bandwidth computed using the subsample
corresponding to $X_{1,i}=x_{1}$, and set $b_{\zeta}$ to be the
same ROT bandwidth as in Algorithm \ref{alg:multiplier}. A JMB UCB
for $\left\{ f_{\varDelta\mid X_{1}}\left(v\mid x_{1}\right):v\in I_{x_{1}}\right\} $,
where $I_{x_{1}}$ denotes an inner closed sub-interval of $\mathscr{S}_{\varDelta\mid X_{1}=x_{1}}$,
can be constructed similarly to $\mathit{CB}_{\mathsf{jmb}}$ in (\ref{eq:confidence band definition})
by adapting Algorithm \ref{alg:multiplier}.

For the unconditional PDF $f_{\varDelta}$ of the ITE, the estimator
is given by $\widehat{f}_{\varDelta}\left(v;b\right)\coloneqq\left(nb\right)^{-1}\sum_{i=1}^{n}K\left(\left(\widehat{\varDelta}_{i}-v\right)/b\right)$.
Let
\begin{eqnarray*}
\widehat{\mathcal{U}}\left(W_{j},W_{i},v;b,b_{\zeta}\right) & \coloneqq & \frac{1}{\sqrt{b}}K\left(\frac{\widehat{\varDelta}_{i}-v}{b}\right)+\frac{1}{b^{3/2}}K'\left(\frac{\widehat{\varDelta}_{j}-v}{b}\right)\widehat{q}_{X_{j}}\left(W_{j},W_{i};b_{\zeta}\right)\widehat{\pi}_{X_{j}}\left(Z_{i},X_{i}\right),\\
\widehat{\mathcal{U}}^{\left[1\right]}\left(W_{i},v;b,b_{\zeta}\right) & \coloneqq & \frac{1}{n-1}\sum_{j\neq i}\widehat{\mathcal{U}}\left(W_{j},W_{i},v;b,b_{\zeta}\right),\\
\widehat{S}_{\mathsf{jmb}}\left(v;b,b_{\zeta}\right) & \coloneqq & \frac{1}{\sqrt{n}}\sum_{i=1}^{n}\nu_{i}\left\{ \widehat{\mathcal{U}}^{\left[1\right]}\left(W_{i},v;b,b_{\zeta}\right)-\sqrt{b}\cdot\widehat{f}_{\varDelta}\left(v;b\right)\right\} .
\end{eqnarray*}
Let $I$ denote an inner closed sub-interval of $\mathscr{S}_{\varDelta}$,
$\widehat{V}_{1}\left(v;b\right)\coloneqq\left(nb\right)^{-1}\sum_{i=1}^{n}K\left(\left(\widehat{\varDelta}_{i}-v\right)/b\right)^{2}-b\cdot\widehat{f}_{\varDelta}\left(v;b\right)^{2}$,
and define $\widehat{V}_{2}\left(v;b,b_{\zeta}\right)$ as in (\ref{eq:V_hat x_dag definition})
but with $\widehat{q}_{x_{1}}\left(W_{j},W_{i};b_{\zeta}\right)$
and $\widehat{q}_{x_{1}}\left(W_{k},W_{i};b_{\zeta}\right)$ replaced
by $\widehat{q}_{X_{j}}\left(W_{j},W_{i};b_{\zeta}\right)$ and $\widehat{q}_{X_{k}}\left(W_{k},W_{i};b_{\zeta}\right)$
respectively. Further, define $\widehat{V}\left(v;b,b_{\zeta}\right)\coloneqq\widehat{V}_{1}\left(v;b\right)+\widehat{V}_{2}\left(v;b,b_{\zeta}\right)$.
A JMB UCB can be constructed similarly to the above. We set $b$ to
be the pseudo-ITE-based version of the Silverman ROT bandwidth computed
using the entire sample, and set $b_{\zeta}$ to be the same ROT bandwidth
as in Algorithm \ref{alg:multiplier}.

\section{Monte Carlo experiments\label{sec:Monte-Carlo-experiments}}

This section evaluates the finite-sample performance of the UCBs proposed
in Section \ref{sec:Robust-inference} for the density $f_{\varDelta}\left(v\right)$
of the ITE. We consider the following experiment design based on FVX.
The outcome and treatment status variables are generated according
to 
\begin{eqnarray}
Y & = & \left(\epsilon+1\right)^{2+D}\nonumber \\
D & = & \mathbbm{1}\left(\gamma_{0}+\gamma_{1}\cdot Z+\eta\geq0\right),\label{eq:model simulation}
\end{eqnarray}
where $\left(\epsilon,\eta\right)=\left(\varPhi\left(U\right),\varPhi\left(V\right)\right)$,
$\left(U,V\right)$ has a mean-zero bivariate normal distribution
with $\mathrm{Var}\left[U\right]=\mathrm{Var}\left[V\right]=1$ and
$\mathrm{Cov}\left[U,V\right]=0.3$, and $\varPhi$ is the standard
normal CDF. The instrument is generated according to $Z=\mathbbm{1}\left(N>0\right)$,
where $N$ is a standard normal random variable independent of $\left(\epsilon,\nu\right)$.
In this design, the ITE satisfies $\varDelta=\epsilon\left(\epsilon+1\right)^{2}$,
where $\epsilon$ is uniformly distributed on $\left[0,1\right]$
and $\varDelta$ is supported on $\left[0,4\right]$. We consider
two sets of values for $(\gamma_{0},\gamma_{1}):$ $(-0.5,0.5)$ and
$(-0.4,0.6).$ We use the triweight kernel for $K$, $K_{\zeta}$
and $K_{\mathsf{b}}$, and the Silverman ROT bandwidths for $\widehat{h}$,
$\widehat{h}_{\mathsf{b}}$ and $\widehat{h}_{\zeta}.$\footnote{To be specific, we take $\widehat{h}=3.15\cdot\widehat{\sigma}_{\varDelta}\cdot n^{-1/5}$,
$\widehat{h}_{\mathsf{b}}=2.7\cdot\widehat{\sigma}_{\varDelta}\cdot n^{-1/9}$.
When it comes to $\widehat{h}_{\zeta}$, we distinguish between the
treated and control subsamples, i.e., $\widehat{h}_{\zeta,d}=3.15\cdot\widehat{\sigma}_{Y\mid D=d}\cdot n_{d}^{-1/5}$
with $n_{d}\coloneqq\sum_{i=1}^{n}\mathbbm{1}\left(D_{i}=d\right)$.} The number of Monte Carlo replications is set to $1,000$.

\begin{table}[t]
\caption{Simultaneous coverage rates for UCBs and the interpolated bootstrap
percentile CIs}
\label{table: uniform_CP}\resizebox{\textwidth}{!}{%
\begin{tabular}{cccccccccccccc}
\toprule 
\multicolumn{1}{c}{} & \multicolumn{1}{c}{} & \multicolumn{6}{c}{$\gamma_{0}=-0.5,\gamma_{1}=0.5$} & \multicolumn{6}{c}{$\gamma_{0}=-0.4,\gamma_{1}=0.6$}\tabularnewline
\cmidrule{3-14} \cmidrule{4-14} \cmidrule{5-14} \cmidrule{6-14} \cmidrule{7-14} \cmidrule{8-14} \cmidrule{9-14} \cmidrule{10-14} \cmidrule{11-14} \cmidrule{12-14} \cmidrule{13-14} \cmidrule{14-14} 
\multicolumn{1}{c}{} & \multicolumn{1}{c}{} & \multicolumn{3}{c}{$v\in\left[0.5,3.5\right]$} & \multicolumn{3}{c}{$v\in\left[0.8,3.2\right]$} & \multicolumn{3}{c}{$v\in\left[0.5,3.5\right]$} & \multicolumn{3}{c}{$v\in\left[0.8,3.2\right]$}\tabularnewline
\cmidrule{3-14} \cmidrule{4-14} \cmidrule{5-14} \cmidrule{6-14} \cmidrule{7-14} \cmidrule{8-14} \cmidrule{9-14} \cmidrule{10-14} \cmidrule{11-14} \cmidrule{12-14} \cmidrule{13-14} \cmidrule{14-14} 
\multicolumn{1}{c}{$n$} & \multicolumn{1}{c}{Methods} & \multicolumn{1}{c}{0.90} & \multicolumn{1}{c}{0.95} & \multicolumn{1}{c}{0.99} & \multicolumn{1}{c}{0.90} & \multicolumn{1}{c}{0.95} & \multicolumn{1}{c}{0.99} & \multicolumn{1}{c}{0.90} & \multicolumn{1}{c}{0.95} & \multicolumn{1}{c}{0.99} & \multicolumn{1}{c}{0.90} & \multicolumn{1}{c}{0.95} & \multicolumn{1}{c}{0.99}\tabularnewline
\midrule 
2000 & Bias-corrected JMB & 0.823 & 0.891 & 0.953 & 0.836 & 0.893 & 0.957 & 0.817 & 0.887 & 0.943 & 0.820 & 0.883 & 0.946\tabularnewline
 & Bias-corrected NPB & 0.852 & 0.921 & 0.980 & 0.869 & 0.930 & 0.980 & 0.851 & 0.930 & 0.986 & 0.861 & 0.926 & 0.987\tabularnewline
 & Interpolated CIs & 0.507 & 0.715 & 0.927 & 0.605 & 0.795 & 0.958 & 0.579 & 0.769 & 0.949 & 0.663 & 0.829 & 0.963\tabularnewline
\midrule 
4000 & Bias-corrected JMB & 0.831 & 0.891 & 0.956 & 0.847 & 0.905 & 0.953 & 0.819 & 0.885 & 0.946 & 0.835 & 0.891 & 0.946\tabularnewline
 & Bias-corrected NPB & 0.868 & 0.932 & 0.985 & 0.876 & 0.936 & 0.987 & 0.875 & 0.936 & 0.989 & 0.870 & 0.932 & 0.989\tabularnewline
 & Interpolated CIs & 0.405 & 0.628 & 0.914 & 0.533 & 0.729 & 0.954 & 0.476 & 0.703 & 0.934 & 0.581 & 0.782 & 0.957\tabularnewline
\midrule 
6000 & Bias-corrected JMB & 0.837 & 0.902 & 0.966 & 0.864 & 0.909 & 0.966 & 0.844 & 0.900 & 0.960 & 0.849 & 0.904 & 0.958\tabularnewline
 & Bias-corrected NPB & 0.860 & 0.944 & 0.994 & 0.894 & 0.954 & 0.996 & 0.876 & 0.946 & 0.991 & 0.884 & 0.944 & 0.990\tabularnewline
 & Interpolated CIs & 0.364 & 0.580 & 0.885 & 0.548 & 0.744 & 0.956 & 0.431 & 0.655 & 0.930 & 0.586 & 0.773 & 0.967\tabularnewline
\bottomrule
\end{tabular}}
\end{table}

Tables \ref{table: uniform_CP} reports the simultaneous coverage
rates of two types of UCBs: the bias-corrected JMB UCB defined in
(\ref{eq:bias corrected CB}) and the bias-corrected nonparametric
bootstrap (NPB) UCB described in Appendix \ref{sec:Nonparametric-bootstrap}.
FVX interpolates pointwise (nonparametric) bootstrap percentile confidence
intervals (CIs) to construct a confidence band for the density of
the ITE. It follows from our results that such intervals are valid
in the pointwise sense.\footnote{Validity of pointwise bootstrap percentile confidence intervals follows
from (\ref{eq:bias corrected linearization}), Lemma \ref{lem:S_star linearization}
in the supplement and standard arguments (see the proof of \citealp[Theorem 4.2]{Ma2019}).} We also report the coverage probability of the confidence band constructed
by interpolating the bootstrap percentile pointwise confidence intervals.
The nominal coverage rates are $0.90$, $0.95$, and $0.99$. We consider
two ranges of $v$: a longer interval $I=[0.5,3.5]$ and a shorter
one $I=[0.8,3.2]$. We use grid search to solve the one-dimensional
optimization problems in (\ref{eq:phi_hat definition}) when estimating
the pseudo ITEs and calculating the supremum of the bootstrap process.
The number of bootstrap replications is set to $5,000$. Table \ref{table: uniform_length}
reports the average widths of the bias-corrected JMB and NPB UCBs
relative to the interpolated pointwise CIs.\footnote{The average width of bias-corrected JMB (or NPB) UCB is computed by
first averaging the widths of the confidence band over all grid points
in the given range $I$ and then averaging over all simulation replications.
The reported number is the ratio of the average width of the UCBs
to that of the interpolated pointwise CIs.}

\begin{table}[t]
\caption{Average widths of the $95\%$ UCBs relative to the interpolated bootstrap
percentile CIs}
\label{table: uniform_length} \centering{}%
\begin{tabular}{cccccc}
\toprule 
\multicolumn{1}{c}{} & \multicolumn{1}{c}{} & \multicolumn{2}{c}{$\gamma_{0}=-0.5,\gamma_{1}=0.5$} & \multicolumn{2}{c}{$\gamma_{0}=-0.4,\gamma_{1}=0.6$}\tabularnewline
\cmidrule{3-6} \cmidrule{4-6} \cmidrule{5-6} \cmidrule{6-6} 
\multicolumn{1}{c}{} & \multicolumn{1}{c}{Methods} & \multicolumn{1}{c}{$v\in\left[0.5,3.5\right]$} & \multicolumn{1}{c}{$v\in\left[0.8,3.2\right]$} & \multicolumn{1}{c}{$v\in\left[0.5,3.5\right]$} & \multicolumn{1}{c}{$v\in\left[0.8,3.2\right]$}\tabularnewline
\midrule 
2000 & Bias-corrected JMB & 1.424 & 1.527 & 1.516 & 1.634\tabularnewline
 & Bias-corrected NPB & 1.454 & 1.595 & 1.576 & 1.754\tabularnewline
\midrule 
4000 & Bias-corrected JMB & 1.384 & 1.480 & 1.430 & 1.527\tabularnewline
 & Bias-corrected NPB & 1.416 & 1.543 & 1.492 & 1.642\tabularnewline
\midrule 
6000 & Bias-corrected JMB & 1.526 & 1.631 & 1.413 & 1.509\tabularnewline
 & Bias-corrected NPB & 1.550 & 1.681 & 1.458 & 1.595\tabularnewline
\bottomrule
\end{tabular}
\end{table}

We make the following observations regarding the simulation results.
First, as expected, interpolation of pointwise CIs exhibits substantial
under-coverage, especially for the nominal coverage probabilities
$0.90$ and $0.95.$ Therefore, appealing as it looks to practitioners,
interpolation of pointwise CIs fails to cover the true density curve
with the desired coverage probability even in large samples. Second,
the bias-corrected JMB and NPB UCBs yield reasonably good coverage
rates across different setups and sample sizes. The JMB UCBs are slightly
narrower than the NPB UCBs; however they are also less accurate. Third,
we focus on the studentized UCB because it has variable width and
thus is narrower than the non-studentized counterpart (Footnote \ref{fn:constant width}).
The additional simulation results in Section \ref{sec:Additional-Monte-Carlo}
of the online supplement confirm that the non-studentized UCBs are
on average wider than the studentized ones. The computation of the
JMB is faster than the NPB, as the former avoids the estimation of
ITEs for each bootstrap sample. Therefore, we recommend the bias-corrected
JMB UCB defined in (\ref{eq:bias corrected CB}) to practitioners
for assessing the shape of the density of ITEs.

\section{Empirical application: Childbearing and labor income\label{sec:Empirical-illustrations}}

In this section, we apply the FVX estimator for the density of the
ITE and our bias-corrected JMB UCB to investigate the effect of family
size on labor income. Understanding the relationship between the two
variables is important for policymakers; however, estimation of the
effect can be complicated due to the simultaneity between the labor
supply and fertility decisions \citep[AE hereafter]{angrist1998children}.

We revisit the 1980 Census Public Use Micro Samples (PUMS) previously
used by AE and other authors. Following AE, we focus on married women
aged 21-35 with at least two children. The focus on households with
at least two children is due to the identification strategy developed
in AE, as explained below. Our outcome variable $Y$ is the sum of
the mother's and father's 1979 labor incomes (in thousands of dollars).
The binary treatment variable $D$ takes the value one if the mother
has more than two children. The instrument proposed in AE is the ``same-sex''
dummy variable that takes the value one when the first two children
are of the same sex. This identification strategy relies on parental
preferences for a mixed sibling-sex composition: parents whose first
two children are of the same sex are more likely to have an additional
child. AE shows that in 1980, the estimated probabilities of having
a third child for women with same-sex and mixed-sex children were
0.432 and 0.372, respectively. Moreover, the difference between the
two groups is highly significant.

Our covariates $X$ include the mother's education level (less than
high school, high school, some college, college), the quartile in
the age distribution, race (white, black, Hispanic, others), and the
sex of the first birth.\footnote{The three quartiles of the distribution of mother's age in our sample
are 28, 31, and 33.} Following FVX, we drop observations in $X$-defined cells that contain
less than $1\%$ of the sample. Our remaining sample has $224,962$
observations.

Using 2SLS with a linear IV regression model, the estimated effect
of having more than two children on parents' labor income is $-3.54$
with a standard error of $1.51$. The estimates suggest that in 1979,
having more than two children reduced parents' labor income by \$3,540
(or \$12,624 in 2020 dollars).\footnote{We used the CPI series from FRED, Federal Reserve Bank of St. Louis,
for the conversion.} The effect is substantial and corresponds to 7.7\% of the average
household labor income in our sample.\footnote{The average household labor income in our sample is \$45,829 in 1979
dollars.}

\begin{table}[t]
\caption{Summary statistics for ITE estimates of the effect of having more
than two children on parents' labor income (in thousands of 1979 dollars).}
\label{table: ITE_stats}
\centering{}\resizebox{\textwidth}{!}{%
\begin{tabular}{ccccccccc}
\toprule 
 & Mean & Std.dev & 1st decile & 1st quartile & Median & 3rd quartile & 9th decile & $\Pr[\varDelta>0]$\tabularnewline
\midrule 
Full sample & -2.67 & 36.72 & -16.99 & -8.71 & -4.14 & 0.25 & 10.37 & 0.264\tabularnewline
($n=224,692$) &  &  &  &  &  &  &  & \tabularnewline
\midrule 
High School, Age> 31 & -5.01 & 30.75 & -10.93 & -7.59 & -5.45 & -3.00 & 1.03 & 0.122\tabularnewline
($n=48,129$) &  &  &  &  &  &  &  & \tabularnewline
College, Age> 31 & 10.88 & 64.20 & -39.99 & -14.00 & -1.42 & 21.69 & 53.22 & 0.462\tabularnewline
($n=19,103$) &  &  &  &  &  &  &  & \tabularnewline
\midrule 
High School, Age $\leq$ 31 & -7.93 & 23.56 & -12.89 & -8.96 & -4.70 & -2.06 & 0 & 0.097\tabularnewline
($n=65,447$) &  &  &  &  &  &  &  & \tabularnewline
College, Age $\leq$ 31 & 25.93 & 68.24 & -16.08 & -5.58 & 9.95 & 27.07 & 178.43 & 0.667\tabularnewline
($n=8,547$) &  &  &  &  &  &  &  & \tabularnewline
\bottomrule
\end{tabular}}
\end{table}

Table \ref{table: ITE_stats} reports summary statistics for the ITE
estimates. According to the results, estimated ITEs display substantial
heterogeneity. For example, the median ITE in our sample is -4.14
with an interquartile range of 9.03.\footnote{The results are consistent with the findings in \citet{Froelich2013}
who also report substantial heterogeneity using the 2000 PUMS data
and quantile treatment effects.} Conditioning on the above median age and college-level education
produces an even wide range of estimated ITEs: -1.42 for the median
effect with the interquartile range of 35.69. While for the below
median age, college-educated mothers, the median effect is positive
(9.95), the corresponding interquartile range is similarly wide (32.65).
The table also shows that in the case of mothers with only high-school-level
education, ITEs tend to be more negative. For example, conditional
on the above median age and only high-school-level education, the
median ITE is -5.45, with an interquartile range of 7.93. Only 12.2\%
of the households in this group have positive estimated ITEs, compared
to 46.2\% of the households with college-educated mothers from the
same age group. The group with the largest fraction of households
with positive estimated ITEs is the below-median-age mothers with
a college education: 66.7\%.

\begin{figure}[t]
\caption{The unconditional PDF of the ITE with the 95\% pointwise and uniform
confidence bands\label{fig:unconditional-distribution}}

\centering{}\includegraphics[scale=0.8]{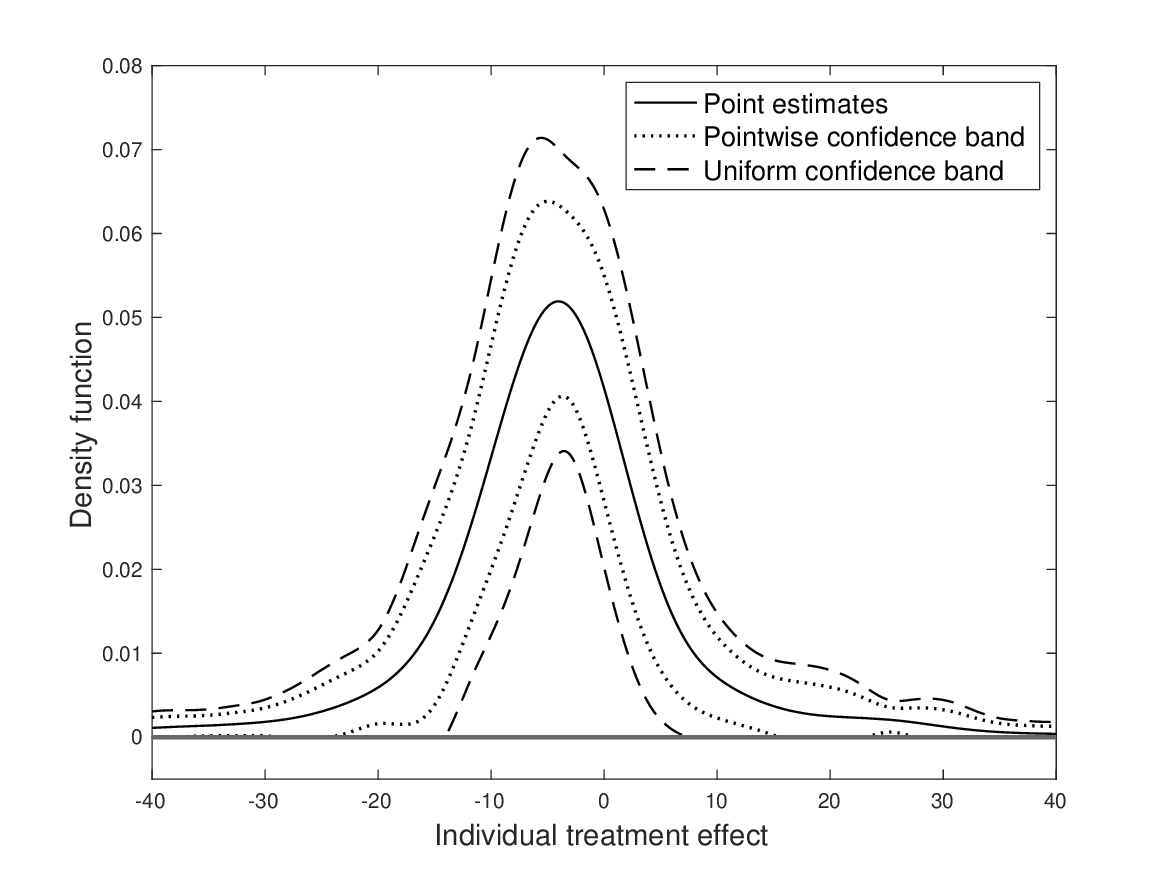}
\end{figure}

Next, we use the FVX estimator with our bias-corrected JMB UCB to
analyze the distribution of the ITE. Figure \ref{fig:unconditional-distribution}
shows the unconditional PDF of the ITE together with the 95\% pointwise
and uniform confidence bands for the density. Following Remark \ref{Rmk: truncation},
the lower bounds of the confidence bands are truncated to zero. One
can see that while the UCB developed in this paper is somewhat wider
than the pointwise, it is still informative. The estimated mode of
the unconditional distribution is -4.08, and according to the UCB
the mode is located between -5.60 and -3.52.

\begin{figure}[t]
\caption{The conditional PDFs of the ITE (solid lines) and their 95\% UCBs
(shaded areas) conditional on the mother's age, high school (red),
or college (black) education: the left panel is (a) Above the median
age, and the right panel is (b) Below the median age}

\noindent \centering{}\label{fig:conditional PDFs }\includegraphics[scale=0.42]{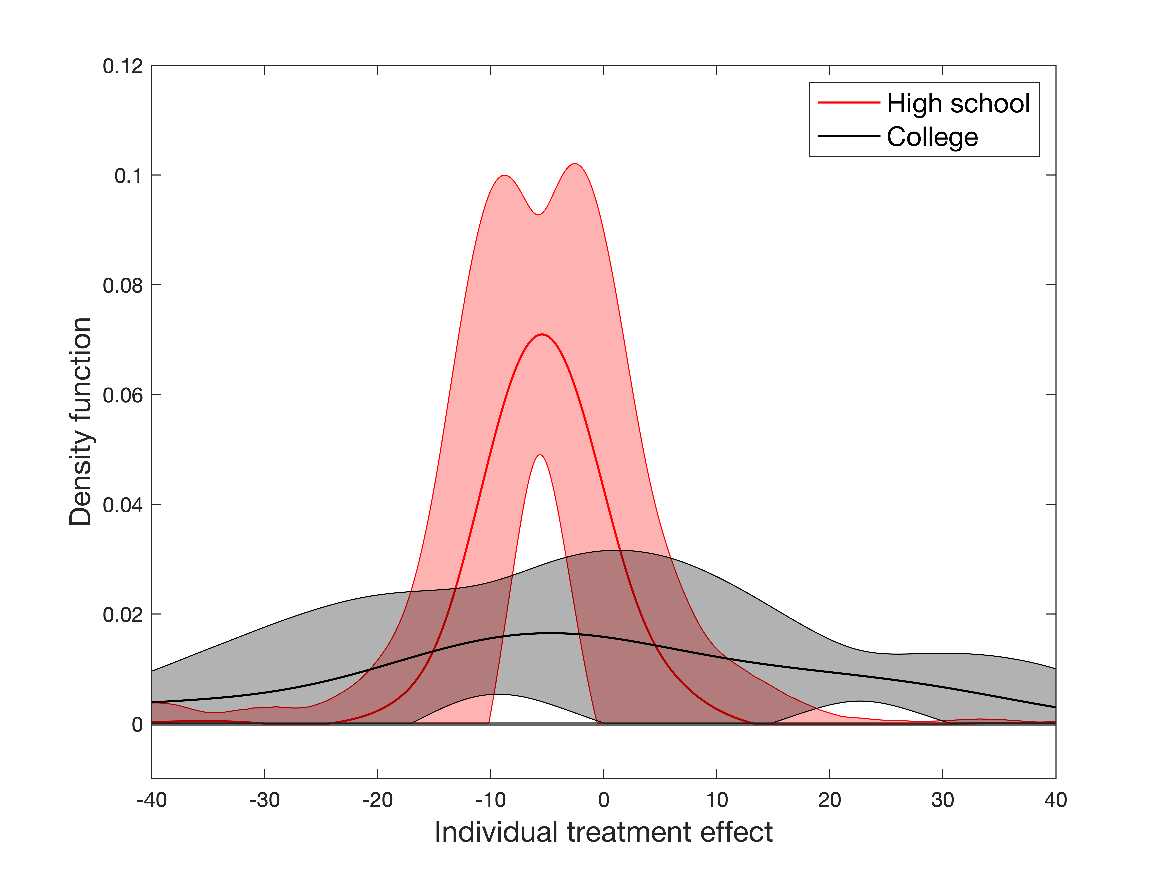}\includegraphics[scale=0.42]{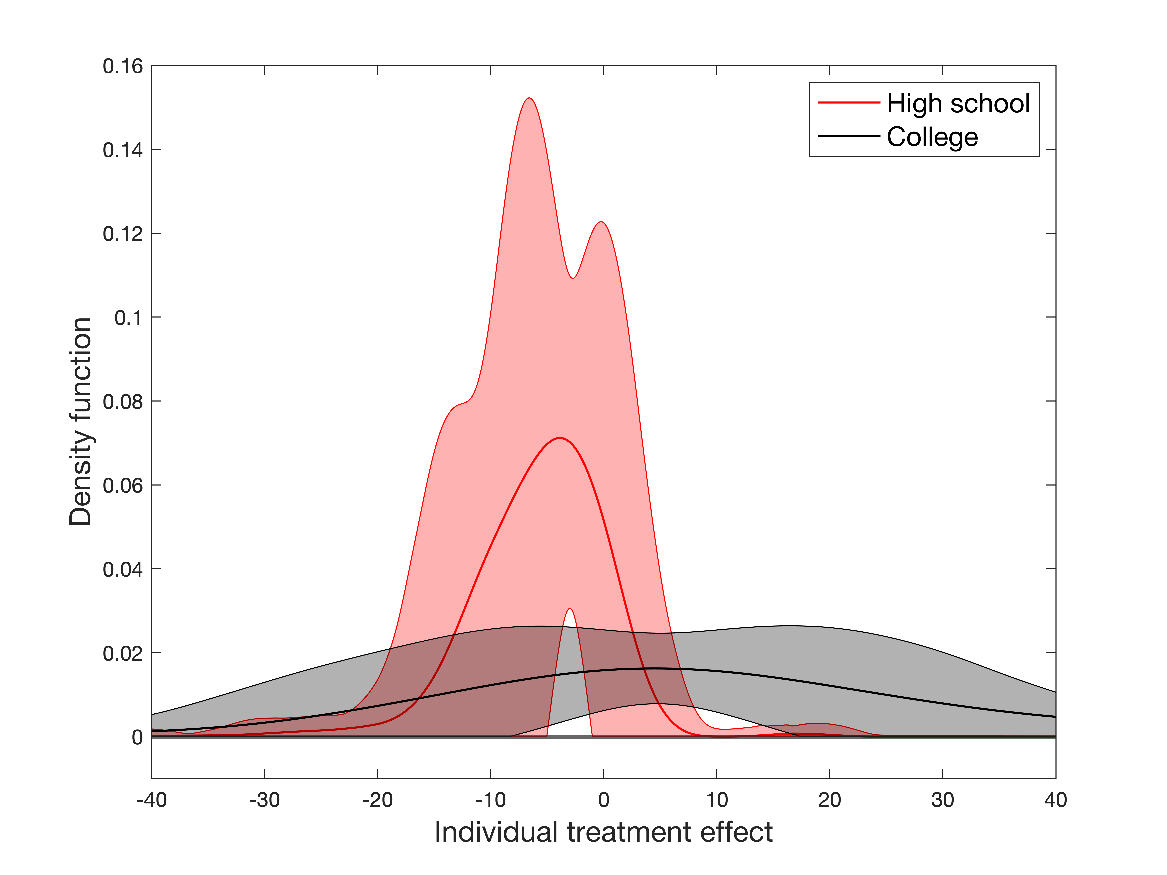}
\end{figure}

Figure \ref{fig:conditional PDFs } shows the conditional PDFs of
the ITE conditional on the mother's age (above or below the median
age in our sample) and education (high school only or college levels)
with their 95\% UCBs. Figure \ref{fig:conditional PDFs }(a) displays
the results conditional on the above median age for high school only
and college education levels. As there are regions where the two UCBs
do not intersect, we can conclude with at least 90.25\% confidence
that the two densities conditional on high school and college are
different.\footnote{The two 95\% UCBs are independent since they are computed on non-overlapping
samples. Hence, the overall confidence level when comparing the two
distributions is $0.95^{2}=0.9025$.} In particular, while the distribution of the ITE conditional on college
is much more dispersed, it also has more probability mass in the positive
range. E.g., on the interval corresponding to ITEs between \$20,000
and \$30,000, the confidence band for the density conditional on high
school is very narrow and close to zero. At the same time, the density
conditional on college is significantly different from zero. Hence,
a non-negligible fraction of households with college-educated mothers
above the median age received a substantial positive effect of a magnitude
between 43.6\% and 65.5\% of the average household labor income in
our sample. There is no evidence that households with only high-school-educated
mothers experienced ITEs of this magnitude.

In the case of high school only, the UCB does not rule out a bimodal
density with the two modes at -8.80 and -2.56. According to these
results, households with high-school-educated mothers above the median
age are likely to experience either a strong negative effect around
19.2\% of the average household labor income or a more moderate negative
effect around 5.6\% of the average household labor income.

Figure \ref{fig:conditional PDFs }(b) shows similar results conditional
on the below median age. We can again conclude with a 90.25\% confidence
that the conditional distributions by education level (high school
or college) are different. The ITE distribution is more dispersed
for households with college-educated mothers than for households with
high-school-educated mothers. Similarly to the previous case, the
distribution conditional on college has more mass in the positive
range than the distribution conditional on high school. The results
conditional on high school again cannot rule out bimodality; however,
this time the first mode at -6.56 corresponds to a more moderate effect,
and the second mode at 0.13 occurs in the positive range.

We conclude that there are significant differences in the distributions
of the ITE across the education levels. For households with high-school-only-educated
mothers, the distribution of the ITE is heavily concentrated in the
negative range and potentially bimodal. Households with college-educated
mothers have a wider range of ITEs. However, such households can also
experience positive effects of a large magnitude. Predicting the effect
of having more than two children on labor income for such households
is difficult as the distribution is thinly spread from large negative
to large positive values.

Recently, \citet{abrevaya2021estimation} studied the distributional
effect of having a third child on female labor supply by applying
a weakly nonseparable model (equipped with the mean-variance-effect
structure) to the PUMS dataset in 2000. Similar to our findings, they
also documented a large amount of heterogeneity in the ITE distributions
(See their Figures 2 to 5). In terms of how the ITE distribution varies
across mothers' education levels, they found less variation with the
2000 data than we document in this paper with the 1980 data.

\bibliographystyle{chicago}
\bibliography{ITE}

\appendix
\counterwithin{thm}{section} 

\renewcommand{\thethm}{\thesection\arabic{thm}}

\section{\label{sec:Appendix A}Proofs of Theorems in Section \ref{sec:Asymptotic-properties}}
\begin{lem}
\label{lem:differentiability of the PDF}Under Assumptions \ref{assu: DGP1}
and \ref{assu: DGP4}, the PDF $f_{\varDelta\mid X}\left(\cdot\mid x\right)$
exists and is $P$-times continuously differentiable on any inner
closed sub-interval $I_{x}$ of $\mathscr{S}_{\varDelta\mid X=x}$.
\end{lem}
\begin{proof}
Let $\left(\varDelta_{x,j}^{-1}\right)'$ denote the derivative of
$\varDelta_{x,j}^{-1}$. By Assumption \ref{assu: DGP4}(a,b) and
\citet[Theorem 7.3]{severini2005elements}, the conditional distribution
of $\varDelta=\varDelta_{x}\left(\epsilon\right)$ given $X=x$ admits
a Lebesgue density $f_{\varDelta\mid X}\left(\cdot\mid x\right)$:
\begin{equation}
f_{\varDelta\mid X}\left(v\mid x\right)=\sum_{j=1}^{m}f_{\epsilon\mid X}\left(\varDelta_{x,j}^{-1}\left(v\right)\mid x\right)\left|\left(\varDelta_{x,j}^{-1}\right)'\left(v\right)\right|\mathbbm{1}\left(v\in\varDelta_{x}\left(\left(\epsilon_{x,j-1},\epsilon_{x,j}\right)\right)\right),\label{eq:f expression}
\end{equation}
for $v\in\varDelta_{x}\left(\bigcup_{j=1}^{m}\left(\epsilon_{x,j-1},\epsilon_{x,j}\right)\right)$,
and $\mathscr{S}_{\varDelta\mid X=x}$ is the closure of $\varDelta_{x}\left(\bigcup_{j=1}^{m}\left(\epsilon_{x,j-1},\epsilon_{x,j}\right)\right)$.
From (\ref{eq:f expression}), the density $f_{\varDelta\mid X}\left(\cdot\mid x\right)$
has a jump at an interior point of $\mathscr{S}_{\varDelta\mid X=x}$
($\varDelta_{x}\left(\epsilon_{x,j-1}\right)$ or $\varDelta_{x}\left(\epsilon_{x,j+1}\right)$)
if $\varDelta_{x}\left(\left(\epsilon_{x,j-1},\epsilon_{x,j}\right)\right)\neq\varDelta_{x}\left(\left(\epsilon_{x,j},\epsilon_{x,j+1}\right)\right)$.
Such cases are ruled out by Assumption \ref{assu: DGP4}(c).
\end{proof}

\subsection{Notations and mathematical definitions}

Let $X_{n},Y_{n}$ be (sequences of) random variables and $\alpha_{n},\beta_{n},\gamma_{n},\alpha_{n}',\beta_{n}',\gamma_{n}'$
be sequences in $\left(0,\infty\right)$. We write $X_{n}=O_{p}^{\star}\left(\alpha_{n},\beta_{n}\right)$
if there exists positive constants $\left(C_{1},C_{2}\right)$ such
that $\mathrm{Pr}\left[\left|X_{n}\right|>C_{1}\alpha_{n}\right]\leq C_{2}\beta_{n}$.
We write $X_{n}=O_{p}^{\star}\left(\alpha_{n}\right)$ for simplicity
if $\beta_{n}=n^{-1}$. It is straightforward to check that the following
properties hold for the $O_{p}^{\star}$ notations. If $X_{n}=O_{p}^{\star}\left(\alpha_{n},\beta_{n}\right)$
and $Y_{n}=O_{p}^{\star}\left(\alpha_{n}',\beta_{n}'\right)$, then
$X_{n}Y_{n}=O_{p}^{\star}\left(\alpha_{n}\alpha_{n}',\beta_{n}+\beta_{n}'\right)$
and $X_{n}+Y_{n}=O_{p}^{\star}\left(\alpha_{n}+\alpha_{n}',\beta_{n}+\beta_{n}'\right)$.
It is easy to see that if $X_{n}=O_{p}^{\star}\left(\alpha_{n},\beta_{n}\right)$
and $\beta_{n}\downarrow0$, then $X_{n}=O_{p}\left(\alpha_{n}\right)$.
We write $\xi\left(v\right)=O_{p}^{\star}\left(\alpha_{n},\beta_{n}\right)$,
uniformly in $v\in A$, if $\mathrm{sup}_{v\in A}\left|\xi\left(v\right)\right|=O_{p}^{\star}\left(\alpha_{n},\beta_{n}\right)$.
We write $X_{n}=O_{p}^{\sharp}\left(\alpha_{n},\beta_{n},\gamma_{n}\right)$
if there exists positive constants $\left(C_{1},C_{2},C_{3}\right)$
such that $\mathrm{Pr}\left[\mathrm{Pr}_{\mid W_{1}^{n}}\left[\left|X_{n}\right|>C_{1}\alpha_{n}\right]>C_{2}\beta_{n}\right]\leq C_{3}\gamma_{n}$.
For simplicity, we write $X_{n}=O_{p}^{\sharp}\left(\alpha_{n}\right)$
if $\beta_{n}=\gamma_{n}=n^{-1}$. It is straightforward to verify
that if $X_{n}=O_{p}^{\sharp}\left(\alpha_{n},\beta_{n},\gamma_{n}\right)$
and $Y_{n}=O_{p}^{\sharp}\left(\alpha_{n}',\beta_{n}',\gamma_{n}'\right)$,
then $X_{n}+Y_{n}=O_{p}^{\sharp}\left(\alpha_{n}+\alpha_{n}',\beta_{n}+\beta_{n}',\gamma_{n}+\gamma_{n}'\right)$
and $X_{n}Y_{n}=O_{p}^{\sharp}\left(\alpha_{n}\alpha_{n}',\beta_{n}+\beta_{n}',\gamma_{n}+\gamma_{n}'\right)$.
We say $\xi\left(v\right)=O_{p}^{\sharp}\left(\alpha_{n},\beta_{n},\gamma_{n}\right)$,
uniformly in $v\in A$, if $\mathrm{sup}_{v\in A}\left|\xi\left(v\right)\right|=O_{p}^{\sharp}\left(\alpha_{n},\beta_{n},\gamma_{n}\right)$.

Let $a\wedge b$ and $a\vee b$ denote $\mathrm{min}\left\{ a,b\right\} $
and $\mathrm{max}\left\{ a,b\right\} $ respectively. Let $C,C_{1},C_{2},...$
denote positive constants that are independent of the sample size
and whose values may change in different places. $\apprle$ denotes
an inequality up to a universal constant, i.e., $a\apprle b$ is understood
as $a\leq C\cdot b$ for some $C>0$ that does not depend on the distribution
of the variables in the model, any unknown quantity related to the
model or the sample size. ``$=_{d}$'' is understood as being equal
in distribution. For some set $A$, let $\ell^{\infty}\left(A\right)$
denote the space of all bounded functions $f:A\rightarrow\mathbb{R}$
endowed with the sup-norm $\left\Vert f\right\Vert _{A}\coloneqq\mathrm{sup}_{t\in A}\left|f\left(t\right)\right|$
of $f$ on $A$. For $f:\mathbb{R}\rightarrow\mathbb{R}$, $\left\Vert f\right\Vert _{\infty}$
is understood as $\mathrm{sup}_{x\in\mathbb{R}}\left|f\left(x\right)\right|$.
$\iota\left(\left[a,b\right]\right)$ denotes the length $b-a$ of
the interval $\left[a,b\right]$. ``$a\eqqcolon b$'' means ``$b$
is defined by $a$''. $f^{\left(k\right)}$ denotes the $k$-th derivative
of $f:\mathbb{R}\rightarrow\mathbb{R}$ and $\left(f',f''\right)$
are understood as $\left(f^{\left(1\right)},f^{\left(2\right)}\right)$.

Let $\mathfrak{F}$ denote a class of $\mathbb{R}$-valued functions
defined on a compact set in a finite-dimensional Euclidean space $\mathscr{S}$.
Let $\mathfrak{F}$ be equipped with a norm $\left\Vert \cdot\right\Vert $.
We say that $\mathfrak{F}^{\circ}\subseteq\mathfrak{F}$ is an $\varepsilon$-net
if the union of the closed $\left\Vert \cdot\right\Vert $-balls of
radius $\varepsilon$ centered at points in $\mathfrak{F}^{\circ}$
covers $\mathfrak{F}$. Let the $\varepsilon$-covering number $N\left(\varepsilon,\mathfrak{F},\left\Vert \cdot\right\Vert \right)$
be given by $N\left(\varepsilon,\mathfrak{F},\left\Vert \cdot\right\Vert \right)\coloneqq\text{inf}\left\{ \#\mathfrak{F}^{\circ}:\textrm{\ensuremath{\mathfrak{F}^{\circ}} is an \ensuremath{\varepsilon}-net of }\mathfrak{F}\right\} $,
where $\#A$ denotes the cardinality of a set $A$. A function $F_{\mathfrak{F}}:\mathscr{S}\rightarrow\mathbb{R}_{+}$
is an envelope of $\mathfrak{F}$ if $\mathrm{sup}_{f\in\mathfrak{F}}\left|f\right|\leq F_{\mathfrak{F}}$.
Some of the function classes that appear later in this paper depend
on the sample size $n$. We suppress the dependence for notational
simplicity. We say that $\mathfrak{F}$ is a (uniform) Vapnik--Chervonenkis-type
(VC-type) class with respect to the envelope $F_{\mathfrak{F}}$ (see,
e.g., \citealp[Definition 3.6.10]{gine2016mathematical}) if there
exist some positive constants (VC characteristics) $A_{\mathfrak{F}}\geq e$
and $V_{\mathfrak{F}}>1$ that are independent of the sample size
$n$ such that 
\begin{equation}
\underset{Q\in\mathcal{Q}_{\mathscr{S}}^{\mathsf{fd}}}{\mathrm{sup}}N\left(\varepsilon\left\Vert F_{\mathfrak{F}}\right\Vert _{Q,2},\mathfrak{F},\left\Vert \cdot\right\Vert _{Q,2}\right)\leq\left(\frac{A_{\mathfrak{F}}}{\varepsilon}\right)^{V_{\mathfrak{F}}},\,\forall\varepsilon\in\left(0,1\right],\label{eq:VC type bound}
\end{equation}
where $\mathcal{Q}_{\mathscr{S}}^{\mathsf{fd}}$ denotes the collection
of all finitely discrete probability measures on $\mathscr{S}$ and
the symbol $\forall$ is understood as ``for all''. We denote $\mathbb{P}_{n}^{W}f\coloneqq n^{-1}\sum_{i=1}^{n}f\left(W_{i}\right)$,
$\mathbb{P}^{W}f\coloneqq\mathrm{E}\left[f\left(W\right)\right]$
and $\mathbb{G}_{n}^{W}\coloneqq\sqrt{n}\left(\mathbb{P}_{n}^{W}-\mathbb{P}^{W}\right)$.
And also $\left\Vert \mathbb{G}_{n}^{W}\right\Vert _{\mathfrak{F}}\coloneqq\mathrm{sup}_{f\in\mathfrak{F}}\left|\mathbb{G}_{n}^{W}f\right|$,
$\left\Vert \mathbb{P}^{W}\right\Vert _{\mathfrak{F}}\coloneqq\mathrm{sup}_{f\in\mathfrak{F}}\left|\mathbb{P}^{W}f\right|$
and $\left\Vert \mathbb{P}_{n}^{W}-\mathbb{P}^{W}\right\Vert _{\mathfrak{F}}\coloneqq\mathrm{sup}_{f\in\mathfrak{F}}\left|\left(\mathbb{P}_{n}^{W}-\mathbb{P}^{W}\right)f\right|$.
Denote $U_{i}\coloneqq\left(\epsilon_{i},D_{i},Z_{i},X_{i}\right)$
and $U\coloneqq\left(\epsilon,D,Z,X\right)$. $\left(\mathbb{P}_{n}^{U},\mathbb{P}^{U},\mathbb{G}_{n}^{U}\right)$
and $\left(\left\Vert \mathbb{G}_{n}^{U}\right\Vert _{\mathfrak{F}},\left\Vert \mathbb{P}^{U}\right\Vert _{\mathfrak{F}},\left\Vert \mathbb{P}_{n}^{U}-\mathbb{P}^{U}\right\Vert _{\mathfrak{F}}\right)$
are defined similarly. For some $f:\mathscr{S}_{U}^{r}\rightarrow\mathbb{R}$
($r\geq2$), let 
\[
\mathbb{U}_{n}^{\left(r\right)}f\coloneqq\sqrt{n}\left(\frac{1}{n_{\left(r\right)}}\sum_{\left(i_{1},...,i_{r}\right)}f\left(U_{i_{1}},...,U_{i_{r}}\right)-\mathrm{E}\left[f\left(U_{1},...,U_{r}\right)\right]\right),
\]
where $n_{\left(r\right)}$ is understood as $n!/\left(n-r\right)!$
and $\sum_{\left(i_{1},...,i_{r}\right)}$ is understood as summation
over indices $\left(i_{1},...,i_{r}\right)\in\left\{ 1,...,n\right\} ^{r}$
which are all distinct, and $\left\Vert \mathbb{U}_{n}^{\left(r\right)}\right\Vert _{\mathfrak{F}}\coloneqq\mathrm{sup}_{f\in\mathfrak{F}}\left|\mathbb{U}_{n}^{\left(r\right)}f\right|$.

\subsection{Proofs}

Recall that $\widehat{\phi}_{dx}\left(y\right)$ is the ``leave-in''
version of $\widehat{\phi}_{dx}^{\left(-i\right)}\left(y\right)$
(see Footnote \ref{fn:phi_hat - phi}). We refine the asymptotic expansion
for $\widehat{\phi}_{dx}\left(y\right)$ in FVX (see Theorem 1 therein)
\begin{equation}
\widehat{\phi}_{dx}\left(y\right)-\phi_{dx}\left(y\right)=\frac{1}{n}\sum_{i=1}^{n}\mathcal{L}_{dx}\left(W_{i},y\right)+o_{p}\left(n^{-1/2}\right),\label{eq:phi_hat - phi expansion}
\end{equation}
uniformly in $y\in\mathscr{S}_{g\left(d',x,\epsilon\right)\mid X=x}=\left[\underline{y}_{d'x},\overline{y}_{d'x}\right]$,
where
\[
\mathcal{L}_{dx}\left(W_{i},y\right)\coloneqq\zeta_{dx}\left(\phi_{dx}\left(y\right)\right)^{-1}\left\{ \mathbbm{1}\left(Y_{i}\leq\phi_{dx}\left(y\right),D_{i}=d\right)+\mathbbm{1}\left(Y_{i}\leq y,D_{i}=d'\right)-R_{d'x}\left(y\right)\right\} \pi_{x}\left(Z_{i},X_{i}\right).
\]
We provide a Bahadur-representation-type result for the estimated
counterfactual mapping with an $O_{p}\left(n^{-3/4}\right)$ (up to
a logarithmic term) estimate for the order of magnitude of the remainder
term in (\ref{eq:phi_hat - phi expansion}). The proof of our Theorem
\ref{thm:uniform rate} (derivation of (\ref{eq:linearization f_hat - f three terms}))
relies on the latter result. Proofs of all lemmas in the rest of the
appendix can be found in the online supplement.
\begin{lem}
\label{lem:Lemma 1 phi linear representation}Suppose that Assumption
\ref{assu: DGP1} holds. Then, 
\begin{equation}
\widehat{\phi}_{dx}\left(y\right)-\phi_{dx}\left(y\right)=\frac{1}{n}\sum_{i=1}^{n}\mathcal{L}_{dx}\left(W_{i},y\right)+O_{p}^{\star}\left(\left(\frac{\mathrm{log}\left(n\right)}{n}\right)^{3/4}\right),\label{eq:phi_hat - phi linearization}
\end{equation}
and the remainder term is uniform in $y\in I_{d'x}\coloneqq\mathscr{S}_{g\left(d',x,\epsilon\right)\mid X=x}=\left[\underline{y}_{d'x},\overline{y}_{d'x}\right]$.
\end{lem}
For any $\left(v,b\right)\in I_{x}\times\left[\underline{h},\overline{h}\right]$,
decompose $\widehat{f}_{\varDelta X}\left(v,x;b\right)-f_{\varDelta X}\left(v,x\right)$
into the sum of $\widehat{f}_{\varDelta X}\left(v,x;b\right)-\widetilde{f}_{\varDelta X}\left(v,x;b\right)$
and $\widetilde{f}_{\varDelta X}\left(v,x;b\right)-f_{\varDelta X}\left(v,x\right)$.
The following lemma shows that the leading term in the asymptotic
expansion of $\widehat{f}_{\varDelta X}\left(v,x;b\right)-\widetilde{f}_{\varDelta X}\left(v,x;b\right)$
can be represented by a \textit{U}-statistic, uniformly in $\left(v,b\right)\in I_{x}\times\left[\underline{h},\overline{h}\right]$.
\begin{lem}
\label{lem:lemma 2}Suppose that the assumptions of Theorem \ref{thm:uniform rate}
hold. Then,
\[
\widehat{f}_{\varDelta X}\left(v,x;b\right)-\widetilde{f}_{\varDelta X}\left(v,x;b\right)=\frac{1}{n_{\left(2\right)}}\sum_{\left(i,j\right)}\mathcal{G}_{x}\left(W_{i},W_{j},v;b\right)+O_{p}^{\star}\left(\frac{\mathrm{log}\left(n\right)}{nh^{2}}+\frac{\mathrm{log}\left(n\right)^{3/4}}{n^{3/4}h}\right),
\]
where the remainder is uniform in $\left(v,b\right)\in I_{x}\times\left[\underline{h},\overline{h}\right]$.
\end{lem}
Note that if $X_{i}=x$ and $D_{i}=d'$, $\phi_{dx}\left(Y_{i}\right)=g\left(d,x,\epsilon_{i}\right)$
and
\[
\left\{ \mathbbm{1}\left(Y_{j}\leq\phi_{dx}\left(Y_{i}\right),D_{j}=d\right)+\mathbbm{1}\left(Y_{j}\leq Y_{i},D_{j}=d'\right)\right\} \mathbbm{1}\left(X_{j}=x\right)=\mathbbm{1}\left(\epsilon_{j}\leq\epsilon_{i}\right)\mathbbm{1}\left(X_{j}=x\right).
\]
It is shown in the proof of Lemma \ref{lem:Lemma 1 phi linear representation}
that $R_{d'x}\left(y\right)=F_{g\left(d',x,\epsilon\right)\mid X}\left(y\mid x\right)\coloneqq\mathrm{Pr}\left[g\left(d',x,\epsilon\right)\leq y\mid X=x\right]$.
Therefore, if $X_{i}=x$ and $D_{i}=d'$, $R_{d'x}\left(Y_{i}\right)=F_{\epsilon\mid X}\left(\epsilon_{i}\mid x\right)$.
And therefore,
\begin{equation}
q_{dx}\left(W_{i},W_{j}\right)\mathbbm{1}\left(X_{j}=x\right)=\frac{\mathbbm{1}\left(D_{i}=d',X_{i}=x\right)}{\zeta_{dx}\left(g\left(d,x,\epsilon_{i}\right)\right)}\left\{ \mathbbm{1}\left(\epsilon_{j}\leq\epsilon_{i}\right)-F_{\epsilon\mid X}\left(\epsilon_{i}\mid x\right)\right\} \mathbbm{1}\left(X_{j}=x\right).\label{eq:q_dx simplify}
\end{equation}
Let 
\begin{eqnarray*}
\varpi_{x}\left(U_{i}\right) & \coloneqq & \frac{\mathbbm{1}\left(D_{i}=0,X_{i}=x\right)}{\zeta_{1x}\left(g\left(1,x,\epsilon_{i}\right)\right)}-\frac{\mathbbm{1}\left(D_{i}=1,X_{i}=x\right)}{\zeta_{0x}\left(g\left(0,x,\epsilon_{i}\right)\right)}\\
\mathcal{C}_{x}\left(U_{i},U_{j}\right) & \coloneqq & \varpi_{x}\left(U_{i}\right)\left\{ \mathbbm{1}\left(\epsilon_{j}\leq\epsilon_{i}\right)-F_{\epsilon\mid X}\left(\epsilon_{i}\mid x\right)\right\} \pi_{x}\left(Z_{j},X_{j}\right)
\end{eqnarray*}
and
\begin{eqnarray*}
\mathcal{H}_{x}\left(U_{i},U_{j},v;b\right) & \coloneqq & \mathcal{G}_{x}\left(\left(g\left(D_{i},X_{i},\epsilon_{i}\right),D_{i},Z_{i},X_{i}\right),\left(g\left(D_{j},X_{j},\epsilon_{j}\right),D_{j},Z_{j},X_{j}\right),v;b\right)\\
 & = & \frac{1}{b^{2}}K'\left(\frac{\varDelta_{x}\left(\epsilon_{i}\right)-v}{b}\right)\mathcal{C}_{x}\left(U_{i},U_{j}\right).
\end{eqnarray*}
Denote $\mathcal{H}_{x}^{\left[1\right]}\left(u,v;b\right)\coloneqq\mathrm{E}\left[\mathcal{H}_{x}\left(U,u,v;b\right)\right]$.
Clearly, we have $\mathcal{G}_{x}\left(W_{i},W_{j},v;b\right)=\mathcal{H}_{x}\left(U_{i},U_{j},v;b\right)$,
$\forall\ensuremath{i\neq j}$ and $\mathcal{G}_{x}^{\left[1\right]}\left(W_{i},v;b\right)=\mathcal{H}_{x}^{\left[1\right]}\left(U_{i},v;b\right)$
$\forall i$. Note that by conditional independence of $\epsilon$
and $Z$ given $X$, $\mathrm{E}\left[\mathcal{G}_{x}\left(w,W,v;b\right)\right]=\mathrm{E}\left[\mathcal{H}_{x}\left(u,U,v;b\right)\right]=0$.
And, $\mathrm{E}\left[\mathcal{G}_{x}\left(W_{i},W_{j},v;b\right)\right]=\mathrm{E}\left[\mathcal{H}_{x}\left(U_{i},U_{j},v;b\right)\right]=0,$
$\forall\ensuremath{i\neq j}$. Then, we have the following result.
\begin{lem}
\label{lem:lemma 3}Suppose that the assumptions of Theorem \ref{thm:uniform rate}
hold. Then,
\[
\frac{1}{n_{\left(2\right)}}\sum_{\left(i,j\right)}\mathcal{G}_{x}\left(W_{i},W_{j},v;b\right)=\frac{1}{n_{\left(2\right)}}\sum_{\left(i,j\right)}\mathcal{H}_{x}\left(U_{i},U_{j},v;b\right)=O_{p}^{\star}\left(\sqrt{\frac{\mathrm{log}\left(n\right)}{nh}},\sqrt{\frac{\mathrm{log}\left(n\right)}{nh^{3}}}\right),
\]
uniformly in $\left(v,b\right)\in I_{x}\times\left[\underline{h},\overline{h}\right]$.
\end{lem}
The following theorem is a stronger version of Theorem \ref{thm:uniform rate}.
It is easy to check that the asymptotic results in Theorem \ref{thm:uniform rate}
are straightforward implications of the non-asymptotic deviation bounds
here.
\begin{thm}
\label{thm: uniform rate appendix}Under the assumptions of Theorem
\ref{thm:uniform rate},
\[
\left\Vert \widehat{f}_{\varDelta\mid X}\left(\cdot\mid x;h\right)-f_{\varDelta\mid X}\left(\cdot\mid x\right)\right\Vert _{I_{x}}=O_{p}^{\star}\left(\sqrt{\frac{\mathrm{log}\left(n\right)}{nh}}+h^{P},\sqrt{\frac{\mathrm{log}\left(n\right)}{nh^{3}}}\right)
\]
and
\[
\left\Vert \widehat{f}_{\varDelta\mid X}\left(\cdot\mid x;\widehat{h}\right)-f_{\varDelta\mid X}\left(\cdot\mid x\right)\right\Vert _{I_{x}}=O_{p}^{\star}\left(\sqrt{\frac{\mathrm{log}\left(n\right)}{nh}}+h^{P},\sqrt{\frac{\mathrm{log}\left(n\right)}{nh^{3}}}+\delta_{n}\right).
\]
\end{thm}
\begin{proof}[Proof of Theorem \ref{thm: uniform rate appendix}]\sloppy
Denote $\mathcal{E}_{x}\left(U_{i},v;b\right)\coloneqq b^{-1}K\left(\left(\varDelta_{x}\left(\epsilon_{i}\right)-v\right)/b\right)\mathbbm{1}\left(X_{i}=x\right)$.
Then we write $\left\Vert \widetilde{f}_{\varDelta X}\left(\cdot,x;\cdot\right)-m_{\varDelta X}\left(\cdot,x;\cdot\right)\right\Vert _{I_{x}\times\left[\underline{h},\overline{h}\right]}=\left\Vert \mathbb{P}_{n}^{U}-\mathbb{P}^{U}\right\Vert _{\mathfrak{E}}$,
where $\mathfrak{E}\coloneqq\left\{ \mathcal{E}_{x}\left(\cdot,v;b\right):\left(v,b\right)\in I_{x}\times\left[\underline{h},\overline{h}\right]\right\} $.
By similar arguments used in the proof of Lemma \ref{lem:lemma 3}
(decomposing $K$ into the difference of two bounded monotone functions,
using \citet[Lemma 9.6]{Kosorok2007}, \citet[Lemma 9.9(viii)]{Kosorok2007},
\citet[Theorem 3.6.9]{gine2016mathematical} and \citet[Lemma B.2]{Chernozhukov2014anti}),
$\mathfrak{E}$ is uniformly VC-type with respect to a constant envelope
$F_{\mathfrak{E}}=O\left(h^{-1}\right)$. By arguments used for proving
$\left\Vert \mathbb{P}_{n}^{U}-\mathbb{P}^{U}\right\Vert _{\mathfrak{I}}=O_{p}^{\star}\left(\sqrt{\mathrm{log}\left(n\right)/\left(nh\right)}\right)$
in the proof of Lemma \ref{lem:lemma 2}, we have $\left\Vert \mathbb{P}_{n}^{U}-\mathbb{P}^{U}\right\Vert _{\mathfrak{E}}=O_{p}^{\star}\left(\sqrt{\mathrm{log}\left(n\right)/\left(nh\right)}\right)$.
By Lemmas \ref{lem:lemma 2} and \ref{lem:lemma 3}, we have 
\[
\left\Vert \widehat{f}_{\varDelta X}\left(\cdot,x;\cdot\right)-\widetilde{f}_{\varDelta X}\left(\cdot,x;\cdot\right)\right\Vert _{I_{x}\times\left[\underline{h},\overline{h}\right]}=O_{p}^{\star}\left(\sqrt{\frac{\mathrm{log}\left(n\right)}{nh}},\sqrt{\frac{\mathrm{log}\left(n\right)}{nh^{3}}}\right).
\]
By Hoeffding's inequality, $\widehat{p}_{x}-p_{x}=O_{p}^{\star}\left(\sqrt{\mathrm{log}\left(n\right)/n}\right)$
($\mathrm{Pr}\left[\left|\widehat{p}_{x}-p_{x}\right|>C_{1}\sqrt{\mathrm{log}\left(n\right)/n}\right]\leq C_{2}n^{-1}$).
Then, $p_{x}/\widehat{p}_{x}-1=O_{p}^{\star}\left(\sqrt{\mathrm{log}\left(n\right)/n}\right)$
follows from $\mathrm{Pr}\left[\left|p_{x}/\widehat{p}_{x}-1\right|>\left(2C_{1}/p_{x}\right)\sqrt{\mathrm{log}\left(n\right)/n}\right]\leq\mathrm{Pr}\left[\left|\widehat{p}_{x}-p_{x}\right|>C_{1}\sqrt{\mathrm{log}\left(n\right)/n}\right]+\mathrm{Pr}\left[\widehat{p}_{x}<p_{x}/2\right]\leq2C_{2}n^{-1}$,
where the last inequality holds when $n$ is sufficiently large. Then,
by these results, the equality $a/b=a/c-a\left(b-c\right)/c^{2}+a\left(b-c\right)^{2}/\left(bc^{2}\right)$
and Lemma \ref{lem:lemma 2},
\begin{multline}
\widehat{f}_{\varDelta\mid X}\left(v\mid x;b\right)-f_{\varDelta\mid X}\left(v\mid x\right)=p_{x}^{-1}\left(\widetilde{f}_{\varDelta X}\left(v,x;b\right)-m_{\varDelta X}\left(v,x;b\right)\right)+p_{x}^{-1}\frac{1}{n_{\left(2\right)}}\sum_{\left(i,j\right)}\mathcal{H}_{x}\left(U_{i},U_{j};v,b\right)\\
+p_{x}^{-1}\left(m_{\varDelta X}\left(v,x;b\right)-f_{\varDelta X}\left(v,x\right)\right)+O_{p}^{\star}\left(\sqrt{\frac{\mathrm{log}\left(n\right)}{n}}+\frac{\mathrm{log}\left(n\right)}{nh^{2}}+\frac{\mathrm{log}\left(n\right)^{3/4}}{n^{3/4}h},\sqrt{\frac{\mathrm{log}\left(n\right)}{nh^{3}}}\right),\label{eq:f_hat - f decompose}
\end{multline}
uniformly in $\left(v,b\right)\in I_{x}\times\left[\underline{h},\overline{h}\right]$.
By this result, Lemma \ref{lem:lemma 3}, (\ref{eq:infeasible decompose})
and $\left\Vert \mathbb{P}_{n}^{U}-\mathbb{P}^{U}\right\Vert _{\mathfrak{E}}=O_{p}^{\star}\left(\sqrt{\mathrm{log}\left(n\right)/\left(nh\right)}\right)$,
we now have 
\begin{equation}
\underset{\left(v,b\right)\in I_{x}\times\left[\underline{h},\overline{h}\right]}{\mathrm{sup}}\left|\widehat{f}_{\varDelta\mid X}\left(v\mid x;b\right)-f_{\varDelta\mid X}\left(v\mid x\right)\right|=O_{p}^{\star}\left(\sqrt{\frac{\mathrm{log}\left(n\right)}{nh}},\sqrt{\frac{\mathrm{log}\left(n\right)}{nh^{3}}}\right)+O\left(h^{P}\right).\label{eq:sup f_hat - f O_p_star bound}
\end{equation}
The first conclusion follows from this result. It follows from the
assumption $\mathrm{Pr}\left[\widehat{h}\in\left[\underline{h},\overline{h}\right]\right]>1-\delta_{n}$
and (\ref{eq:sup f_hat - f O_p_star bound}) that
\begin{multline*}
\mathrm{Pr}\left[\left\Vert \widehat{f}_{\varDelta\mid X}\left(\cdot\mid x;\widehat{h}\right)-f_{\varDelta\mid X}\left(\cdot\mid x\right)\right\Vert _{I_{x}}>C_{1}\left(\sqrt{\frac{\mathrm{log}\left(n\right)}{nh}}+h^{P}\right)\right]\\
\leq\mathrm{Pr}\left[\underset{\left(v,b\right)\in I_{x}\times\left[\underline{h},\overline{h}\right]}{\mathrm{sup}}\left|\widehat{f}_{\varDelta\mid X}\left(v\mid x;b\right)-f_{\varDelta\mid X}\left(v\mid x\right)\right|>C_{1}\left(\sqrt{\frac{\mathrm{log}\left(n\right)}{nh}}+h^{P}\right)\right]+\mathrm{Pr}\left[\widehat{h}\notin\left[\underline{h},\overline{h}\right]\right]\\
\leq\sqrt{\frac{\mathrm{log}\left(n\right)}{nh^{3}}}+\delta_{n}.
\end{multline*}
The second conclusion follows from this result.\end{proof}

The following lemma is a refinement of Lemma \ref{lem:lemma 3} and
the result $\left\Vert \widetilde{f}_{\varDelta X}\left(\cdot,x;\cdot\right)-m_{\varDelta X}\left(\cdot,x;\cdot\right)\right\Vert _{I_{x}\times\left[\underline{h},\overline{h}\right]}=O_{p}^{\star}\left(\sqrt{\mathrm{log}\left(n\right)/\left(nh\right)}\right)$.
It provides an estimate of the effect of a random bandwidth that satisfies
Assumption \ref{assu:h_hat}.
\begin{lem}
\label{lem:lemma random btw}Suppose that the assumptions of Theorem
\ref{thm:asymptotic normality} hold. Then, (a)
\[
\sqrt{n}\left(\frac{1}{n_{\left(2\right)}}\sum_{\left(i,j\right)}\mathcal{H}_{x}^{\vartriangle}\left(U_{i},U_{j},v;b,h\right)\right)=O_{p}^{\star}\left(\varepsilon_{n}\sqrt{\mathrm{log}\left(n\right)},\sqrt{\frac{\mathrm{log}\left(n\right)}{nh^{3}}}\right),
\]
uniformly in $\left(v,b\right)\in I_{x}\times\left[\underline{h},\overline{h}\right]$,
where $\mathcal{H}_{x}^{\vartriangle}\left(U_{i},U_{j},v;b,h\right)\coloneqq\sqrt{b}\cdot\mathcal{H}_{x}\left(U_{i},U_{j},v;b\right)-\sqrt{h}\cdot\mathcal{H}_{x}\left(U_{i},U_{j},v;h\right)$.
(b)
\[
\sqrt{nb}\left(\widetilde{f}_{\varDelta X}\left(v,x;b\right)-m_{\varDelta X}\left(v,x;b\right)\right)-\sqrt{nh}\left(\widetilde{f}_{\varDelta X}\left(v,x;h\right)-m_{\varDelta X}\left(v,x;h\right)\right)=O_{p}^{\star}\left(\varepsilon_{n}\sqrt{\mathrm{log}\left(n\right)}\right),
\]
uniformly in $\left(v,b\right)\in I_{x}\times\left[\underline{h},\overline{h}\right]$.
\end{lem}
Then by using these lemmas, we prove the asymptotic normality result
with either a deterministic bandwidth or a random bandwidth that satisfies
Assumption \ref{assu:h_hat}. For simplicity, denote 
\[
\varPsi\left(v\mid x;b\right)\coloneqq\sqrt{nb}\left(\widehat{f}_{\varDelta\mid X}\left(v\mid x;b\right)-f_{\varDelta\mid X}\left(v\mid x\right)-f_{\varDelta\mid X}^{\left(P\right)}\left(v\mid x\right)\mu_{K,P}b^{P}\right).
\]

\begin{proof}[Proof of Theorem \ref{thm:asymptotic normality}]The
Hoeffding decomposition (\ref{eq:G Hoeffding}) can be equivalently
written as 
\begin{eqnarray*}
\frac{1}{n_{\left(2\right)}}\sum_{\left(i,j\right)}\mathcal{G}_{x}\left(W_{i},W_{j},v;h\right) & = & \frac{1}{n_{\left(2\right)}}\sum_{\left(i,j\right)}\mathcal{H}_{x}\left(U_{i},U_{j},v;h\right)\\
 & = & \frac{1}{n}\sum_{i=1}^{n}\mathcal{H}_{x}^{\left[1\right]}\left(U_{i},v;h\right)+\frac{1}{n_{\left(2\right)}}\sum_{\left(i,j\right)}\left\{ \mathcal{H}_{x}\left(U_{i},U_{j},v;h\right)-\mathcal{H}_{x}^{\left[1\right]}\left(U_{j},v;h\right)\right\} .
\end{eqnarray*}
Then we show that the second term in the Hoeffding decomposition is
negligible uniformly in $v\in I_{x}$. It is shown in the proof of
Lemma \ref{lem:lemma 3} that $\mathfrak{H}\coloneqq\left\{ \mathcal{H}_{x}\left(\cdot,v;b\right):\left(v,b\right)\in I_{x}\times\left[\underline{h},\overline{h}\right]\right\} $
is uniformly VC-type with respect to a constant envelope $F_{\mathfrak{H}}=O\left(h^{-2}\right)$.
Then, by Corollary 5.6 of CK, 
\[
\mathrm{E}\left[\underset{\left(v,b\right)\in I_{x}\times\left[\underline{h},\overline{h}\right]}{\mathrm{sup}}\left|\frac{1}{n_{\left(2\right)}}\sum_{\left(i,j\right)}\left\{ \mathcal{H}_{x}\left(U_{i},U_{j},v;b\right)-\mathcal{H}_{x}^{\left[1\right]}\left(U_{j},v;b\right)\right\} \right|\right]=O\left(\left(nh^{2}\right)^{-1}\right).
\]
Then, by Lemma \ref{lem:lemma 2}, $\widehat{f}_{\varDelta X}\left(v,x;h\right)-\widetilde{f}_{\varDelta X}\left(v,x;h\right)=n^{-1}\sum_{i=1}^{n}\mathcal{H}_{x}^{\left[1\right]}\left(U_{i},v;h\right)+o_{p}\left(\left(nh\right)^{-1/2}\right)$.
By this result, (\ref{eq:infeasible decompose}) and (\ref{eq:f_hat - f decompose}),
$\varPsi\left(v\mid x;h\right)=p_{x}^{-1}\sum_{i=1}^{n}J_{i}+o_{p}\left(1\right)$,
where 
\[
J_{i}\coloneqq n^{-1/2}\left\{ h^{1/2}\left(\frac{1}{h}K\left(\frac{\varDelta_{i}-v}{h}\right)\mathbbm{1}\left(X_{i}=x\right)-m_{\varDelta X}\left(v,x;h\right)\right)+h^{1/2}\mathcal{H}_{x}^{\left[1\right]}\left(U_{i},v;h\right)\right\} .
\]
Let $\sigma_{J}^{2}\coloneqq\mathrm{E}\left[\left(\sum_{i=1}^{n}J_{i}\right)^{2}\right]$.
Denote
\begin{eqnarray*}
\rho_{x}\left(e\right) & \coloneqq & \frac{f_{\epsilon DX}\left(e,0,x\right)}{\zeta_{1x}\left(g\left(1,x,e\right)\right)}-\frac{f_{\epsilon DX}\left(e,1,x\right)}{\zeta_{0x}\left(g\left(0,x,e\right)\right)}\\
\varGamma_{x}\left(\epsilon,v;b\right) & \coloneqq & \int_{\underline{\epsilon}_{x}}^{\overline{\epsilon}_{x}}\frac{1}{b^{2}}K'\left(\frac{\varDelta_{x}\left(e\right)-v}{b}\right)\rho_{x}\left(e\right)\left\{ \mathbbm{1}\left(\epsilon\leq e\right)-F_{\epsilon\mid X}\left(e\mid x\right)\right\} \mathrm{d}e.
\end{eqnarray*}
Since $\mathcal{H}_{x}^{\left[1\right]}\left(U,v;h\right)=\varGamma_{x}\left(\epsilon,v;h\right)\pi_{x}\left(Z,X\right)$,
we have
\begin{multline}
\mathrm{E}\left[\mathcal{H}_{x}^{\left[1\right]}\left(U,v;h\right)\left(\frac{1}{h}K\left(\frac{\varDelta-v}{h}\right)\mathbbm{1}\left(X=x\right)-m_{\varDelta X}\left(v,x;h\right)\right)\right]=\\
\mathrm{E}\left[\varGamma_{x}\left(\epsilon,v;h\right)\left(\frac{1}{h}K\left(\frac{\varDelta_{x}\left(\epsilon\right)-v}{h}\right)\mathbbm{1}\left(X=x\right)-m_{\varDelta X}\left(v,x;h\right)\right)\pi_{x}\left(Z,X\right)\right]=0,\label{eq:cross term zero}
\end{multline}
where the second equality follows from LIE, the fact that $\epsilon$
is conditionally independent of $Z$ given $X$ and the fact that
$\mathrm{E}\left[\pi_{x}\left(Z,X\right)\mid X\right]=0$. Therefore,
\begin{eqnarray}
\sigma_{J}^{2} & = & \mathrm{E}\left[h\left(\frac{1}{h}K\left(\frac{\varDelta-v}{h}\right)\mathbbm{1}\left(X=x\right)-m_{\varDelta X}\left(v,x;h\right)\right)^{2}\right]+\mathrm{E}\left[h\cdot\mathcal{H}_{x}^{\left[1\right]}\left(U,v;h\right)\right]\nonumber \\
 & = & \mathscr{V}_{1}\left(v,x\right)+\mathscr{V}_{2}\left(v,x\right)+o\left(1\right),\label{eq:sigma_L limit}
\end{eqnarray}
where it follows from standard arguments for kernel density estimators
(\citealp{Newey:1994jb}) and (\ref{eq:infeasible decompose}) that
the first term on the right hand side of the second equality is $\mathscr{V}_{1}\left(v,x\right)+o\left(1\right)$
and in the proof of Lemma \ref{lem:lemma 3}, we show that $\mathrm{E}\left[h\cdot\mathcal{H}_{x}^{\left[1\right]}\left(U,v;h\right)\right]=\mathscr{V}_{2}\left(v,x\right)+o\left(1\right)$.
Then, we verify Lyapunov's condition. By Lo$\grave{\mathrm{e}}$ve's
$c_{r}$ inequality,
\begin{multline}
\sum_{i=1}^{n}\mathrm{E}\left[\left|\frac{J_{i}}{\sigma_{J}}\right|^{3}\right]\apprle\\
\sigma_{J}^{-3}n^{-1/2}h^{3/2}\left\{ \mathrm{E}\left[\left|\mathcal{H}_{x}^{\left[1\right]}\left(U,v;h\right)\right|^{3}\right]+\mathrm{E}\left[\left|\frac{1}{h}K\left(\frac{\varDelta-v}{h}\right)\mathbbm{1}\left(X=x\right)\right|^{3}\right]+\left|m_{\varDelta X}\left(v,x;h\right)\right|^{3}\right\} ,\label{eq:Lyapunov upper bound}
\end{multline}
where by change of variables, the second term in the bracket on the
right hand side of (\ref{eq:Lyapunov upper bound}) is $O\left(h^{-2}\right)$
and the third term is $O\left(1\right)$. Then by the $c_{r}$ inequality,
\begin{multline}
\mathrm{E}\left[\left|\mathcal{H}_{x}^{\left[1\right]}\left(U,v;h\right)\right|^{3}\right]\apprle\mathrm{E}\left[\left|\int_{\underline{\epsilon}_{x}}^{\overline{\epsilon}_{x}}\frac{1}{h^{2}}K'\left(\frac{\varDelta_{x}\left(e\right)-v}{h}\right)\rho_{x}\left(e\right)\mathbbm{1}\left(\epsilon\leq e\right)\mathrm{d}e\right|^{3}\right]\\
+\left|\int_{\underline{\epsilon}_{x}}^{\overline{\epsilon}_{x}}\frac{1}{h^{2}}K'\left(\frac{\varDelta_{x}\left(e\right)-v}{h}\right)\rho_{x}\left(e\right)F_{\epsilon\mid X}\left(e\mid x\right)\mathrm{d}e\right|^{3}.\label{eq:E=00005B|H^3|=00005D upper bound}
\end{multline}
By change of variables, 
\begin{multline*}
\mathrm{E}\left[\left|\int_{\underline{\epsilon}_{x}}^{\overline{\epsilon}_{x}}\frac{1}{h^{2}}K'\left(\frac{\varDelta_{x}\left(e\right)-v}{h}\right)\rho_{x}\left(e\right)\mathbbm{1}\left(\epsilon\leq e\right)\mathrm{d}e\right|^{3}\right]\\
\leq\left(\int_{\underline{\epsilon}_{x}}^{\overline{\epsilon}_{x}}\left|\frac{1}{h^{2}}K'\left(\frac{\varDelta_{x}\left(e\right)-v}{h}\right)\rho_{x}\left(e\right)\right|\mathrm{d}e\right)^{3}=O\left(h^{-3}\right).
\end{multline*}
It is shown in the proof of Lemma \ref{lem:lemma 3} that the second
term on the right hand side of (\ref{eq:E=00005B|H^3|=00005D upper bound})
is $O\left(1\right)$. Therefore, $\mathrm{E}\left[\left|\mathcal{H}_{x}^{\left[1\right]}\left(U,v;h\right)\right|^{3}\right]=O\left(h^{-3}\right)$.
Then, $\sum_{i=1}^{n}\mathrm{E}\left[\left|J_{i}/\sigma_{J}\right|^{3}\right]=O\left(\left(nh^{3}\right)^{-1/2}\right)$
follows from this result, (\ref{eq:sigma_L limit}) and (\ref{eq:Lyapunov upper bound}).
By Lyapunov's central limit theorem, $\sum_{i=1}^{n}J_{i}/\sigma_{J}\rightarrow_{d}\mathrm{N}\left(0,1\right).$
The first assertion $\varPsi\left(v\mid x;h\right)\rightarrow_{d}\mathrm{N}\left(0,\mathscr{V}\left(v\mid x\right)\right)$
follows from this result, $\varPsi\left(v\mid x;h\right)=p_{x}^{-1}\sum_{i=1}^{n}J_{i}+o_{p}\left(1\right)$,
(\ref{eq:sigma_L limit}) and Slutsky's lemma.

Recall that $S\left(v\mid x;b\right)$ is defined by (\ref{eq:S Z definition}).
For the second part, by (\ref{eq:f_hat - f decompose}),
\begin{multline}
S\left(v\mid x;b\right)-S\left(v\mid x;h\right)=\\
p_{x}^{-1}\left\{ \sqrt{nb}\left(\widetilde{f}_{\varDelta X}\left(v,x;b\right)-m_{\varDelta X}\left(v,x;b\right)\right)-\sqrt{nh}\left(\widetilde{f}_{\varDelta X}\left(v,x;h\right)-m_{\varDelta X}\left(v,x;h\right)\right)\right\} \\
+p_{x}^{-1}\left\{ \sqrt{nb}\left(m_{\varDelta X}\left(v,x;b\right)-f_{\varDelta X}\left(v,x\right)\right)-\sqrt{nh}\left(m_{\varDelta X}\left(v,x;h\right)-f_{\varDelta X}\left(v,x\right)\right)\right\} \\
+p_{x}^{-1}\left\{ \sqrt{n}\left(\frac{1}{n_{\left(2\right)}}\sum_{i,j}\mathcal{H}_{x}^{\vartriangle}\left(U_{i},U_{j},v;b\right)\right)\right\} +O_{p}^{\star}\left(\upsilon_{n},\sqrt{\frac{\mathrm{log}\left(n\right)}{nh^{3}}}\right),\label{eq:S_b - S_h}
\end{multline}
uniformly in $\left(v,b\right)\in I_{x}\times\left[\underline{h},\overline{h}\right]$,
where $\upsilon_{n}\coloneqq\sqrt{\mathrm{log}\left(n\right)h}+\mathrm{log}\left(n\right)/\sqrt{nh^{3}}+\left(\mathrm{log}\left(n\right)^{3}/\left(nh^{2}\right)\right)^{1/4}$.
It then follows from Lemma \ref{lem:lemma random btw} and (\ref{eq:infeasible decompose})
that 
\[
\varPsi\left(v\mid x;b\right)-\varPsi\left(v\mid x;h\right)=O_{p}^{\star}\left(\upsilon_{n}+\varepsilon_{n}\sqrt{\mathrm{log}\left(n\right)},\sqrt{\frac{\mathrm{log}\left(n\right)}{nh^{3}}}\right)+o\left(\sqrt{nh}h^{P}\right),
\]
uniformly in $\left(v,b\right)\in I_{x}\times\left[\underline{h},\overline{h}\right]$.
Then, by using $\mathrm{Pr}\left[\widehat{h}\in\left[\underline{h},\overline{h}\right]\right]>1-\delta_{n}$,
\[
\left\Vert \varPsi\left(\cdot\mid x;\widehat{h}\right)-\varPsi\left(\cdot\mid x;h\right)\right\Vert _{I_{x}}\leq\underset{\left(v,b\right)\in I_{x}\times\left[\underline{h},\overline{h}\right]}{\mathrm{sup}}\left|\varPsi\left(v\mid x;b\right)-\varPsi\left(v\mid x;h\right)\right|=o_{p}\left(1\right),
\]
where the inequality holds with probability $1-O\left(\delta_{n}\right)$,
we have $\varPsi\left(v\mid x;\widehat{h}\right)=\varPsi\left(v\mid x;h\right)+o_{p}\left(1\right)$.
The second assertion follows from this result, $\varPsi\left(v\mid x;h\right)\rightarrow_{d}\mathrm{N}\left(0,\mathscr{V}\left(v\mid x\right)\right)$
and Slutsky's lemma.\end{proof}

\section{\label{sec:Proofs B}Proofs of Theorems in Section \ref{sec:Robust-inference}}

The following lemma gives rates of convergence of $\widehat{R}_{d'x}$
and $\widehat{\zeta}_{dx}$.
\begin{lem}
\label{lem:Lemma 4}Suppose that the assumptions of Theorem \ref{thm:variance estimator}
hold. Then, (a) $\left\Vert \widehat{R}_{d'x}-R_{d'x}\right\Vert _{I_{d'x}}$
$=O_{p}^{\star}\left(\sqrt{\mathrm{log}\left(n\right)/n}\right)$.
(b) $\widehat{\zeta}_{dx}\left(y;b_{\zeta}\right)-\zeta_{dx}\left(y\right)=O_{p}^{\star}\left(\sqrt{\mathrm{log}\left(n\right)/\left(nh_{\zeta}\right)}\right)+O\left(h_{\zeta}^{2}\right)$,
uniformly in $\left(y,b_{\zeta}\right)\in\dot{I}_{dx}\times\left[\underline{h}_{\zeta},\overline{h}_{\zeta}\right]$,
where $\dot{I}_{dx}$ is any inner closed sub-interval of $I_{dx}$,
$\underline{h}_{\zeta}\coloneqq\left(1-\varepsilon_{n}^{\zeta}\right)h_{\zeta}$
and $\overline{h}_{\zeta}\coloneqq\left(1+\varepsilon_{n}^{\zeta}\right)h_{\zeta}$.
\end{lem}
Denote 
\[
\kappa_{1}^{V}\left(\gamma\right)\coloneqq\sqrt{\frac{\mathrm{log}\left(n\right)}{nh_{\zeta}}}+h_{\zeta}^{2}+\frac{\mathrm{log}\left(n\right)^{2/3}}{\left(nh\right)^{2/3}h\gamma^{1/3}}+\frac{\mathrm{log}\left(n\right)}{nh^{4}\gamma}+\sqrt{\frac{\mathrm{log}\left(n\right)}{nh^{2}}}\textrm{ and }\kappa_{2}^{V}\left(\gamma\right)\coloneqq\gamma+\sqrt{\frac{\mathrm{log}\left(n\right)}{nh^{3}}},
\]
for $\gamma\in\left(0,1\right)$. The following lemma is useful in
proving Theorem \ref{thm:variance estimator}.
\begin{lem}
\label{lem:variance estimator 1}Under the assumptions of Theorem
\ref{thm:variance estimator}, (a) for some constants $C_{1},C_{2}>0$,
when $n$ is sufficiently large,
\[
\mathrm{Pr}\left[\underset{\left(v,b,b_{\zeta}\right)\in I_{x}\times\left[\underline{h},\overline{h}\right]\times\left[\underline{h}_{\zeta},\overline{h}_{\zeta}\right]}{\mathrm{sup}}\left|\widehat{V}\left(v\mid x;b,b_{\zeta}\right)-V\left(v\mid x;b\right)\right|>C_{1}\kappa_{1}^{V}\left(\gamma\right)\right]\leq C_{2}\kappa_{2}^{V}\left(\gamma\right),\,\forall\gamma\in\left(0,1\right).
\]
(b) $V\left(v\mid x;b\right)-V\left(v\mid x;h\right)=O\left(\varepsilon_{n}h\right)$,
uniformly in $\left(v,b\right)\in I_{x}\times\left[\underline{h},\overline{h}\right]$.
\end{lem}
\sloppy Then, in the following theorem, we present a non-asymptotic
deviation bound for $\left\Vert \widehat{V}\left(\cdot\mid x;\widehat{h},\widehat{h}_{\zeta}\right)-V\left(\cdot\mid x;h\right)\right\Vert _{I_{x}}$.
It is easy to see that this result is stronger than Theorem \ref{thm:variance estimator}
presented in the main text. Its proof is relegated to the online supplement.
\begin{thm}
\label{thm:variance theorem appendix}Under the assumptions of Theorem
\ref{thm:variance estimator}, for some constants $C_{1},C_{2}>0$,
when $n$ is sufficiently large,
\[
\mathrm{Pr}\left[\left\Vert \widehat{V}\left(\cdot\mid x;\widehat{h},\widehat{h}_{\zeta}\right)-V\left(\cdot\mid x;h\right)\right\Vert _{I_{x}}>C_{1}\left(\kappa_{1}^{V}\left(\gamma\right)+\varepsilon_{n}h\right)\right]\leq C_{2}\left(\kappa_{2}^{V}\left(\gamma\right)+\delta_{n}+\delta_{n}^{\zeta}\right),\,\forall\gamma\in\left(0,1\right).
\]
\end{thm}
The following lemma states a useful property of the $O_{p}^{\sharp}$
notation.
\begin{lem}
\label{lem:O_p_bound O_p_star}Let $\alpha_{n},\beta_{n},\gamma_{n},\delta_{n}$
be deterministic sequences in $\left(0,\infty\right)$. Suppose that
$Y_{n}\geq0$ depends only on the data ($W_{1}^{n}$). Suppose that
$Y_{n}=O_{p}^{\star}\left(\alpha_{n},\beta_{n}\right)$ and $\mathrm{Pr}_{\mid W_{1}^{n}}\left[\left|X_{n}\right|>C_{1}Y_{n}\right]=O_{p}^{\star}\left(\gamma_{n},\delta_{n}\right)$.
Then, (a) $X_{n}=O_{p}^{\sharp}\left(\alpha_{n},\gamma_{n},\beta_{n}+\delta_{n}\right)$.
(b) $Y_{n}=O_{p}^{\sharp}\left(\alpha_{n},\varepsilon_{n},\beta_{n}\right)$,
where $\varepsilon_{n}$ is an arbitrary deterministic sequence in
$\left(0,\infty\right)$ that decays to zero as $n\uparrow\infty$.
\end{lem}
Let the infeasible JMB process be given by
\begin{equation}
S_{\mathsf{jmb}}\left(v\mid x;b\right)\coloneqq\frac{1}{\sqrt{n}}\sum_{i=1}^{n}\nu_{i}p_{x}^{-1}\left\{ \left(\widetilde{\mathcal{U}}_{x}^{\left[1\right]}\left(W_{i},v;b\right)-\widetilde{\mu}_{\mathcal{U}_{x}}\left(v;b\right)\right)+\left(\widetilde{\mathcal{U}}_{x}^{\left[2\right]}\left(W_{i},v;b\right)-\widetilde{\mu}_{\mathcal{U}_{x}}\left(v;b\right)\right)\right\} \label{eq:infeasible multiplier process definition}
\end{equation}
where $\widetilde{\mathcal{U}}_{x}^{\left[1\right]}\left(W_{i},v;b\right)\coloneqq\left(n-1\right)^{-1}\sum_{j\neq i}\mathcal{U}_{x}\left(W_{j},W_{i},v;b\right)$
is the jackknife estimator of $\mathcal{U}_{x}^{\left[1\right]}\left(W_{i},v;b\right)$,
$\mathcal{U}_{x}^{\left[2\right]}\left(W_{i},v;b\right)$ is defined
similarly and $\widetilde{\mu}_{\mathcal{U}_{x}}\left(v;b\right)\coloneqq n_{\left(2\right)}^{-1}\sum_{\left(i,j\right)}\mathcal{U}_{x}\left(W_{i},W_{j},v;b\right)$.
The following lemma also shows that the difference of $\widehat{Z}_{\mathsf{jmb}}\left(v\mid x;\widehat{h},\widehat{h}_{\zeta}\right)$
and 
\[
Z_{\mathsf{jmb}}\left(v\mid x;h\right)\coloneqq\frac{S_{\mathsf{jmb}}\left(v\mid x;h\right)}{\sqrt{V\left(v\mid x;h\right)}}
\]
is negligible.
\begin{lem}
\label{lem:bootstrap equivalence}Suppose that the assumptions in
the statement of Theorem \ref{thm:confidence band} hold. Then,
\begin{multline*}
\widehat{Z}_{\mathsf{jmb}}\left(v\mid x;\widehat{h},\widehat{h}_{\zeta}\right)-Z_{\mathsf{jmb}}\left(v\mid x;h\right)\\
=O_{p}^{\sharp}\left(\kappa_{1,n}^{V}\sqrt{\mathrm{log}\left(n\right)}+\left(\frac{\mathrm{log}\left(n\right)^{3}}{nh^{2}}\right)^{1/4}+\varepsilon_{n}\sqrt{\mathrm{log}\left(n\right)},n^{-1},\kappa_{2,n}^{V}+\delta_{n}+\delta_{n}^{\zeta}\right),
\end{multline*}
uniformly in $v\in I_{x}$, where 
\[
\kappa_{1,n}^{V}\coloneqq\sqrt{\frac{\mathrm{log}\left(n\right)}{nh_{\zeta}}}+h_{\zeta}^{2}+\sqrt{\frac{\mathrm{log}\left(n\right)}{nh^{4}}}\textrm{ and }\kappa_{2,n}^{V}\coloneqq\sqrt{\frac{\mathrm{log}\left(n\right)}{nh^{4}}}.
\]
\end{lem}
Denote
\begin{eqnarray}
\mathcal{M}_{x}\left(U_{i},U_{j},v;b\right) & \coloneqq & p_{x}^{-1}\mathcal{U}_{x}\left(\left(g\left(D_{i},X_{i},\epsilon_{i}\right),D_{i},Z_{i},X_{i}\right),\left(g\left(D_{j},X_{j},\epsilon_{j}\right),D_{j},Z_{j},X_{j}\right),v;b\right)\nonumber \\
 & = & p_{x}^{-1}\sqrt{b}\left\{ \mathcal{E}_{x}\left(U_{i},v;b\right)+\mathcal{H}_{x}\left(U_{i},U_{j},v;b\right)\right\} ,\label{eq:M =00003D E + H}
\end{eqnarray}
$\mathcal{M}_{x}^{\left[1\right]}\left(u,v;b\right)\coloneqq\mathrm{E}\left[\mathcal{M}_{x}\left(U,u,v;b\right)\right]$
and $\mathcal{M}_{x}^{\left[2\right]}\left(u,v;b\right)\coloneqq\mathrm{E}\left[\mathcal{M}_{x}\left(u,U,v;b\right)\right]$.
Note that $\mathcal{M}_{x}^{\left[2\right]}\left(\cdot,v;b\right)$
is constant and equal to $\mu_{\mathcal{M}_{x}}\left(v;b\right)\coloneqq\mathrm{E}\left[\mathcal{M}_{x}\left(U_{1},U_{2},v;b\right)\right]$
and $V\left(v\mid x,b\right)=\mathrm{Var}\left[\mathcal{M}_{x}^{\left[1\right]}\left(U,v;b\right)\right]$.
By (\ref{eq:f_hat - f decompose}) and (\ref{eq:M =00003D E + H}),

\begin{equation}
S\left(v\mid x;h\right)=\sqrt{n}\left\{ \frac{1}{n_{\left(2\right)}}\sum_{\left(i,j\right)}\mathcal{M}_{x}\left(U_{i},U_{j},v;h\right)-\mu_{\mathcal{M}_{x}}\left(v;h\right)\right\} +O_{p}^{\star}\left(\upsilon_{n},\sqrt{\frac{\mathrm{log}\left(n\right)}{nh^{3}}}\right)+O\left(\sqrt{nh^{5}}\right),\label{eq:S leading}
\end{equation}
uniformly in $v\in I_{x}$. Note that by (\ref{eq:M =00003D E + H}),
\[
S_{\mathsf{jmb}}\left(v\mid x;b\right)=\frac{1}{\sqrt{n}}\sum_{i=1}^{n}\nu_{i}\left\{ \left(\widetilde{\mathcal{M}}_{x}^{\left[1\right]}\left(U_{i},v;b\right)-\widetilde{\mu}_{\mathcal{M}_{x}}\left(v;b\right)\right)+\left(\widetilde{\mathcal{M}}_{x}^{\left[2\right]}\left(U_{i},v;b\right)-\widetilde{\mu}_{\mathcal{M}_{x}}\left(v;b\right)\right)\right\} ,
\]
where $\widetilde{\mathcal{M}}_{x}^{\left[1\right]}\left(U_{i},v;b\right)\coloneqq\left(n-1\right)^{-1}\sum_{j\neq i}\mathcal{M}_{x}\left(U_{j},U_{i},v;b\right)$
is the jackknife estimator of $\mathcal{M}_{x}^{\left[1\right]}\left(U_{i},v;b\right)$,
$\widetilde{\mathcal{M}}_{x}^{\left[2\right]}\left(U_{i},v;b\right)$
is defined similarly and $\widetilde{\mu}_{\mathcal{M}_{x}}\left(v;b\right)\coloneqq n_{\left(2\right)}^{-1}\sum_{\left(i,j\right)}\mathcal{M}_{x}\left(U_{i},U_{j},v;b\right)$.

The proof of Theorem \ref{thm:confidence band} hinges on the \textit{U}-process
representation (\ref{eq:S leading}), the coupling theorem for \textit{U}-process
suprema (Proposition 2.1 of CK) and the JMB coupling theorem (Theorem
3.1 of CK). Denote $\bar{\mathfrak{M}}^{\left[1\right]}\coloneqq\left\{ \mathcal{M}_{x}^{\left[1\right]}\left(\cdot,v;h\right)/\sqrt{V\left(v\mid x;h\right)}:v\in I_{x}\right\} $.
Let $\left\{ G^{U}\left(f\right):f\in\bar{\mathfrak{M}}^{\left[1\right]}\right\} $
be a centered Gaussian process with the covariance structure $\mathrm{E}\left[G^{U}\left(f_{1}\right)G^{U}\left(f_{2}\right)\right]=\mathrm{Cov}\left[f_{1}\left(U\right),f_{2}\left(U\right)\right]$,
$\forall\left(f_{1},f_{2}\right)\in\bar{\mathfrak{M}}^{\left[1\right]}\times\bar{\mathfrak{M}}^{\left[1\right]}$.\footnote{The existence of $\left\{ G^{U}\left(f\right):f\in\mathfrak{M}^{\left[1\right]}\right\} $
is guaranteed by the Kolmogorov extension theorem.} One can show that this Gaussian process admits a version that is
a tight random element in $\ell^{\infty}\left(\bar{\mathfrak{M}}^{\left[1\right]}\right)$,
and we use $\left\{ G^{U}\left(f\right):f\in\bar{\mathfrak{M}}^{\left[1\right]}\right\} $
to denote the tight version. By the coupling theorems of CK and the
anti-concentration inequality of \citet{Chernozhukov2014anti}, the
distribution of $\left\Vert Z\left(\cdot\mid x;\widehat{h},\widehat{h}_{\zeta}\right)\right\Vert _{I_{x}}$
and the conditional distribution of $\left\Vert \widehat{Z}_{\mathsf{jmb}}\left(\cdot\mid x;\widehat{h},\widehat{h}_{\zeta}\right)\right\Vert _{I_{x}}$
given the original sample can be approximated by that of $\left\Vert G^{U}\right\Vert _{\mathfrak{\bar{M}}^{\left[1\right]}}$.
I.e., there are positive constants $C_{1},C_{2},C_{3}$ such that
\begin{equation}
\underset{t\in\mathbb{R}}{\mathrm{sup}}\left|\mathrm{Pr}\left[\left\Vert Z\left(\cdot\mid x;\widehat{h},\widehat{h}_{\zeta}\right)\right\Vert _{I_{x}}\leq t\right]-\mathrm{Pr}\left[\left\Vert G^{U}\right\Vert _{\mathfrak{\bar{M}}^{\left[1\right]}}\leq t\right]\right|\leq C_{1}\bar{\kappa}_{1,n}\label{eq:Kolmogorov Z}
\end{equation}
and
\begin{equation}
\mathrm{Pr}\left[\underset{t\in\mathbb{R}}{\mathrm{sup}}\left|\mathrm{Pr}_{\mid W_{1}^{n}}\left[\left\Vert \widehat{Z}_{\mathsf{jmb}}\left(\cdot\mid x;\widehat{h},\widehat{h}_{\zeta}\right)\right\Vert _{I_{x}}\leq t\right]-\mathrm{Pr}\left[\left\Vert G^{U}\right\Vert _{\mathfrak{\bar{M}}^{\left[1\right]}}\leq t\right]\right|\leq C_{2}\bar{\kappa}_{2,n}^{\sharp}\right]>1-C_{3}\bar{\kappa}_{3,n}^{\sharp},\label{eq:Kolmogorov Z_hat}
\end{equation}
where 
\begin{gather*}
\bar{\kappa}_{1,n}\coloneqq\mathrm{log}\left(n\right)\kappa_{1,n}^{V}+\sqrt{\mathrm{log}\left(n\right)}\sqrt{nh^{5}}+\mathrm{log}\left(n\right)\varepsilon_{n}+\left(\frac{\mathrm{log}\left(n\right)^{7}}{nh^{3}}\right)^{1/8}+\delta_{n}+\delta_{n}^{\zeta},\\
\bar{\kappa}_{2,n}^{\sharp}\coloneqq\left(\frac{\mathrm{log}\left(n\right)^{5}}{nh^{3}}\right)^{1/16}+\mathrm{log}\left(n\right)\kappa_{1,n}^{V}+\mathrm{log}\left(n\right)\varepsilon_{n}\textrm{ and }\bar{\kappa}_{3,n}^{\sharp}\coloneqq\left(\frac{\mathrm{log}\left(n\right)^{5}}{nh^{3}}\right)^{1/16}+\delta_{n}+\delta_{n}^{\zeta}.
\end{gather*}
 Then we show that $\left|\mathrm{Pr}\left[\left\Vert Z\left(\cdot\mid x;\widehat{h},\widehat{h}_{\zeta}\right)\right\Vert _{I_{x}}\leq z_{1-\alpha}^{\mathsf{jmb}}\right]-\left(1-\alpha\right)\right|$
is bounded by $C_{1}\bar{\kappa}_{1,n}+C_{2}\bar{\kappa}_{2,n}^{\sharp}+C_{3}\bar{\kappa}_{3,n}^{\sharp}$.
This result gives an estimate of the coverage error decay rate of
the JMB UCB. We state it as the following theorem, which implies the
conclusion of Theorem \ref{thm:confidence band}. It is clear that
the coverage error decay rate is $O\left(n^{-\varrho}\right)$ if
the assumptions of Theorem \ref{thm:confidence band} are satisfied.
\begin{thm}
\label{thm:confidence band appendix}Suppose that the assumptions
of Theorem \ref{thm:confidence band} hold. Then,
\begin{multline}
\mathrm{Pr}\left[f_{\varDelta\mid X}\left(v\mid x\right)\in\mathit{CB}_{\mathsf{jmb}}\left(v\mid x;\widehat{h},\widehat{h}_{\zeta}\right),\,\forall v\in I_{x}\right]=\left(1-\alpha\right)\\
+O\left(\left(\frac{\mathrm{log}\left(n\right)^{5}}{nh^{3}}\right)^{1/16}+\mathrm{log}\left(n\right)\kappa_{1,n}^{V}+\sqrt{\mathrm{log}\left(n\right)}\sqrt{nh^{5}}+\mathrm{log}\left(n\right)\varepsilon_{n}+\delta_{n}+\delta_{n}^{\zeta}\right).\label{eq:coverage error decay rate Z}
\end{multline}
\end{thm}
\begin{proof}[Proof of Theorem \ref{thm:confidence band appendix}]It
follows from (\ref{eq:infeasible decompose}), (\ref{eq:S_b - S_h})
and Lemma \ref{lem:lemma random btw} that 
\[
S\left(v\mid x;b\right)-S\left(v\mid x;h\right)=O_{p}^{\star}\left(\varepsilon_{n}\sqrt{\mathrm{log}\left(n\right)}+\upsilon_{n},\sqrt{\frac{\mathrm{log}\left(n\right)}{nh^{3}}}\right)+O\left(\sqrt{nh}\left(h^{3}+\varepsilon_{n}h^{2}\right)\right)
\]
uniformly in $\left(v,b\right)\in I_{x}\times\left[\underline{h},\overline{h}\right]$.
Then, by this result and $\mathrm{Pr}\left[\widehat{h}\in\left[\underline{h},\overline{h}\right]\right]>1-\delta_{n}$,
we have
\begin{equation}
S\left(v\mid x;\widehat{h}\right)-S\left(v\mid x;h\right)=O_{p}^{\star}\left(\varepsilon_{n}\sqrt{\mathrm{log}\left(n\right)}+\upsilon_{n}+\sqrt{nh}\left(h^{3}+\varepsilon_{n}h^{2}\right),\sqrt{\frac{\mathrm{log}\left(n\right)}{nh^{3}}}+\delta_{n}\right).\label{eq:S_hat - S_h}
\end{equation}
Write
\begin{multline}
Z\left(v\mid x;\widehat{h},\widehat{h}_{\zeta}\right)-\frac{S\left(v\mid x;h\right)}{\sqrt{V\left(v\mid x;h\right)}}=\\
\frac{S\left(v\mid x;\widehat{h}\right)}{\sqrt{V\left(v\mid x;h\right)}}\left(\frac{\sqrt{V\left(v\mid x;h\right)}}{\sqrt{\widehat{V}\left(v\mid x;\widehat{h},\widehat{h}_{\zeta}\right)}}-1\right)+\frac{S\left(v\mid x;\widehat{h}\right)-S\left(v\mid x;h\right)}{\sqrt{V\left(v\mid x;h\right)}}.\label{eq:Z-S/V}
\end{multline}
Taking $\gamma=\sqrt{\mathrm{log}\left(n\right)/\left(nh^{4}\right)}$
in Theorem \ref{thm:variance theorem appendix}, we get
\[
\left\Vert \widehat{V}\left(\cdot\mid x;\widehat{h},\widehat{h}_{\zeta}\right)-V\left(\cdot\mid x;h\right)\right\Vert _{I_{x}}=O_{p}^{\star}\left(\kappa_{1,n}^{V}+\varepsilon_{n}h,\kappa_{2,n}^{V}+\delta_{n}+\delta_{n}^{\zeta}\right).
\]
Note that the second equality of (\ref{eq:sigma_L limit}) holds uniformly
in $v\in I_{x}$ so that we have $V\left(v\mid x;h\right)\rightarrow\mathscr{V}\left(v\mid x\right)$
uniformly in $v\in I_{x}$, as $h\downarrow0$ and therefore, $\underline{V}\coloneqq\mathrm{inf}_{v\in I_{x}}V\left(v\mid x;h\right)\rightarrow\mathrm{inf}_{v\in I_{x}}\mathscr{V}\left(v\mid x\right)$
as $h\downarrow0$. By (\ref{eq:V_2 > 0}), (\ref{eq:f expression})
and continuity of $f_{\epsilon\mid X}\left(\cdot\mid x\right)$, $\varDelta_{x,j}^{-1}$,
$\varDelta_{x,j}'$, $f_{\epsilon DX}\left(\cdot,d,x\right)$, $g\left(d,x,\cdot\right)$
and $f_{dx\mid C_{x}}$, we have $\mathrm{inf}_{v\in I_{x}}\mathscr{V}\left(v\mid x\right)>0$.
Therefore, when $h$ is sufficiently small, $\underline{V}>\mathrm{inf}_{v\in I_{x}}\mathscr{V}\left(v\mid x\right)/2>0$.
By these results, we also have $\mathrm{Pr}\left[\mathrm{inf}_{v\in I_{x}}\widehat{V}\left(v\mid x;\widehat{h},\widehat{h}_{\zeta}\right)>\mathrm{inf}_{v\in I_{x}}\mathscr{V}\left(v\mid x\right)/2\right]=1-O\left(\kappa_{2,n}^{V}+\delta_{n}+\delta_{n}^{\zeta}\right)$.
By these results, 
\begin{equation}
\left|\frac{\sqrt{V\left(v\mid x;h\right)}}{\sqrt{\widehat{V}\left(v\mid x;\widehat{h},\widehat{h}_{\zeta}\right)}}-1\right|\leq\left|\frac{V\left(v\mid x;h\right)}{\widehat{V}\left(v\mid x;\widehat{h},\widehat{h}_{\zeta}\right)}-1\right|=O_{p}^{\star}\left(\kappa_{1,n}^{V}+\varepsilon_{n}h,\kappa_{2,n}^{V}+\delta_{n}+\delta_{n}^{\zeta}\right),\label{eq:sqrt(V)/sqrt(V_hat)-1 bound}
\end{equation}
uniformly in $v\in I_{x}$. Then, by this result, (\ref{eq:S_hat - S_h}),
(\ref{eq:Z-S/V}) and $S\left(v\mid x,h\right)=O_{p}^{\star}\left(\sqrt{\mathrm{log}\left(n\right)},\sqrt{\mathrm{log}\left(n\right)/\left(nh^{3}\right)}\right)$,
we have
\begin{multline*}
Z\left(v\mid x;\widehat{h},\widehat{h}_{\zeta}\right)-\frac{S\left(v\mid x;h\right)}{\sqrt{V\left(v\mid x;h\right)}}\\
=O_{p}^{\star}\left(\varepsilon_{n}\sqrt{\mathrm{log}\left(n\right)}+\upsilon_{n}+\sqrt{nh}\left(h^{3}+\varepsilon_{n}h^{2}\right)+\kappa_{1,n}^{V}\sqrt{\mathrm{log}\left(n\right)},\kappa_{2,n}^{V}+\delta_{n}+\delta_{n}^{\zeta}\right),
\end{multline*}
uniformly in $v\in I_{x}$. By (\ref{eq:S leading}) and $\underline{V}>\mathrm{inf}_{v\in I_{x}}\mathscr{V}\left(v\mid x\right)/2>0$
when $h$ is sufficiently small, we have the deviation bound:
\begin{multline}
\mathrm{Pr}\left[\left|\left\Vert Z\left(\cdot\mid x;\widehat{h},\widehat{h}_{\zeta}\right)\right\Vert _{I_{x}}-\left\Vert \mathbb{U}_{n}^{\left(2\right)}\right\Vert _{\bar{\mathfrak{M}}}\right|>C_{1}\left(\kappa_{1,n}^{V}\sqrt{\mathrm{log}\left(n\right)}+\upsilon_{n}+\varepsilon_{n}\sqrt{\mathrm{log}\left(n\right)}+\sqrt{nh^{5}}\right)\right]\\
\leq C_{2}\left(\kappa_{2,n}^{V}+\delta_{n}+\delta_{n}^{\zeta}\right).\label{eq:deviation bound 1}
\end{multline}
By similar arguments used in the proof of Lemma \ref{lem:lemma 3}
and the fact that when $h$ is sufficiently small, $\underline{V}>\mathrm{inf}_{v\in I_{x}}\mathscr{V}\left(v\mid x\right)/2>0$,
$\bar{\mathfrak{M}}\coloneqq\left\{ \mathcal{M}_{x}\left(\cdot,v;h\right)/\sqrt{V\left(v\mid x;h\right)}:v\in I_{x}\right\} $
is uniformly VC-type with respect to a constant envelope $F_{\bar{\mathfrak{M}}}=O\left(h^{-3/2}\right)$.
By Lemma A.3 of CK, $\bar{\mathfrak{M}}^{\left[1\right]}$ is also
uniformly VC-type with respect to a constant envelope $F_{\bar{\mathfrak{M}}^{\left[1\right]}}=F_{\bar{\mathfrak{M}}}$.
Let $\overline{\sigma}_{\bar{\mathfrak{M}}^{\left[1\right]}}^{2}\coloneqq\mathrm{sup}_{f\in\bar{\mathfrak{M}}^{\left[1\right]}}\mathbb{P}^{U}f^{2}$
and $\sigma_{\bar{\mathfrak{M}}}^{2}\coloneqq\mathrm{sup}_{f\in\bar{\mathfrak{M}}}\mathrm{E}\left[f\left(U_{1},U_{2}\right)^{2}\right]$.
By $\mathrm{E}\left[\mathcal{M}_{x}^{\left[1\right]}\left(U,v;h\right)^{2}\right]/V\left(v\mid x;h\right)=1+h\left(m_{\varDelta X}\left(v,x;h\right)/p_{x}\right)^{2}/V\left(v\mid x;h\right)$
and (\ref{eq:infeasible decompose}), $\overline{\sigma}_{\bar{\mathfrak{M}}^{\left[1\right]}}^{2}\leq1+h\cdot\left\Vert m_{\varDelta X}\left(\cdot,x;h\right)\right\Vert _{I_{x}}^{2}/\left(p_{x}^{2}\underline{V}\right)=O\left(1\right)$.
By calculations in the proof of Lemma \ref{lem:lemma 3}, $\sigma_{\bar{\mathfrak{M}}}^{2}=O\left(h^{-2}\right)$.
Denote $\bar{\mathfrak{M}}_{\pm}\coloneqq\bar{\mathfrak{M}}\cup\left(-\bar{\mathfrak{M}}\right)$
($-\bar{\mathfrak{M}}\coloneqq\left\{ -f:f\in\bar{\mathfrak{M}}\right\} $)
and $\bar{\mathfrak{M}}_{\pm}^{\left[1\right]}\coloneqq\bar{\mathfrak{M}}^{\left[1\right]}\cup\left(-\bar{\mathfrak{M}}^{\left[1\right]}\right)$.
The coupling theorem (CK Proposition 2.1 with $\mathcal{H}=\bar{\mathfrak{M}}_{\pm}$,
$\overline{\sigma}_{\mathfrak{g}}=\overline{\sigma}_{\bar{\mathfrak{M}}^{\left[1\right]}}$,
$\sigma_{\mathfrak{h}}=\sigma_{\mathfrak{\bar{M}}}$, $b_{\mathfrak{g}}=b_{\mathfrak{h}}=F_{\bar{\mathfrak{M}}}$,
$\chi_{n}=0$ and $q=\infty$) implies that when $n$ is sufficiently
large ($n\geq\left(V_{\bar{\mathfrak{M}}_{\pm}}\mathrm{log}\left(A_{\bar{\mathfrak{M}}_{\pm}}\vee n\right)\right)^{3}\vee3$),
for each coupling error $\gamma\in\left(0,1\right)$, one can construct
a random variable $Z_{\bar{\mathfrak{M}}_{\pm},\gamma}$ that satisfies
the following conditions: $Z_{\bar{\mathfrak{M}}_{\pm},\gamma}=_{d}\mathrm{sup}_{f\in\bar{\mathfrak{M}}_{\pm}^{\left[1\right]}}G^{U}\left(f\right)=\left\Vert G^{U}\right\Vert _{\bar{\mathfrak{M}}^{\left[1\right]}}$,
where $\left\{ G^{U}\left(f\right):f\in\bar{\mathfrak{M}}_{\pm}^{\left[1\right]}\right\} $
is a centered separable Gaussian process that has the same covariance
structure as the H$\mathrm{\acute{a}}$jek process $\left\{ \mathbb{G}_{n}^{U}f:f\in\bar{\mathfrak{M}}_{\pm}^{\left[1\right]}\right\} $
($\mathrm{E}\left[G^{U}\left(f\right)G^{U}\left(g\right)\right]=\mathrm{Cov}\left[f\left(U\right),g\left(U\right)\right]$,
$\forall f,g\in\bar{\mathfrak{M}}_{\pm}^{\left[1\right]}$), and the
difference between $\mathrm{sup}_{f\in\bar{\mathfrak{M}}_{\pm}}\mathbb{U}_{n}^{\left(2\right)}f=\left\Vert \mathbb{U}_{n}^{\left(2\right)}\right\Vert _{\bar{\mathfrak{M}}}$
and $Z_{\bar{\mathfrak{M}}_{\pm},\gamma}$ satisfies the deviation
bound:
\begin{equation}
\mathrm{Pr}\left[\left|\left\Vert \mathbb{U}_{n}^{\left(2\right)}\right\Vert _{\bar{\mathfrak{M}}}-Z_{\bar{\mathfrak{M}}_{\pm},\gamma}\right|>C_{1}\kappa_{\bar{\mathfrak{M}}_{\pm}}\left(\gamma\right)\right]\leq C_{2}\left(\gamma+n^{-1}\right),\label{eq:deviation bound 2}
\end{equation}
where $\kappa_{\bar{\mathfrak{M}}_{\pm}}\left(\gamma\right)\coloneqq\mathrm{log}\left(n\right)^{2/3}/\left(\gamma^{1/3}\left(nh^{3}\right)^{1/6}\right)+\mathrm{log}\left(n\right)/\left(\gamma\sqrt{nh^{3}}\right)$.
By the Gaussian anti-concentration inequality (\citealp[Corollary 2.1]{Chernozhukov2014anti}),
since $\mathrm{E}\left[G^{U}\left(f\right)^{2}\right]=1$ $\forall f\in\bar{\mathfrak{M}}^{\left[1\right]}$,
\begin{equation}
\underset{t\in\mathbb{R}}{\mathrm{sup}}\,\mathrm{Pr}\left[\left|\left\Vert G^{U}\right\Vert _{\bar{\mathfrak{M}}^{\left[1\right]}}-t\right|\leq\varepsilon\right]\apprle\varepsilon\left(\mathrm{E}\left[\left\Vert G^{U}\right\Vert _{\bar{\mathfrak{M}}^{\left[1\right]}}\right]+1\right),\,\forall\varepsilon>0.\label{eq:anti-concentration bound}
\end{equation}
By Dudley's metric entropy bound (\citealp[Theorem 2.3.7]{gine2016mathematical}),
Lemma A.2 of CK and calculations (see the proof of Lemma \ref{lem:lemma 3}
for details), 
\begin{equation}
\mathrm{E}\left[\left\Vert G^{U}\right\Vert _{\bar{\mathfrak{M}}^{\left[1\right]}}\right]\apprle\left(\overline{\sigma}_{\bar{\mathfrak{M}}^{\left[1\right]}}\vee n^{-1/2}\left\Vert F_{\bar{\mathfrak{M}}^{\left[1\right]}}\right\Vert _{\mathbb{P}^{U},2}\right)\sqrt{\mathrm{log}\left(n\right)},\label{eq:metric entropy bound}
\end{equation}
when $n$ is sufficiently large. Then, since $Z_{\bar{\mathfrak{M}}_{\pm},\gamma}=_{d}\left\Vert G^{U}\right\Vert _{\bar{\mathfrak{M}}^{\left[1\right]}}$,
by \citet[Lemma 2.1]{Chernozhukov2016}, (\ref{eq:deviation bound 1})
and (\ref{eq:deviation bound 2}), when $n$ is sufficiently large,
$\forall\gamma\in\left(0,1\right)$, 
\begin{multline}
\underset{t\in\mathbb{R}}{\mathrm{sup}}\left|\mathrm{Pr}\left[\left\Vert Z\left(\cdot\mid x;\widehat{h},\widehat{h}_{\zeta}\right)\right\Vert _{I_{x}}\leq t\right]-\mathrm{Pr}\left[\left\Vert G^{U}\right\Vert _{\bar{\mathfrak{M}}^{\left[1\right]}}\leq t\right]\right|\leq\\
\underset{t\in\mathbb{R}}{\mathrm{sup}}\,\mathrm{Pr}\left[\left|\left\Vert G^{U}\right\Vert _{\bar{\mathfrak{M}}^{\left[1\right]}}-t\right|\leq C_{1}\left(\kappa_{1,n}^{V}\sqrt{\mathrm{log}\left(n\right)}+\upsilon_{n}+\varepsilon_{n}\sqrt{\mathrm{log}\left(n\right)}+\sqrt{nh^{5}}+\kappa_{\bar{\mathfrak{M}}_{\pm}}\left(\gamma\right)\right)\right]\\
+C_{2}\left(\gamma+\kappa_{2,n}^{V}+\delta_{n}+\delta_{n}^{\zeta}\right).\label{eq:Kolmogorov distance 1}
\end{multline}
By (\ref{eq:anti-concentration bound}), (\ref{eq:metric entropy bound})
and optimally choosing $\gamma$ that gives the fastest rate of convergence
of the right hand side of (\ref{eq:Kolmogorov distance 1}), which
should balance $\gamma$ and $\kappa_{\bar{\mathfrak{M}}_{\pm}}\left(\gamma\right)\sqrt{\mathrm{log}\left(n\right)}$
and set $\gamma=\mathrm{log}\left(n\right)^{7/8}/\left(nh^{3}\right)^{1/8}$,
we have (\ref{eq:Kolmogorov Z}).

We apply the JMB coupling theorem (Theorem 3.1 of CK) with $\mathcal{H}=\bar{\mathfrak{M}}_{\pm}$,
$\overline{\sigma}_{\mathfrak{g}}=\overline{\sigma}_{\bar{\mathfrak{M}}^{\left[1\right]}}$,
$\sigma_{\mathfrak{h}}=\sigma_{\bar{\mathfrak{M}}}$, $b_{\mathfrak{g}}=b_{\mathfrak{h}}=\nu_{\mathfrak{h}}=F_{\bar{\mathfrak{M}}}$,
$\chi_{n}=0$ and $q=\infty$. When $n$ is sufficiently large (so
that the assumptions in Equation (9) of CK are satisfied and $n\geq3$),
for any coupling error $\gamma\in\left(0,1\right)$, there exists
a random variable $Z_{\bar{\mathfrak{M}}_{\pm},\gamma}^{\sharp}$
such that (1) $Z_{\bar{\mathfrak{M}}_{\pm},\gamma}^{\sharp}$ is independent
of the data; (2) $Z_{\bar{\mathfrak{M}}_{\pm},\gamma}^{\sharp}$ has
the same distribution as $\left\Vert G^{U}\right\Vert _{\bar{\mathfrak{M}}^{\left[1\right]}}$;
(3) $Z_{\bar{\mathfrak{M}}_{\pm},\gamma}^{\sharp}$ and $\left\Vert Z_{\mathsf{jmb}}\left(\cdot\mid x;h\right)\right\Vert _{I_{x}}$
satisfies the deviation bound: when $n$ is sufficiently large, $\forall\gamma\in\left(0,1\right)$,
\[
\mathrm{Pr}\left[\left|\left\Vert Z_{\mathsf{jmb}}\left(\cdot\mid x;h\right)\right\Vert _{I_{x}}-Z_{\bar{\mathfrak{M}}_{\pm},\gamma}^{\sharp}\right|>C_{1}\kappa_{\bar{\mathfrak{M}}_{\pm}}^{\sharp}\left(\gamma\right)\right]\leq C_{2}\left(\gamma+n^{-1}\right),
\]
where $\kappa_{\bar{\mathfrak{M}}_{\pm}}^{\sharp}\left(\gamma\right)\coloneqq\mathrm{log}\left(n\right)^{3/4}/\left(\gamma^{3/2}\left(nh^{3}\right)^{1/4}\right)$
and then, by Markov's inequality, 
\[
\mathrm{Pr}\left[\mathrm{Pr}_{\mid W_{1}^{n}}\left[\left|\left\Vert Z_{\mathsf{jmb}}\left(\cdot\mid x;h\right)\right\Vert _{I_{x}}-Z_{\bar{\mathfrak{M}}_{\pm},\gamma}^{\sharp}\right|>C_{1}\kappa_{\bar{\mathfrak{M}}_{\pm}}^{\sharp}\left(\gamma\right)\right]>\sqrt{C_{2}\left(\gamma+n^{-1}\right)}\right]\leq\sqrt{C_{2}\left(\gamma+n^{-1}\right)}.
\]
By this result and Lemma \ref{lem:bootstrap equivalence}, when $n$
is sufficiently large, $\forall\gamma\in\left(0,1\right)$, with probability
at least $1-C_{3}\left(\sqrt{\gamma}+\kappa_{2,n}^{V}+\delta_{n}+\delta_{n}^{\zeta}\right)$,
\begin{multline*}
\mathrm{Pr}_{\mid W_{1}^{n}}\left[\left|\left\Vert \widehat{Z}_{\mathsf{jmb}}\left(\cdot\mid x;\widehat{h},\widehat{h}_{\zeta}\right)\right\Vert _{I_{x}}-Z_{\bar{\mathfrak{M}}_{\pm},\gamma}^{\sharp}\right|>C_{1}\left(\kappa_{1,n}^{V}\sqrt{\mathrm{log}\left(n\right)}+\varepsilon_{n}\sqrt{\mathrm{log}\left(n\right)}+\kappa_{\bar{\mathfrak{M}}_{\pm}}^{\sharp}\left(\gamma\right)\right)\right]\\
\leq C_{2}\left(\sqrt{\gamma}+n^{-1/2}\right).
\end{multline*}
Then, since $Z_{\bar{\mathfrak{M}}_{\pm},\gamma}^{\sharp}$ is independent
of the data and $Z_{\bar{\mathfrak{M}}_{\pm},\gamma}^{\sharp}=_{d}\left\Vert G^{U}\right\Vert _{\bar{\mathfrak{M}}^{\left[1\right]}}$,
by the above deviation bound and \citet[Lemma 2.1]{Chernozhukov2016},
with probability at least $1-C_{3}\left(\sqrt{\gamma}+\kappa_{2,n}^{V}+\delta_{n}+\delta_{n}^{\zeta}\right)$,
\begin{multline}
\underset{t\in\mathbb{R}}{\mathrm{sup}}\left|\mathrm{Pr}_{\mid W_{1}^{n}}\left[\left\Vert \widehat{Z}_{\mathsf{jmb}}\left(\cdot\mid x;\widehat{h},\widehat{h}_{\zeta}\right)\right\Vert _{I_{x}}\leq t\right]-\mathrm{Pr}\left[\left\Vert G^{U}\right\Vert _{\bar{\mathfrak{M}}^{\left[1\right]}}\leq t\right]\right|\leq\\
\underset{t\in\mathbb{R}}{\mathrm{sup}}\,\mathrm{Pr}\left[\left|\left\Vert G^{U}\right\Vert _{\bar{\mathfrak{M}}^{\left[1\right]}}-t\right|\leq C_{1}\left(\kappa_{1,n}^{V}\sqrt{\mathrm{log}\left(n\right)}+\varepsilon_{n}\sqrt{\mathrm{log}\left(n\right)}+\kappa_{\bar{\mathfrak{M}}_{\pm}}^{\sharp}\left(\gamma\right)\right)\right]+C_{2}\left(\sqrt{\gamma}+n^{-1/2}\right).\label{eq:Z_hat_pound Z_til_pound Kolmogorov}
\end{multline}
Then (\ref{eq:Kolmogorov Z_hat}) follows from (\ref{eq:anti-concentration bound}),
(\ref{eq:metric entropy bound}) and optimally choosing $\gamma$
that gives the fastest rate of convergence of the upper bound in (\ref{eq:Z_hat_pound Z_til_pound Kolmogorov}),
which should balance $\sqrt{\gamma}$ and $\kappa_{\bar{\mathfrak{M}}_{\pm}}^{\sharp}\left(\gamma\right)\sqrt{\mathrm{log}\left(n\right)}$
and set $\gamma=\mathrm{log}\left(n\right)^{5/8}/\left(nh^{3}\right)^{1/8}$.

\sloppy Let $F_{G}\left(t\right)\coloneqq\mathrm{Pr}\left[\left\Vert G^{U}\right\Vert _{\bar{\mathfrak{M}}^{\left[1\right]}}\leq t\right]$
and $F_{G}^{-1}\left(1-\alpha\right)\coloneqq\mathrm{inf}\left\{ t\in\mathbb{R}:F_{G}\left(t\right)\geq1-\alpha\right\} $
denote the (unconditional) CDF and quantile of $\left\Vert G^{U}\right\Vert _{\bar{\mathfrak{M}}^{\left[1\right]}}$.
It is easy to check that (\ref{eq:anti-concentration bound}) implies
that $F_{G}$ is continuous everywhere. Suppose that $\mathrm{sup}_{t\in\mathbb{R}}\left|\mathrm{Pr}_{\mid W_{1}^{n}}\left[\left\Vert \widehat{Z}_{\mathsf{jmb}}\left(\cdot\mid x;\widehat{h},\widehat{h}_{\zeta}\right)\right\Vert _{I_{x}}\leq t\right]-F_{G}\left(t\right)\right|\leq C_{2}\bar{\kappa}_{2,n}^{\sharp}$
and (\ref{eq:Kolmogorov Z_hat}) implies that this event happens with
probability greater than $1-C_{3}\bar{\kappa}_{3,n}^{\sharp}$. Then,
$F_{G}\left(z_{1-\alpha}^{\mathsf{jmb}}\right)\geq1-\alpha-C_{2}\bar{\kappa}_{2,n}^{\sharp}$
and $\mathrm{Pr}_{\mid W_{1}^{n}}\left[\left\Vert \widehat{Z}_{\mathsf{jmb}}\left(\cdot\mid x;\widehat{h},\widehat{h}_{\zeta}\right)\right\Vert _{I_{x}}\leq F_{G}^{-1}\left(1-\alpha+C_{2}\bar{\kappa}_{2,n}^{\sharp}\right)\right]\geq1-\alpha$,
by \citet[Lemma 21.1(ii)]{van2000asymptotic}. Then we have $F_{G}^{-1}\left(1-\alpha-C_{2}\bar{\kappa}_{2,n}^{\sharp}\right)\leq z_{1-\alpha}^{\mathsf{jmb}}\leq F_{G}^{-1}\left(1-\alpha+C_{2}\bar{\kappa}_{2,n}^{\sharp}\right)$
and such an event happens with probability greater than $1-C_{3}\bar{\kappa}_{3,n}^{\sharp}$.
Now using $\mathrm{Pr}\left[z_{1-\alpha}^{\mathsf{jmb}}\leq F_{G}^{-1}\left(1-\alpha+C_{2}\bar{\kappa}_{2,n}^{\sharp}\right)\right]>1-C_{3}\bar{\kappa}_{3,n}^{\sharp}$,
we have
\begin{eqnarray}
\mathrm{Pr}\left[\left\Vert Z\left(\cdot\mid x;\widehat{h},\widehat{h}_{\zeta}\right)\right\Vert _{I_{x}}\leq z_{1-\alpha}^{\mathsf{jmb}}\right] & \leq & \mathrm{Pr}\left[\left\Vert Z\left(\cdot\mid x;\widehat{h},\widehat{h}_{\zeta}\right)\right\Vert _{I_{x}}\leq F_{G}^{-1}\left(1-\alpha+C_{2}\bar{\kappa}_{2,n}^{\sharp}\right)\right]+C_{3}\bar{\kappa}_{3,n}^{\sharp}\nonumber \\
 & \leq & F_{G}\left(F_{G}^{-1}\left(1-\alpha+C_{2}\bar{\kappa}_{2,n}^{\sharp}\right)\right)+C_{1}\bar{\kappa}_{1,n}+C_{3}\bar{\kappa}_{3,n}^{\sharp}\nonumber \\
 & = & 1-\alpha+C_{1}\bar{\kappa}_{1,n}+C_{2}\bar{\kappa}_{2,n}^{\sharp}+C_{3}\bar{\kappa}_{3,n}^{\sharp},\label{eq:S coverage error upper bound}
\end{eqnarray}
where the second inequality follows from (\ref{eq:Kolmogorov Z})
and the equality follows from continuity of $F_{G}$ and \citet[Lemma 21.1(ii)]{van2000asymptotic}.
By using $\mathrm{Pr}\left[F_{G}^{-1}\left(1-\alpha-C_{2}\bar{\kappa}_{2,n}^{\sharp}\right)\leq z_{1-\alpha}^{\mathsf{jmb}}\right]>1-C_{3}\bar{\kappa}_{3,n}^{\sharp}$
and similar arguments, we have $\mathrm{Pr}\left[\left\Vert Z\left(\cdot\mid x;\widehat{h},\widehat{h}_{\zeta}\right)\right\Vert _{I_{x}}>z_{1-\alpha}^{\mathsf{jmb}}\right]\leq\alpha+C_{1}\bar{\kappa}_{1,n}+C_{2}\bar{\kappa}_{2,n}^{\sharp}+C_{3}\bar{\kappa}_{3,n}^{\sharp}$.
Then, the conclusion of the theorem follows.\end{proof}

\section{Nonparametric bootstrap confidence band\label{sec:Nonparametric-bootstrap}}

A nonparametric bootstrap sample $\left\{ W_{1}^{*},...,W_{n}^{*}\right\} $
consists of $n$ independent draws from the original sample $W_{1}^{n}\coloneqq\left\{ W_{1},...,W_{n}\right\} $
($W_{i}\coloneqq\left(Y_{i},D_{i},X_{i},Z_{i}\right)$) with replacement.
Let $\widehat{\phi}_{dx}^{\left(-i\right)*}\left(y\right)$ be the
bootstrap analogue of $\widehat{\phi}_{dx}^{\left(-i\right)}\left(y\right)$,
i.e., $\widehat{\phi}_{dx}^{\left(-i\right)*}\left(y\right)$ is the
minimizer of the bootstrap analogue of (\ref{eq:Q_hat definition})
computed using the bootstrap sample. Similarly, we construct the bootstrap
analogues $\widehat{\varDelta}_{i}^{*}$ and $\widehat{f}_{\varDelta\mid X}^{*}\left(v\mid x;b\right)$
of $\widehat{\varDelta}_{i}$ and $\widehat{f}_{\varDelta\mid X}\left(v\mid x;b\right)$
respectively by computing the corresponding terms in (\ref{eq:pseudo ITE definition})
and (\ref{eq:f_hat definition}) using the bootstrap sample. Let 
\begin{equation}
S_{\mathsf{npb}}\left(v\mid x;b\right)\coloneqq\sqrt{nb}\left(\widehat{f}_{\varDelta\mid X}^{*}\left(v\mid x;b\right)-\widehat{f}_{\varDelta\mid X}\left(v\mid x;b\right)\right)\textrm{ and }Z_{\mathsf{npb}}\left(v\mid x;b,b_{\zeta}\right)\coloneqq\frac{S_{\mathsf{npb}}\left(v\mid x;b\right)}{\sqrt{\widehat{V}\left(v\mid x;b,b_{\zeta}\right)}}\label{eq:S_npb Z_npb}
\end{equation}
be the nonparametric bootstrap analogues of $S\left(v\mid x;b\right)$
and $Z\left(v\mid x;b,b_{\zeta}\right)$ respectively. We define $z_{1-\alpha}^{\mathsf{npb}}$
as the $1-\alpha$ quantile of the conditional distribution of $\left\Vert Z_{\mathsf{npb}}\left(\cdot\mid x;\widehat{h},\widehat{h}_{\zeta}\right)\right\Vert _{I_{x}}$.
A nonparametric bootstrap confidence band can be constructed similarly
to (\ref{eq:confidence band definition}):
\begin{equation}
\mathit{CB}_{\mathsf{npb}}\left(v\mid x;\widehat{h},\widehat{h}_{\zeta}\right)\coloneqq\left[\widehat{f}_{\varDelta\mid X}\left(v\mid x;\widehat{h}\right)\pm z_{1-\alpha}^{\mathsf{npb}}\sqrt{\frac{\widehat{V}\left(v\mid x;\widehat{h},\widehat{h}_{\zeta}\right)}{n\widehat{h}}}\right].\label{eq:nonparametric bootstrap confidence band}
\end{equation}
The bootstrap critical value $z_{1-\alpha}^{\mathsf{npb}}$ can be
estimated by Monte Carlo simulations. The procedure is summarized
below. Undersmoothing in Step 3 is not needed if one takes the bias
correction approach by replacing the kernel $K$ with $M\left(\cdot;b,b_{\mathsf{b}}\right)$
in (\ref{eq:S_npb Z_npb}). It is also straightforward to adapt the
algorithm to construct nonparametric bootstrap confidence bands for
the PDF conditioning on sub-vectors.
\begin{lyxalgorithm}[Nonparametric bootstrap]
\textbf{Steps 1-4}: Same as those in Algorithm \ref{alg:multiplier}.
\textbf{Step 5}: In each of the iterations $r=1,...,B$, independently
draw $\left\{ W_{1}^{*\left(r\right)},...,W_{n}^{*\left(r\right)}\right\} $
with replacement from the original sample; compute the pseudo ITEs
$\left\{ \widehat{\varDelta}_{1}^{*\left(r\right)},...,\widehat{\varDelta}_{n}^{*\left(r\right)}\right\} $
by applying (\ref{eq:Q_hat definition}), (\ref{eq:phi_hat definition}),
and (\ref{eq:pseudo ITE definition}) to the bootstrap sample $\left\{ W_{1}^{*\left(r\right)},...,W_{n}^{*\left(r\right)}\right\} $;
compute $\widehat{f}_{\varDelta\mid X}^{*\left(r\right)}\left(v\mid x;\widehat{h}\right)$
by applying (\ref{eq:f_hat definition}) to $\left\{ \widehat{\varDelta}_{1}^{*\left(r\right)},...,\widehat{\varDelta}_{n}^{*\left(r\right)}\right\} $
and $\left\{ W_{1}^{*\left(r\right)},...,W_{n}^{*\left(r\right)}\right\} $;
and compute $Z_{\mathsf{npb}}^{\left(r\right)}\left(v\mid x;\widehat{h},\widehat{h}_{\zeta}\right)\coloneqq S_{\mathsf{npb}}^{\left(r\right)}\left(v\mid x;\widehat{h}\right)/\sqrt{\widehat{V}\left(v\mid x;\widehat{h},\widehat{h}_{\zeta}\right)}$.
\textbf{Step 6}: Compute the critical value :
\begin{equation}
z_{1-\alpha}^{\mathsf{npb}}=\mathrm{inf}\left\{ t\in\mathbb{R}:\frac{1}{B}\sum_{r=1}^{B}\mathbbm{1}\left(\underset{v\in I_{x}^{G}}{\mathrm{max}}\left|Z_{\mathsf{npb}}^{\left(r\right)}\left(v\mid x;\widehat{h},\widehat{h}_{\zeta}\right)\right|\leq t\right)\geq1-\alpha\right\} .\label{eq:nonparametric bootstrap critical value}
\end{equation}
\textbf{Step 7}: Compute the nonparametric bootstrap confidence band
$\mathit{CB}_{\mathsf{npb}}$ using (\ref{eq:nonparametric bootstrap critical value})
over $v\in I_{x}^{G}$.
\end{lyxalgorithm}
Theorem \ref{thm:nonparametric bootstrap} below extends the result
of Theorem \ref{thm:confidence band appendix} to the nonparametric
bootstrap. It shows that the coverage probability errors of the nonparametric
bootstrap UCB in (\ref{eq:nonparametric bootstrap confidence band})
decay at a polynomial rate. The proof is found in the online supplement.
In the proof, we show a nonparametric bootstrap version of (\ref{eq:bias corrected linearization}).
By the nonparametric bootstrap coupling theorem of \citet{Chernozhukov2016},
we have
\begin{equation}
\mathrm{Pr}\left[\underset{t\in\mathbb{R}}{\mathrm{sup}}\left|\mathrm{Pr}_{\mid W_{1}^{n}}\left[\left\Vert Z_{\mathsf{npb}}\left(\cdot\mid x;\widehat{h},\widehat{h}_{\zeta}\right)\right\Vert _{I_{x}}\leq t\right]-\mathrm{Pr}\left[\left\Vert G^{U}\right\Vert _{\bar{\mathfrak{M}}^{\left[1\right]}}\leq t\right]\right|\leq C_{2}\bar{\kappa}_{2,n}^{*}\right]>1-C_{3}\bar{\kappa}_{3,n}^{*},\label{eq:Z_star G_U Kolmogorov distance}
\end{equation}
where 
\[
\bar{\kappa}_{2,n}^{*}\coloneqq\left(\frac{\mathrm{log}\left(n\right)^{5}}{nh^{3}}\right)^{1/12}+\mathrm{log}\left(n\right)\kappa_{1,n}^{V}+\mathrm{log}\left(n\right)\varepsilon_{n}\textrm{ and }\bar{\kappa}_{3,n}^{*}\coloneqq\left(\frac{\mathrm{log}\left(n\right)^{5}}{nh^{3}}\right)^{1/12}+\delta_{n}+\delta_{n}^{\zeta}.
\]
The result is then implied by (\ref{eq:Kolmogorov Z}) and (\ref{eq:Z_star G_U Kolmogorov distance}).
\begin{thm}
\label{thm:nonparametric bootstrap}Suppose that the assumptions of
Theorem \ref{thm:confidence band} hold. Then,
\begin{multline}
\mathrm{Pr}\left[f_{\varDelta\mid X}\left(v\mid x\right)\in\mathit{CB}_{\mathsf{npb}}\left(v\mid x;\widehat{h},\widehat{h}_{\zeta}\right),\,\forall v\in I_{x}\right]=\left(1-\alpha\right)\\
+O\left(\left(\frac{\mathrm{log}\left(n\right)^{5}}{nh^{3}}\right)^{1/12}+\kappa_{1,n}^{V}\sqrt{\mathrm{log}\left(n\right)}+\sqrt{\mathrm{log}\left(n\right)}\sqrt{nh^{5}}+\varepsilon_{n}\sqrt{\mathrm{log}\left(n\right)}+\delta_{n}+\delta_{n}^{\zeta}\right).\label{eq:coverage error decay rate 2 nonparametric bootstrap}
\end{multline}
\end{thm}
A non-studentized constant-width nonparametric bootstrap confidence
band can be constructed as 
\[
\widetilde{\mathit{CB}}_{\mathsf{npb}}\left(v\mid x;\widehat{h}\right)\coloneqq\left[\widehat{f}_{\varDelta\mid X}\left(v\mid x;\widehat{h}\right)\pm\frac{s_{1-\alpha}^{\mathsf{npb}}}{\sqrt{n\widehat{h}}}\right],
\]
where $s_{1-\alpha}^{\mathsf{npb}}$ denotes the $1-\alpha$ quantile
of the conditional distribution of $\left\Vert S_{\mathsf{npb}}\left(\cdot\mid x;\widehat{h}\right)\right\Vert _{I_{x}}$.
Note that $\widetilde{\mathit{CB}}_{\mathsf{npb}}\left(v\mid x;\widehat{h}\right)$
does not require estimation of $\zeta_{dx}$ and the additional tuning
parameters $b_{\zeta}$ and $K_{\zeta}$. One can show deviation bounds
similar to (\ref{eq:Kolmogorov Z}) and (\ref{eq:Kolmogorov Z_hat}):
there are positive constants $C_{1},C_{2},C_{3}$ such that
\begin{equation}
\underset{t\in\mathbb{R}}{\mathrm{sup}}\left|\mathrm{Pr}\left[\left\Vert S\left(\cdot\mid x;\widehat{h}\right)\right\Vert _{I_{x}}\leq t\right]-\mathrm{Pr}\left[\left\Vert G^{U}\right\Vert _{\mathfrak{M}^{\left[1\right]}}\leq t\right]\right|\leq C_{1}\kappa_{1,n}\label{eq:Kolmogorov S}
\end{equation}
and
\begin{equation}
\mathrm{Pr}\left[\underset{t\in\mathbb{R}}{\mathrm{sup}}\left|\mathrm{Pr}_{\mid W_{1}^{n}}\left[\left\Vert S_{\mathsf{npb}}\left(\cdot\mid x;\widehat{h}\right)\right\Vert _{I_{x}}\leq t\right]-\mathrm{Pr}\left[\left\Vert G^{U}\right\Vert _{\mathfrak{M}^{\left[1\right]}}\leq t\right]\right|\leq C_{2}\kappa_{2,n}^{*}\right]>1-C_{3}\kappa_{3,n}^{*},\label{eq:S_star G_U Kolmogorov distance}
\end{equation}
where 
\begin{gather*}
\kappa_{1,n}\coloneqq\left(\frac{\mathrm{log}\left(n\right)^{7}}{nh^{3}}\right)^{1/8}+\sqrt{\mathrm{log}\left(n\right)}\sqrt{nh^{5}}+\mathrm{log}\left(n\right)\varepsilon_{n}+\delta_{n},\\
\kappa_{2,n}^{*}\coloneqq\left(\frac{\mathrm{log}\left(n\right)^{5}}{nh^{3}}\right)^{1/12}+\mathrm{log}\left(n\right)\varepsilon_{n}\textrm{ and }\kappa_{3,n}^{*}\coloneqq\left(\frac{\mathrm{log}\left(n\right)^{5}}{nh^{3}}\right)^{1/12}+\delta_{n}.
\end{gather*}
Then by (\ref{eq:Kolmogorov S}) and (\ref{eq:S_star G_U Kolmogorov distance}),
we have the following result.
\begin{thm}
\label{thm:nonparametric bootstrap non-studentized}Suppose that Assumptions
\ref{assu: DGP1}-\ref{assu:kernel} hold with $P=2$, the third-order
derivative functions in Assumption \ref{assu: DGP4}(a) are Lipschitz
continuous and $h\wasypropto n^{-\lambda}$ with $1/5<\lambda<1/3$,
\begin{multline*}
\mathrm{Pr}\left[f_{\varDelta\mid X}\left(v\mid x\right)\in\widetilde{\mathit{CB}}_{\mathsf{npb}}\left(v\mid x;\widehat{h}\right),\,\forall v\in I_{x}\right]=\left(1-\alpha\right)\\
+O\left(\left(\frac{\mathrm{log}\left(n\right)^{5}}{nh^{3}}\right)^{1/12}+\sqrt{\mathrm{log}\left(n\right)}\sqrt{nh^{5}}+\mathrm{log}\left(n\right)\varepsilon_{n}+\delta_{n}\right).
\end{multline*}
\end{thm}

\end{document}